\documentclass[10pt]{article}

\usepackage{amsmath,amssymb,amsthm,graphicx,hyperref}
\usepackage{physics} 
\usepackage{cite}
\usepackage{orcidlink}
\usepackage{float}
\usepackage{geometry}
\usepackage{mathrsfs}
\usepackage[english]{babel}
\usepackage{ragged2e}

\theoremstyle{definition}
\newtheorem{Definition}{Definition}[section]

\theoremstyle{proposition}

\theoremstyle{lemma}
\newtheorem{Lemma}{Lemma}[section]

\theoremstyle{theorem}
\newtheorem{Theorem}{Theorem}[section]

\theoremstyle{remark}
\newtheorem{Remark}{Remark}[section]

\theoremstyle{example}
\newtheorem{Example}{Example}[section]

\title{Notes on a Gaussian-Based Distribution Algebra for the Non-linear Wave Equation of the Shift Vector in Quantum Foam}
\date{November 30, 2025}
\author{Claes Cramer\thanks{Email: claes.cramer@algoarts.com}\space\space\orcidlink{0000-0001-5639-545X}}

\begin{document}
\maketitle

\begin{abstract}
We develop a non-linear distributional renormalisation algebra for Gaussian Quantum Foam, built from sequences of scaled Gaussians on spacelike hypersurfaces of homotopic, globally hyperbolic spacetimes and their distributional limits. The algebra is closed under multiplication and second-order differentiation, with all non-linear operations defined on smooth representatives before taking the limit. 

Applied to the non-linear scalar-field wave equation for the shift vector, the wave operator converges to a linear combination of the Dirac measure and its second-order derivative, encoding a sharply localised curvature impulse that displaces the vacuum; in the correspondence limit, the equation reduces to the massless Klein–Gordon equation. 

Classical singularities are replaced by a well-defined distributional structure: the scalar Ricci projection is non-negative on the singular support and converges to a positive linear combination of the Dirac measure and its second-order derivative while away from the support, in the emerging classical spacetime, the strong energy condition is violated on open sets. The trace of the extrinsic curvature, the mean curvature, and the null expansions vanish on the support (no trapped surfaces). For finite values of the sequence index, there exist open neighbourhoods in which both the inward and outward null expansions are strictly negative; thus, locally and in a classical context, trapped surfaces can occur in those regions.

The level sets of the global time function, together with their normal, become asymptotically null, yielding a limiting unstable characteristic hypersurface that fixes evolution by null data and forbids any extension
into chronology-violating regions.

Finally, it is argued that, within this framework, a gravity-induced spontaneous state reduction restores the Equivalence Principle in the emerging classical spacetimes.
\end{abstract}
\begin{itemize} \tiny
     \item[] \textbf{Keywords}: Distribution Algebra, Distribution Geometry, Quantum Gravity, Quantum Foam, Wave Equation, Emergence of Spacetime
\end{itemize}
\pagebreak
\section{Introduction}

In an early exploratory phase, including scaffolding, and in an attempt to model a foamy spacetime~\cite{cramer20251,cramer20252}, the shift vector field, which displaces spatial coordinates across Cauchy surfaces, was extended to the space of distributions. This displacement property suggested a quantum relation to coherent states and, together with its role as a proxy for the diffeomorphic deformation of local spatial grids across successive hypersurfaces, led to its identification as the geometrical remnant of Planck–scale quantum field oscillations that displace the vacuum. The present \emph{notes}\footnote{\tiny The term \emph{notes} is used to emphasise that the style of the work is a deliberate choice, coherently interweaving precise mathematical statements with physical reasoning and narrative to make the work more accessible.} develops these ideas into a full, self-contained framework of Quantum Foam and therefore supersedes those notes.

An intrinsic feature of the framework is that a wide class of globally hyperbolic spacetimes emerges from quantum fluctuations, without invoking an auxiliary inflaton field or modifying Einstein's equations. The framework admits a Gelfand triple and allows the shift vector to be quantised as a displacement operator, generating coherent states that drive inflation through vacuum geometry alone, from which classical spacetime emerges. This is aligned with a conjecture by Einstein that general relativity might provide the key to a complete quantum theory\footnote{\tiny To be somewhat more precise, here phrased in modern terms, Einstein stated in the introduction to \emph{Physicien et Penseur} (1953), a commemorative volume in honour of Louis de Broglie published in the series \emph{Les Savants et le Monde} by Albin Michel (Paris), that his efforts to complete general relativity were motivated by the conjecture that a meaningful classical field theory might hold the key to quantum gravity. See Pais for details~\cite{pais1982}.}.

However, a critical issue remained open in the framework for the Quantum Foam: the identification of a consistent non-linear field equation governing the dynamics of the shift vector field within its singular support, where the classical spacetime description breaks down and the notions of continuity and differentiability apply only in the distributional sense \footnote{\tiny In a naïve, fragmented, and in some parts incorrect attempt to model the Quantum Foam in 2022, I realised that the shift vector indeed satisfied a linearised massless Klein–Gordon equation and that it should be possible to quantise it as a bosonic field \href{https://arxiv.org/abs/2206.10417}{arXiv:2206.10417}.}. The core difficulty in addressing this question lies in the fact that products of distributions are generally ill-defined within classical Schwartz distribution theory; see, e.g.,~\cite{hormander1983, strichartz2003}. In general relativity, non-linearities and hence products arise, for example, in the curvature tensor through products of terms from the Levi-Civita connection. Analogously, when the wave operator acts on the shift vector in a \(3{+}1\) decomposition, it generates non-linear terms that involve products of the shift vector and its derivatives. These non-linearities present serious challenges for any direct formulation at the level of classical distribution theory, where such products are generally not well-defined.

Nevertheless, given that the construction of Quantum Foam is based on extending the shift vector, or more precisely its components, into a family of smooth functions that converge, in the distributional sense, to a limiting distribution, it is natural to expect the existence of a commutative and associative differential algebra for such distributions. This expectation is closely aligned with the framework developed by Colombeau~\cite{colombeau2000}. When applied to general relativity, the Colombeau approach relies not only on sequences of smooth functions but also on a local definition of distributions on manifolds.

The Quantum Foam framework likewise adopts a local definition of distributions, though it is explicitly constrained to the class of globally hyperbolic spacetimes. A result by Bernal and Sánchez~\cite{bernal2005} establishes that any such spacetime admits a smooth, regular time function, and hence a function with an everywhere timelike gradient. The level surfaces of this time function provide a foliation of the spacetime. Each such level surface is a Cauchy surface. The restriction of the metric field to the level surfaces is Riemannian, and the evolution of successive Cauchy surfaces is described by a positive lapse function.

The formulation of the Bernal–Sánchez theorem~\cite{bernal2005} shows that a smooth foliation can be chosen so that the shift vanishes. Nevertheless, the dependence on the shift vector in diffeomorphic scalar quantities, such as the trace of the extrinsic curvature, is preserved in Gaussian Quantum Foam even if one gauges the shift vector to zero by a foliation-preserving relative velocity (towards the hypersurface normal observers) diffeomorphism on the whole spacetime minus the singular support. In particular, the stretching and contraction of spatial grids, expressed through the relative velocity gauge, remain effects of the shift vector both in Quantum Foam and in the emerging classical spacetime. Consequently, the shift vector functions as a proxy for the Quantum Foam. Therefore, any objection that the Quantum Foam framework lacks full diffeomorphism invariance must be understood in this context.

In conclusion, we have a model that can be formulated with a foliation structure that is both geometrically meaningful and physically justified, since general relativity, when restricted to globally hyperbolic spacetimes, necessarily admits such a decomposition. One may ask whether the model can be extended to non–globally hyperbolic settings—for example, those admitting closed timelike curves? The answer is, no since, as we will see, the level surfaces converge to characteristics (null surfaces), while the slowness in the geometrical-optics sense converges to zero. Consequently, as the level sets of the global time function become asymptotically characteristic (null), there is an unstable pile-up of almost characteristic level surfaces at the limiting null hypersurface in the distributional geometry, which is highly susceptible to back-reaction from any polarisation effect. That is, evolution is then determined up to the initial characteristic, unstable, compactly generated Cauchy horizon by null data. Consequently, within this framework, no region of closed timelike curves can develop within the domain of the model; that is, the theory is effectively "frozen" at the characteristic in distribution geometry.

Returning to the question of distributions, Nigsch and Vickers have developed a fully diffeomorphism invariant theory of distributions and distribution geometry, based on the fundamental concept of smoothing operators~\cite{nigsch20201, nigsch20202}. Although the framework used here does not explicitly adopt the Colombeau algebra or the Nigsch-Vickers theory, it nevertheless provides a consistent mathematical foundation for defining products of distributions. This is achieved in a manner aligned with the Colombeau strategy, specifically the model delta net strategy, while ensuring compatibility with physical principles such as causality and locality. Specifically, by employing sequences of localised test functions that converge in the distributional sense, and, in particular, by selecting sequences of localised and normalised Gaussian functions and as such in a restriction of the Schwartz space, these difficulties can be effectively circumvented. Since the product of two Gaussian functions is again a Gaussian, the product of such sequences can be constructed to remain well-defined at every step (by modulating the product with the sequence index), and in the distributional limit, the product converges to a scaled Dirac measure. This ensures that the product remains meaningful throughout the procedure.

To be more explicit, this work defines a restricted and localised non-linear distributional renormalisation algebra on the hypersurfaces of a sequence of globally hyperbolic and homotopic spacetimes. The algebra is generated by sequences of localised Gaussian test functions and is closed under differentiation, multiplication, and products involving distributions of order at least two. All non-linear operations are carried out at the level of smooth representatives before taking the distributional limit. In this way, consistency with respect to non-regular directed points—and hence with the wave front sets of the resulting distributions—is maintained.

This setting provides the necessary structure for formulating the non-linear wave equation governing the shift vector field in the Quantum Foam regime, specifically, at Planck-scale geometries.

It is then shown that, within this framework, the non-linear wave operator applied to the components of the shift vector in Gaussian Quantum Foam (modelled as a Gaussian sequence) converges, in the sense of distributions, to a linear combination of the Dirac measure and its second-order distributional derivative. This result effectively identifies the Dirac measure as a fundamental solution to the non-linear wave equation in Gaussian Quantum Foam, as anticipated.

Crucially, the second-order derivative of the Dirac measure acts as a sharply localised curvature impulse: a singular quantum geometric source that initiates the oscillatory dynamics of the shift vector. From a physical standpoint, it functions as the catalyst, an acceleration impulse, that displaces the vacuum, triggers inflation, and drives the emergence of classical spacetime. The residual imprint of this displacement is encoded in a geometric shift across hypersurfaces, consistent with Wheeler’s conception of vacuum fluctuations~\cite{wheeler1955, mtw1973}.

In the weak-field regime, the non-linear wave equation reduces to the classical vacuum Klein–Gordon equation, thereby recovering standard scalar field dynamics as a limiting case. Thus, both the quantum foam and its classical geometric limit are unified within a single coherent framework.

It is important to note that, within this framework, the notion of a classical singularity does not retain operational meaning. The reason for this is due to the construction wherein the Quantum Foam element is the distributional limit of a sequence of globally hyperbolic and homotopic spacetimes, each endowed with smooth geometry and well-defined dynamics. The resulting Gaussian Quantum Foam element inherits a sharply localised but well-defined distributional structure in the singular support and hence in a distribution geometry. In this setting, curvature remains finite in the distributional sense, and the singular support serves not as an indicator of geometric breakdown but as a precise localisation of field content.

Nevertheless, as will be shown in a theorem, a precise statement about the sign of the timelike scalar projection of the Ricci curvature, and hence the strong energy condition, can still be made. At the singular support, the scalar projection of the Ricci curvature is non-negative for all finite values of the sequence index, and in the distributional limit it converges to a linear combination of a Dirac measure and its second derivative, positive in the singular support. Away from the singular support and for finite values of the sequence index, and consequently for classical spacetime elements, there necessarily exist open regions where the strong energy condition is locally violated. These regions are not pathological; they are essential. It is precisely this sign-structured curvature, governed by the non-linear dynamics of the shift vector, that drives the emergence of classical spacetime geometry from the quantum foam.

In connection with this, it will also be shown, in a second theorem on singularities, that the trace of the extrinsic curvature for the Cauchy hypersurfaces and for the embedded two-surfaces vanishes at the singular support, as does the mean curvature of the embedded two-surfaces. Consequently, the outward and inward null expansions vanish at the singular support, and no trapped surfaces arise or can exist in a Gaussian Quantum Foam. However, for finite values of the sequence index, there exist open neighbourhoods in which both the inward and outward null expansions are strictly negative. The implication is that, locally and in a classical context, trapped surfaces can occur within those regions, despite the absence of closed trapped surfaces in the limit.

Taken together, this allows one to argue that concepts such as naked singularities or naked black holes are without significance, at least within this framework of distributional geometric description of spacetime. What takes their place are the transient trapped surfaces, or primordial black holes, that arise at finite values for the sequence index in the emerging classical spacetime.

In summary, this work defines the algebraic and distributional foundation required to make sense of non-linear field dynamics in the context of Gaussian Quantum Foam. It constructs and analyses the associated wave equation, identifies its distributional fundamental solution, and shows how the emergence of classical spacetime follows naturally from the dynamics of the shift vector field. This makes the picture or rather the framework of vacuum fluctuations complete, offering a realisation of the curvature-generating mechanism at the heart of the Quantum Foam hypothesis.

Before proceeding to the construction of the algebra and after the notion of a Gaussian Quantum Foam has been discussed at some length to make this work self-consistent, we should also mention that the distributional limit of the Ricci curvature scalar, and hence the trace of the stress-energy tensor, for a Gaussian Quantum Foam takes the form of a linear combination of a Dirac measure and its second-order distributional derivative. This is consistent with the result that the scalar projection of the Ricci curvature takes the same form, since, as we shall see, the scalar projection of the stress-energy tensor vanishes at the singular support. 

This curvature structure is reminiscent of the findings of Nigsch and Vickers~\cite{nigsch20202}, who applied their non-linear theory of distribution geometry to compute the generalised scalar curvature in the context of a conical spacetime. Under their low regularity assumptions, that is, that the metric is continuously differentiable with locally Lipschitz continuous first derivatives, they showed that the resulting curvature is associated with a Dirac measure. 

In contrast, the Gaussian Quantum Foam model imposes no regularity constraints that would not allow it to entirely be well-defined in both a differential and distribution geometric setting sufficient for a formulation of General Relativity in both a classical and distributional context, since it is constructed from scaled sequences of Gaussians within Schwartz space that converge in the sense of distributions, and where the scaling enables well-defined non-linear products. This approach allows for curvature distributions beyond the Dirac measure, thereby capturing sharper local curvature impulses associated with vacuum displacement. Such fluctuations, though initially intense, decay over time, leaving behind residual imprints in the form of gravitational waves.

\section{Gaussian Quantum Foam} \label{sec:review}

While this work, in its early stages, was mostly concerned with the construction of the renormalised distributional algebra, designed specifically to formulate a non-linear wave equation for the shift vector and thereby capture the dynamical evolution of a spacetime-foam (Quantum Foam) element, it rather quickly evolved to a much more general purpose development of a complete distribution geometry for distributions with point support and hence distributions that can be expanded in the Dirac measure and its distributional derivatives. This, in turn, made it possible to explore Gaussian Quantum Foam in much greater depth and from a broader range of perspectives.

Therefore, it is reasonable to here establish the notion of quantum foam and more specifically \emph{Gaussian Quantum Foam}, including its definition and key properties. This establishes the geometric and distributional setting—a \emph{distribution geometry}—within which the field equation unfolds, setting the shift vector into oscillatory motion and, through it, displacing the vacuum.

More precisely, we revisit some earlier work on quantum foam and vacuum fluctuations, and then proceed to introduce the definition of Gaussian Quantum Foam\footnote{The notion was first introduced in \cite{cramer20251}, albeit in an incomplete, fragmented, and exploratory form. It was subsequently quantised in \cite{cramer20252}. The present work is complete and self-contained, so the earlier work is not required for this work.}. We discuss its essential properties and show that, even if the shift vector is gauged away, its effects are preserved under diffeomorphisms. It therefore serves as a proxy—a geometric remnant carrying the quantum information of a Quantum Foam element. Moreover, we show that this proxy property holds not only classically but also at the level of quantum fields, since the definition of Gaussian Quantum Foam, by construction, admits a Gelfand triple. This structure enables the formulation of a correspondence principle for quantum gravity and allows quantisation of the model as a bosonic field, by virtue of the existence of a Hilbert space within the Gelfand triple.

We then discuss the quantised model within the context of coherent states, chosen because they represent configurations as close as possible to classical geometries—thus respecting the correspondence principle encoded by the Gelfand triple—while their vacuum-displacement property naturally aligns with the role of the shift vector in displacing coordinates relative to the normal and to the vacuum configuration of the Quantum Foam element.

\medskip

Carlip has remarked that the concept of quantum fluctuations, and consequently vacuum fluctuations of the metric in quantum gravity remains vague and lacks a precise formulation~\cite{carlip2023}. In fact, no consensus has yet been reached regarding the topology or type of foamy spacetime on which such fluctuations should be defined. Nevertheless, as Carlip notes, at least two approaches have been explored: one based on Euclidean path integrals, and another on canonical gravity. The work herein offers yet another alternative by the construction of a distribution geometry from a sequence of globally hyperbolic spacetimes converging in distributions.

It may well be that we must wait for the completion of the still unfinished revolution of constructing a self-consistent and complete theory of quantum gravity (see Kiefer for comprehensive overviews~\cite{kiefer2023, kiefer2025}) before the concept attains a rigorous definition.
Nevertheless, vacuum fluctuations as a concept remain closely tied to Quantum Foam, originally introduced by Wheeler in the form of \emph{geons}: gravitational–electromagnetic entities whose mass arises solely from photons held in unstable circular orbits by the very gravity generated by their own energy~\cite{wheeler1955}. Although such configurations are unstable, they were proposed to resolve a deep dilemma: either one adheres strictly to classical general relativity—treating a “body” as a singularity in the metric field—or one postulates regularity and hopes that quantum gravity will ultimately explain it.  
If, however, a precise, stable, and self-consistent framework for Quantum Foam and hence the geon can be formulated, it might clarify the meaning of vacuum fluctuations and provide a physically consistent structure underlying the microscopic dynamics of spacetime.  

In more recent work, Carlip~\cite{carlip2019} examined how macroscopic averaging of Planck-scale fluctuations could conceal an enormous cosmological constant. He argued that Planck-scale fluctuations generate regions of extreme curvature which, when averaged, yield near-zero macroscopic curvature. More precisely, Carlip demonstrated the existence of a class of initial data with a local Hubble parameter that is huge at the Planck scale but small at macroscopic scales. This phenomenology closely matches the behaviour expected from a Quantum Foam exhibiting fluctuating extrinsic curvature—a transient structure that builds up and decays exponentially, reaching enormous values during its growth phase.

This insight aligns naturally with the Gaussian Quantum Foam framework developed here. We construct a sequence of homotopic  \(k\)-spacetimes that converge to the distributional structure of Quantum Foam, providing a mathematically rigorous bridge between the microscopic dynamics of spacetime fluctuations and the emergence of smooth classical geometries.  
In particular, we demonstrate that there exists a class of solutions to general relativity in which the local Hubble parameter—or equivalently, the extrinsic curvature-is enormous at the Planck scale, but becomes negligible at macroscopic scales. A similar relation holds for the projected energy density along Eulerian worldlines and for the Ricci scalar.  

A central distinction between Carlip’s approach and Gaussian Quantum Foam lies in the treatment of the shift vector. Carlip assumes a vanishing shift, attributing Planck-scale curvature to a hidden cosmological constant that remains unobservable macroscopically. In contrast, Gaussian Quantum Foam allows for a non-zero shift vector, which acts as a gravitational displacement operator. This generates significant local expansion, shear, and curvature at the Planck scale—effects that nevertheless average to zero macroscopically. The result is a natural mechanism for the emergence of a smooth spacetime without introducing a fundamental cosmological constant.

As has been discussed, the concept of Quantum Foam, remains somewhat vague, largely because the notion of vacuum fluctuations lacks precision. Clarifying this requires either an approximate theory describing Planck-scale physics or a self-consistent, testable theory of quantum gravity. The former is represented by the theory of quantum fields in curved spacetime, but where and crucially the metric field remains classical, which possesses a rigorous algebraic foundation in Lorentzian spacetimes \cite{kay2023}; the latter remains as we have remarked an unfinished revolution with many avenues.  

However, there exists a third alternative: implementing Wheeler’s idea of a self-gravitating entity that resolves the question of how a body acquires its observable mass without singularities in the metric field. Such a construction would make both the notion of vacuum fluctuation and that of Quantum Foam mathematically precise. It would also naturally constitute a formulation within quantum gravity.  

Modelling a geon requires satisfying the Cooperstock–Faraoni–Perry conditions~\cite{perry1999}:  
Einstein’s field equations must be solved self-consistently; the solution must be regular; the spatial geometry should remain stable across all timescales; and the gravitational field should approach asymptotic flatness at infinity.

Gaussian Quantum Foam satisfies all these requirements. Each spacetime in the convergent sequence is a smooth, globally hyperbolic solution, and it remains 'regular' even at the singular support—though now in the sense of distribution geometry.
Indeed, as we later show, the classical notion of a singularity loses meaning in differential geometry and is replaced by the rigorous concept of singular support and characteristics, well-defined microlocally. The Gaussian Quantum Foam construction is also stable across all timescales, and the smooth, regular time function foliates every element of the sequence. Furthermore, the construction itself guarantees asymptotic flatness.  

Beyond these structural properties, the model also reproduces the foamy characteristics Wheeler originally envisioned~\cite{wheeler1981} and that Jack~Ng~\cite{jackng2006}, in his phenomenological analysis, later emphasised: spacetime appears smooth at large scales but increasingly ``bubbly'' and ``foamy'' as one approaches the Planck scale.  These characteristics were summarised again in Carlip’s review~\cite{carlip2023}, which outlines a consistent phenomenological picture of spacetime’s microscopic structure.

Finally, any viable framework must be testable. Gaussian Quantum Foam satisfies this requirement in a straightforward way: it is defined as a sequence of smooth, classically and globally hyperbolic spacetimes that explicitly describe the evolution of spatial hypersurfaces in the early universe. Within this family, inflation and the formation of trapped surfaces, including configurations that are arbitrarily close to naked black holes, arise without additional assumptions.

To be somewhat more precise, the versatility of the framework, its ability to produce a wide class of globally hyperbolic spacetimes and the possibility of modulating the Gaussian with polynomials, to model features of the early universe, while not being tied to a theory whose domain stops at the Planck scale—naturally provides the means for both theoretical and numerical investigations. Concretely, to test the framework, one treats the sequence index  \(k\) as a phenomenological scale parameter to be constrained by observational data. For a fixed finite  \(k\), the corresponding spacetime element carries an observable geon monopole mass generated by vacuum polarisation in the Quantum Foam. One can then compute the relevant quantum and geometric quantities for the physical scenario: the Ricci trace; the projected scalar Ricci curvature along timelike and null directions and hence the corresponding trace and projections of the stress–energy tensor; the hypersurface geometry, including the extrinsic curvature and scalars such as the local Hubble parameter; and the induced two–metric on closed spacelike 2–surfaces and the associated null congruences. From these data one may determine the Hawking (or other quasi–local) mass, the horizon area and the associated entropy, and thereby map the formation and evaporation of transient trapped surfaces, study the distribution of positive and negative energy densities, and analyse high–frequency geometrical–optics propagation in the early universe. In particular, the finite– \(k\) trapped regions provide natural initial data for the study of early compact–object formation.

Moreover, as remarked by Ralegankar, Perri, and Kobayashi, an early matter–dominated era can arise from a coherent scalar–field condensate (often modelled as an inflaton) \cite{ralegankar2025}. In the present setting, the analogous role is played instead by the expectation values of the quantised shift–vector field in coherent states: they generate inflation without an inflaton, displace the vacuum, and create neighbouring regions of positive and negative energy density through harmonic oscillations driven by the non–linear field equation that is central to this work. Hence, it should in principle be possible to use the framework introduced here as a model for the onset of the early matter–dominated era.

On a final note concerning the experimental side, and in particular the discussion of singularities and transient trapped surfaces, there may also be scope to use the present framework to model and understand the rapid emission of energetic Hawking radiation from primordial black holes (PBHs). This would be along the lines suggested in the recent work of Klipfel and Kaiser \cite{klipfel2025}, where ultra–high–energy cosmic rays are proposed as signatures of Hawking evaporation in the final stage of a PBH’s transient existence, and may be related to the singularity structure discussed in Section~\ref{sec:singularities}.

This concludes the background discussion of a foamy spacetime at the Planck scale. In the remainder of this work, we construct the mathematical framework that renders the notion of Quantum Foam precise.
\newline
\newline
\noindent The definition of Gaussian Quantum Foam is based on a sequence of globally hyperbolic and homotopic (in the sense of linear interpolation in the chosen coordinates) spacetimes in which the shift-vector components are smooth and localised Gaussian functions in Schwartz space, converging in the sense of distributions to Dirac measures. 

To understand the meaning of the notion of a localisation in relation to Schwartz space on a local open set of a sequence of spacetimes, recall that a Schwartz function is a rapidly decreasing smooth function on the whole of \(\mathbb{R}^n\) such that the product of any polynomial with any derivative of the function remains bounded. The space of such functions, denoted \(\mathcal{S}(\mathbb{R}^n)\), is a space of smooth, real-valued functions on \(\mathbb{R}^n\) with bounded semi-norms. Equipped with these semi-norms, \(\mathcal{S}(\mathbb{R}^n)\) becomes a Fréchet space~\cite{hormander1983}.  

To restrict and localise the Schwartz space to any local open set
 \(U_{\kappa^{(k)}_t} \subset \mathbb{R}^3\) for a local coordinate system  \(\kappa\) on a
Cauchy surface  \(\Sigma^{(k)}_t\), we use Theorem~1.4.1 in Hörmander~\cite{hormander1983},
which guarantees that for any open set  \(U_{\kappa^{(k)}_t}\) and any compact set
 \(K_{\kappa^{(k)}_t} \subset U_{\kappa^{(k)}_t}\), there exists a compactly supported test
function  \(\phi^{(k)}_c \in C^\infty_0(U_{\kappa^{(k)}_t})\) with
 \(0 \leq \phi^{(k)}_c \leq 1\), such that  \(\phi^{(k)}_c = 1\) in a neighbourhood of
 \(K_{\kappa^{(k)}_t}\). Such a function is called a \emph{cut-off function}.

Furthermore, by Lemma~7.1.8 in~\cite{hormander1983}, the space of compactly supported test
functions is dense in the Schwartz space. Using a cut-off function, we may therefore localise
elements of the Schwartz space, and in particular the set of Gaussians  \(\beta^i_{(k)}\), to
the Schwartz space on the open set  \(U_{\kappa^{(k)}_t}\) as follows:
\[
\beta^i_{(k),\phi^{(k)}_c} := \phi^{(k)}_c \,\beta^i_{(k)}.
\]
This construction provides a rigorous definition of the localisation of elements in a
restriction of the Schwartz space to  \(\mathcal{S}_{\mathcal{G}}(U_{\kappa^{(k)}_t})\) in
Definition~\ref{def:quantumfoam} below. Having established this, we will henceforth drop the
index  \(\phi^{(k)}_c\) and simply write  \(\beta^i_{(k)}\), with the understanding that each
Gaussian in~\eqref{eq:gaussian} is modulated by a compactly supported cut-off function. This
modulation resolves any concern as to whether the sequence of Schwartz functions used here is
well defined, in the limit, as a local distribution on a manifold in the sense of
Definition~6.3.3 in~\cite{hormander1983}.

We are now ready to proceed to the formal definition of Gaussian Quantum Foam.

\begin{Definition}[Gaussian Quantum Foam]\label{def:quantumfoam}
Let  \(\kappa^{(k)}: O_{\kappa^{(k)}} \subset M_{(k)} \to U_{\kappa^{(k)}} \subset \mathbb{R}^4\)
be a local coordinate chart on any homotopic  \(k\)-spacetime  \((M_{(k)}, g_{\mu\nu}^{(k)})\) in
a sequence  \(\{(M_{(k)}, g^{(k)}_{\mu\nu})\}_{k \in \mathbb{N}}\) of homotopic, globally
hyperbolic spacetimes, and denote by  \(\kappa^{(k)}_t = \kappa^{(k)}|_{\Sigma_t^{(k)}}\) its
restriction to a Cauchy surface
 \(\Sigma_t^{(k)}=\{p\in M_{(k)}: t_{(k)}(p)=t, \, t\in \mathbb{R}\}\).

\noindent Let the line element for any spacetime element in the sequence be given by
\begin{equation}\label{eq:gaussianmetric}
ds_{(k)}^2 = -N_{(k)}^2\, dt^2 + \eta_{ij} (dx^i + \beta_{(k)}^i\, dt)(dx^j + \beta_{(k)}^j\, dt),
\end{equation}
in the local coordinate chart with coordinates  \((t, x^1, x^2, x^3)\) (that is, Cartesian
coordinates  \((t, x, y, z)\)). Here,  \(N_{(k)}\) is the lapse function, which may depend on  \(k\)
but is independent of spacetime coordinates, and  \(\eta_{ij}\) is the induced Euclidean metric on
 \(\Sigma_t^{(k)}\). The vector  \(\beta_{(k)}\) is the shift vector field tangent to
 \(\Sigma_t^{(k)}\). In the chart  \(\kappa_t^{(k)}\), its components
\[
\beta_{(k)} := (\beta_{(k)}^1,\beta_{(k)}^2,\beta_{(k)}^3)
\]
are functions on  \(U_{\kappa_t^{(k)}}\) and each belongs to the Schwartz space of localised
Gaussians  \(\mathcal{S}_{\mathcal{G}}(U_{\kappa_t^{(k)}})\), with
\begin{equation}\label{eq:gaussian}
\beta_{(k)}^i:=\beta_{(k)}(x^i) := \frac{1}{\sqrt{4\sigma^2/k^2\,\pi}}\,
\exp\!\left(-\frac{(x^i)^2}{4\sigma^2/k^2}\right),\ \sigma\geq 1.
\end{equation}
Here  \(\sigma\) is a localisation parameter controlling the range in which the effect of the
shift vector cannot be neglected.

\noindent The distributional limit of the sequence
 \(\{(M_{(k)}, g^{(k)}_{\mu\nu})\}_{k \in \mathbb{N}}\), defined by the convergence
 \(\beta_{(k)}^i \to \tilde{\beta}^i\) in the sense of distributions on  \(U_{\kappa_t^{(k)}}\) as
 \(k\to\infty\) (i.e., each  \(\beta_{(k)}^i\) converges to the Dirac measure  \(\delta_{x^i}\)),
determines a localised, distribution-valued spacetime element
\[
(\tilde{M},\tilde{g}_{\mu\nu})_{\mathcal{G}} := \lim_{k\to\infty}(M_{(k)},g^{(k)}_{\mu\nu}),
\]
which is called a \emph{Gaussian Quantum Foam}.
\end{Definition}
\begin{Remark}[Dirac Measure in Level Surfaces]
\noindent It is important to emphasise that, in Definition~\ref{def:quantumfoam}, each component
 \(\beta_{(k)}^i\) is defined from a \emph{one-dimensional} Gaussian in the corresponding spatial
coordinate  \(x^i\). Thus, after localisation to the space  \(\mathcal{S}(U_{\kappa^{(k)}_t})\)
defined above, we have  \(\beta_{(k)}^i \to \delta_{\Sigma_i}\) in  \(\mathcal{S}'(U_{\kappa^{(k)}_t})\),
where
\[
\Sigma_i:=\{x^i=0\}\cap U_{\kappa^{(k)}_t},
\]
and  \(\delta_{\Sigma_i}\) denotes the Dirac measure supported on the level surface  \(\Sigma_i\).

For notational simplicity, we have in the definition made the choice to use \(\delta_{x^i}\) in place of \(\delta_{\Sigma_i}\) (the hypersurface Dirac supported on \(\Sigma_i\)). Hence the \emph{singular support} of each limit distribution is the coordinate plane \(\Sigma_i\). When all three components are considered together in the vector \(\beta\), the \emph{joint singular support} (meaning the union of the component singular supports) is
\[
\operatorname{sing\,supp}\beta = \Sigma_x \,\cup\, \Sigma_y \,\cup\, \Sigma_z,
\]
while the common intersection is the single-point set \(\{0\}=\Sigma_x\cap\Sigma_y\cap\Sigma_z\). This should not be confused with the case where each component is defined from a spherically symmetric three-dimensional Gaussian, which would converge to \(\delta^{(3)}_{\mathbf{x}}\) and thereby alter both the scaling and the singular-support structure.
\end{Remark}
The choice of one-dimensional Gaussians also ensures that the quantised theory, as we will see, naturally factorises into three independent Hilbert spaces and a corresponding Fock space built as their tensor product.
\medskip
We now show that any odd test function \(\phi\in\mathcal{S}(U_{\kappa^{(k)}_t})\) lies in the orthogonal complement of the Gaussian sector in the Gelfand triple
\[
\mathcal{S}_{\mathcal{G}}(U_{\kappa^{(k)}_t})\;\subset\;\mathcal{H}_{(k)}\;\subset\;\mathcal{S}_{\mathcal{G}}'(U_{\kappa^{(k)}_t})
\]
naturally admitted by the Gaussian Quantum Foam. In other words, odd test functions cannot be used, within the distribution geometry of the Gaussian Quantum Foam, to measure the remnant of the quantum information proxied by the shift vector, neither in the emerging classical spacetimes in Definition~\ref{def:quantumfoam} nor in the quantum field theory constructed later in this section, where the shift is an operator-valued distribution acting on states in the Fock space built from \(\mathcal{H}_{(k)}\).

Indeed, for each finite \(k\), each component of the Gaussian shift vector \(\beta^i_{(k)}\) is positive and even in \(x^i\) (and, in particular, locally concave in \(x^i\) on a small neighbourhood of the hypersurface \(\Sigma_i\)). It then follows that
\[
\langle \beta^i_{(k)},\phi\rangle=0 \qquad\text{for all odd }\phi\in\mathcal{S}(U_{\kappa^{(k)}_t}).
\]
The same holds for any scaled Gaussian product in the restricted space: defining
\[
g^{\,n}_{(k)}:=\frac{(\beta^i_{(k)})^n}{k^{\,n-1}}\in\mathcal{S}_{\mathcal{G}}(U_{\kappa^{(k)}_t}),
\]
we have 
\[
\big\langle g^{\,n}_{(k)},\phi\big\rangle
=\big\langle \tfrac{1}{k^{\,n-1}}(\beta^i_{(k)})^n,\phi\big\rangle
=0 \qquad\text{for all odd } \phi\in\mathcal{S}(U_{\kappa^{(k)}_t}),
\]
since the product of Gaussians is Gaussian and even. The scaling by \(1/k^{\,n-1}\) ensures convergence in distributions: \(g^{\,n}_{(k)}\to \delta_{x^i}\) as \(k\to\infty\) (a hypersurface Dirac distribution supported on \(\Sigma_i\)). This observation will be crucial for the development of the renormalised distribution algebra in Section~\ref{sec:algebra}. 

Passing to the distributional limit that defines \(\mathcal{S}_{\mathcal{G}}'(U_{\kappa^{(k)}_t})\) and using continuity of distributions yields
\[
\lim_{k\to\infty}\big\langle g^{\,n}_{(k)},\phi\big\rangle=0
\qquad\text{for all odd }\phi.
\]
In conclusion, all odd test functions belong to the orthogonal complement of each space in the triple with respect to the (even) Gaussian sector. In this sense, “measurement by a test function” is trivial for odd probes: any odd \(\phi\) returns zero on the Quantum Foam and hence in the distribution geometry. Consequently, any non-trivial measurement necessarily requires the test function to have a non-zero projection onto the Gaussian sector \(\mathcal{S}_{\mathcal{G}}(U_{\kappa^{(k)}_t})\). In conjunction with this, and to obtain sensible results from any measurement, we expect an admissible probe (measurement device) not only to be positive and even but also locally concave near the singular support of each component \(\beta^i_{(k)}\), since each component is positive, even, and locally concave in \(x^i\) near \(\Sigma_i\). Thus, henceforth we restrict the space of test functions to elements that mirror these properties. Specifically, when probing component \(i\) we take \(\phi\in\mathcal{S}(U_{\kappa^{(k)}_t})\) to be \emph{positive}, \emph{even} in \(x^i\), and \emph{concave} in \(x^i\) on a small neighbourhood of the singular support (so \(\partial_i\phi|_{x^i=0}=0\) and \(\partial_i^2\phi|_{x^i=0}\le 0\)). 

\medskip
\noindent This view captures the essence of applied distribution theory. To measure any physical quantity we must use a class of measurement devices whose very structure is tailored to the quantity being measured. Any such device has a characteristic size and hence non-zero support, and locally it must possess properties that do not distort the quantity. Crucially, since any device occupies a non-zero volume element, any measurement is a functional (average) over that element. In the context of the Gaussian Quantum Foam, there is therefore no alternative but to use test functions with the same qualitative characteristics as the foam to probe its properties, both at Planck scales and as well in the later emergence of classical spacetime. This perspective aligns with Strichartz’s view on measurement, test functions, and physical quantities as local linear functionals~\cite{strichartz2003}\footnote{\tiny This also resonates with the view I put forward in early work from 2022; see \href{https://arxiv.org/abs/2206.10417}{arXiv:2206.10417}.}.
To formalise this view we introduce the following definition of test function space in Gaussian Quantum Foam that will be used throughout this work.
\begin{Definition}[Admissible test functions in the static Gaussian sector] \label{def:test-functions}
Fix \(i\in\{1,2,3\}\) and \(\Sigma_i:=\{x^i=0\}\cap U_{\kappa^{(k)}_t}\). 
Let \(\mathcal{S}(U_{\kappa^{(k)}_t})\) be the localised restricted spatial Schwartz space on the slice.
A test function \(\psi(t,\mathbf{x})=\chi(t)\phi(\mathbf{x})\in\mathcal{S_G}(U_{\kappa^{(k)}})\) is said to be \emph{admissible for component \(i\)} if
\[
\chi\in C_0^\infty((0,1)),\quad \int_0^1\chi(t)\,dt=1,\qquad 
\phi\in\mathcal{S}(U_{\kappa^{(k)}_t})
\]
and the spatial factor \(\phi\) satisfies, in a neighbourhood of \(\Sigma_i\),
\[
\phi\ge 0,\qquad \phi(\ldots,-x^i,\ldots)=\phi(\ldots,x^i,\ldots),\qquad 
\partial_i^2\phi\le 0\ (\text{local concavity in }x^i).
\]
\end{Definition}
The sequence index \(k\) serves two purposes: it drives the shift vector toward its distributional limit, and it simultaneously scales the geometry down to the Planck scale. In this way, \(k\) encodes both the convergence process and the underlying dynamical scale. The components of the shift vector converge to Dirac measures in the sense of distributions as \(k \to \infty\).

By contrast, the sequence of lapse functions \(\{N_{(k)}\}_{k \in \mathbb{N}}\) is treated differently. Rather than requiring convergence, we allow the lapse to scale with \(k\), both to preserve the consistency of the distributional limit and to probe the physical effects of temporal variation on hypersurface separation. In particular, if the lapse scales as \(N_{(k)} = k\), then the proper-time (Eulerian) separation between neighbouring hypersurfaces in the \(3{+}1\) decomposition of the sequence defined in Definition~\ref{def:quantumfoam} is \(d\tau_k = N_{(k)}\,d t = k\,dt\). This choice of scaling consistently guarantees distributional integrity for scalar curvature quantities, such as the projected Ricci scalar, ensuring that these remain well defined in the Gaussian Quantum Foam model.

With this scaling, the Lorentzian signature of each spacetime element \((M_{(k)}, g^{(k)}_{\mu\nu})\) in Definition~\ref{def:quantumfoam} is preserved for finite  \(k\). Indeed,
\[
N_{(k)}^2>\beta^{(k)}_i\beta_{(k)}^{\,i},\quad \forall k\in\mathbb{N},
\]
since from \eqref{eq:gaussian} we have
\[
\beta^{(k)}_i\beta_{(k)}^{\,i}
=\frac{k^2}{4\sigma^2\pi}\!\left(\exp\!\left(-\frac{k^2x^2}{2\sigma^2}\right)+\exp\!\left(-\frac{k^2y^2}{2\sigma^2}\right)+\exp\!\left(-\frac{k^2z^2}{2\sigma^2}\right)\right)
\leq \frac{3k^2}{4\sigma^2\pi}
< k^2,
\]
and \(\sigma\ge 1\) by definition of the foam.

In the distributional limit, this expression becomes singular. However, since the line element has no operational meaning in the singular support  \(\{0\}\), this does not introduce physical inconsistency.

We can achieve a further understanding of this and the role of the lapse  \(N_{(k)}\) by recalling that, in a suitable Quantum-Foam–preserving diffeomorphism, that will be introduced later in this section, its square is equal to the inverse of the squared norm of the gradient of the global and regular time function  \(t_{(k)}\) (see, e.g., \cite{bernal2005}), which provides the foliation of each spacetime element in the sequence converging to a Gaussian Quantum Foam element in Definition~\ref{def:quantumfoam}. That is,
\begin{equation}
    N_{(k)}^2 = \frac{1}{\|\nabla t_{(k)}\|_{g^{(k)}}^{\,2}}.
\end{equation}
When the lapse scales as  \(N_{(k)} = k\), this imposes the following constraint on the time function:
\begin{equation} \label{eq:sqreikon}
    \|\nabla t_{(k)}\|_{g^{(k)}}^{\,2} = \frac{1}{k^2}.
\end{equation}
Equation~\eqref{eq:sqreikon} is the squared form of the eikonal equation from geometrical optics. In analogy with Fermat's principle, the solution represents the evolution of the emerging spacetime from one hypersurface to the next along a path where its slowness \(v^{-1}_{(k)}\) is the least, here \(k^{-1}\), see e.g., \cite{leuhoucq2025} for a more comprehensive and pedagogical discussion of the notion of slowness in physics. 

Besides this, and considering the evolution of the sequence of spacetimes converging to Gaussian Quantum Foam, will the eikonal equation allow us to understand the hypersurface evolution towards this limit without actually solving the non-linear partial differential eikonal equation. Specifically, we note that the slowness decreases as the index  \(k\) increases and in conjunction with that the timelike normal gradient one-form  \(\partial_\mu t_{(k)}dx^\mu\) to each level surface and hence to each spacelike Cauchy surface  \(\Sigma_t^{(k)} = \{ p \in M_{(k)} : t_{(k)}(p) = t \in \mathbb{R} \}\) converges to null in the limit as  \(k\to\infty\). The same is true for the level surfaces, which converge to characteristic (null) hypersurfaces. From the perspective of the Chronological future  \(I^{+}_{g^{(k)}}(p)\), that is events reachable by timelike curves starting at a point \(p\in\Sigma^{(k)}_t\) while  \(J^{+}_{g^{(k)}}(p)\) denotes the causal future then for  \(k>\ell\) and all  \(p\), we have
\[
I^{+}_{g^{(\ell)}}(p)\;\subset\; I^{+}_{g^{(k)}}(p)\;\subset\; J^{+}_{g^{(k)}}(p).
\]
To summarise, the smooth and regular time function  \(t_{(k)}\) that defines a foliation, through its level surfaces, of each spacetime element in the sequence satisfies an eikonal equation. In the Quantum Foam limit and hence in distribution geometry, these level surfaces converge to characteristic (null) surfaces and the slowness converge to zero. That is, the lapse function  \(N_{(k)}\), which arises naturally in the foliation of any globally hyperbolic spacetime and, in the quantum-gravitational context, within the  \(3{+}1\) decomposition of general relativity, not only determines the dynamical evolution of the hypersurfaces in quantum foam, but also guarantees distributional integrity, controls the slowness in the eikonal equation of the time function, and in fact the formation of the geon.

From this perspective, the evolution of time is not assumed but \emph{triggered} at the singular support, where the shift vector, as will be shown, is set into motion by the second-order distributional derivative  \(\delta^{\prime\prime}\) appearing in the wave equation. Thus, the emergence of time is kick-started by a sharply localised curvature impulse.

In the present model, the Quantum Foam element, and hence a distributional structure with singular support, represents the definition of a physical body: a locus of non-zero field content, encoded entirely in the distributional structure of a shift vector supported on a single-point set. This structure realises Wheeler’s notion of a geon~\cite{wheeler1955}, the third option and a resolution to his dilemma: either accept singularities in the metric, or postulate field regularity everywhere (even in the neighbourhood of a physical body with mass and spatial position), and hope that a full quantum theory of gravity will explain how this is possible.

This stands in contrast to the classical view of a body as a singularity, where the notion of a singularity cannot be expressed as a point or as a specific location in a Lorentzian spacetime, but only implicitly, e.g.\ via the existence of incomplete and inextensible causal geodesics; see Wald~\cite[Chapter 9]{wald1984}. In Section~\ref{sec:singularities} we return to this issue and analyse whether the notion can be given any meaning in the locus defined above, and thus within the distributional structure, where both continuity and differentiability remain well defined in the sense of distributions.

Finally, note that by allowing the Gaussian to depend solely on spatial coordinates, keeping the lapse independent of time (though scaled with \(k\)), and endowing the three-dimensional submanifold with a Euclidean metric, we obtain a model whose geometry is fully determined by the spatial coordinates, the shift vector and its derivatives, and the lapse scaling.

Before closing this section, we need to discuss in some length why the shift vector is an adequate proxy for Quantum Foam, representing Wheeler's Geon in a form that is regular and well defined in the sense of distributions. Although the shift vector is not coordinate-independent in general relativity, nor required in the foliation of a globally hyperbolic spacetime as shown by Bernal and Sánchez~\cite{bernal2005}, the stretching and contraction of any spatial grid across the foliation is nevertheless diffeomorphically preserved. Indeed, the dependence of the shift vector in the trace of the extrinsic curvature carries over unchanged under the diffeomorphism that gauges the shift vector to zero, yielding a metric in the Bernal–Sánchez form~\cite{bernal2005}. This statement remains valid both classically and in the context of local quantum fields. We will demonstrate this explicitly by considering a sequence of spacetimes converging to a Gaussian Quantum Foam element, and constructing a diffeomorphism in the relative velocity gauge that sets the shift vector to zero. Despite this, the stretching and contraction of spatial grids remain governed by the shift vector. To see this, recall that the line element \eqref{eq:gaussianmetric} in Definition~\ref{def:quantumfoam} in a coordinate chart \(\kappa_{(k)}\) with coordinates \((t,x,y,z)\) is given by
\[
ds_{(k)}^2 = -N_{(k)}^2\, dt^2 + \eta_{ij} (dx^i + \beta_{(k)}^i\, dt)(dx^j + \beta_{(k)}^j\, dt).
\]
Introduce a diffeomorphism \(\Phi_{(k)}\) by 
\[
 \kappa_{(k)} \circ \Phi_{(k)}^{-1} :
 \Phi_{(k)}(O_{\Phi_{(k)}}\cap O_{\kappa_{(k)}}) \;\longrightarrow\;
 \kappa_{(k)}(O_{\Phi_{(k)}}\cap O_{\kappa_{(k)}}).
\]
Explicitly, take \(\Phi_{(k)}\) to be a foliation-preserving diffeomorphism
\begin{equation} \label{eq:diffeomorphism}
\Phi_{(k)} : (t,x^1,x^2,x^3)\;\mapsto\;(t',x'^1,x'^2,x'^3),
\end{equation}
as a coordinate transformation representing a relative velocity gauge toward orthonormal observers
\begin{equation}\label{eq:dtransformation}
t'=t, \qquad x'^i =  \Phi^i_{(k)}:=\Phi_{(k)}(x^i, t),\quad i=1,2,3.
\end{equation}
with the flow equation,
\begin{equation}\label{eq:dflow}
\partial_t \Phi^{-1 \, i}_{(k)} =
 -\beta^i_{(k)}\big(\Phi_{(k)}^{-1 \ i}(x^{\prime},t)\big)=-\beta_{(k)}^i.
\end{equation}
Here we exclude the singular support \(\{0\}\) from any open set, as this ensures the existence of the diffeomorphism; otherwise the flow equation \eqref{eq:dflow} is ill-defined at \(\{0\}\) in the distributional limit. 
Using the coordinate transformation \eqref{eq:dtransformation} and the flow equation \eqref{eq:dflow} it follows that
\[
dx^i= \partial_{x^{\prime j}}\Phi_{(k)}^{-1\ i}dx^{\prime j} + \partial_t\Phi^{-1\ i}_{(k)}\,dt 
   = \partial_{x^{\prime j}}\Phi_{(k)}^{-1\ i}dx^{\prime j} - \beta^i_{(k)}\big(\Phi_{(k)}^{-1}(x',t)\big)\,dt,
\]
and hence
\[
\partial_{x^{\prime j}}\Phi_{(k)}^{-1\ i}dx^{\prime j} = dx^i + \beta^i_{(k)}\,dt.
\]
The line element \eqref{eq:gaussianmetric} then takes the form
\[
ds_{(k)}^2 = -N_{(k)}^2\, dt^2 + \eta_{mn}\,\partial_{x^{\prime i}}\Phi_{(k)}^{-1\ m}\,\partial_{x^{\prime j}}\Phi_{(k)}^{-1\ n}\,dx^{\prime i}dx^{\prime j}.
\]
The form here is the same as in the theorem of Bernal and Sánchez~\cite{bernal2005}, and we also notice that in the diffeomorphism \(\Phi_{(k)}\), where the shift vector has been gauged to zero, the induced Riemannian metric \(h^{\prime (k)}_{ij}\) is given by
\begin{equation} \label{eq:shiftfreehij}
    h^{\prime \, (k)}_{ij}=\eta_{mn}\,\partial_{x^{\prime i}}\Phi_{(k)}^{-1\ m}\,\partial_{x^{\prime j}}\Phi_{(k)}^{-1\ n}.
\end{equation}
\noindent Any stretching or contraction of any spatial grid on the hypersurfaces is described by the trace of the extrinsic curvature, where the extrinsic curvature, in the relative velocity gauge, where the shift vector is gauged away is given by
\[
K^{\prime(k)}_{ij}= -\frac{1}{2N_{(k)}}\partial_t h^{\prime(k)}_{ij}.
\]
That leads to the following trace
\begin{equation} \label{eq:ktrace}
    K^{\prime (k)}=h^{\prime \ ij}_{(k)}K^{\prime \ (k)}_{ij}
    = -\frac{1}{2N_{(k)}}\,h^{\prime \ ij}_{(k)}\partial_t h^{\prime \ (k)}_{ij}.
\end{equation}
Now, using the diffeomorphism \eqref{eq:dtransformation} and the flow equation \eqref{eq:dflow} it follows that
\[
\partial_t\big(\partial_{x^{\prime i}}\Phi_{(k)}^{-1\ j}\big) 
=-\partial_{x^{\prime i}}\Phi_{(k)}^{-1\ p}\,\partial_p\beta^j_{(k)},
\]
then we have that
\[
\partial_t h^{\prime (k)}_{ij}=-\eta_{mn}\left(\partial_{x^{\prime j}}\Phi_{(k)}^{-1\ n}\partial_{x^{\prime i}}\Phi_{(k)}^{-1\ p}\,\partial_p\beta^m_{(k)}+\partial_{x^{\prime i}}\Phi_{(k)}^{-1\ m}\partial_{x^{\prime j}}\Phi_{(k)}^{-1\ p}\,\partial_p\beta^n_{(k)} \right).
\]
Recall that the induced metric and its inverse are diagonal and that the shift vector \(\beta_{(k)}^i\) is dependent only on \(x^i\). Using these properties then it follows that the trace of the above and hence the trace of the extrinsic curvature \eqref{eq:ktrace} is given by
\begin{equation}\label{eq:tracekprime}
K^{\prime (k)} = \frac{1}{N_{(k)}}\Big(\partial_x \beta_{(k)}^x + \partial_y \beta_{(k)}^y + \partial_z \beta_{(k)}^z\Big)=\frac{1}{N_{(k)}}\partial_i\beta_{(k)}^i.
\end{equation}
This should be compared with the trace in the coordinates in the definition \ref{def:quantumfoam} of Gaussian Quantum Foam, where the induced metric field is just the Euclidean metric \(\eta_{ij}\) and hence static, and thus where the trace reduces to
\begin{equation}\label{eq:tracek}
K^{(k)}=\frac{1}{2N_{(k)}}\eta^{ij}\left(\partial_i\beta^{(k)}_j+\partial_j\beta^{(k)}_i\right)=\frac{1}{N_{(k)}}\partial_i\beta_{(k)}^i=K^{\prime (k)}.
\end{equation}
Thus, we conclude that if \(\{(M_{(k)},g^{(k)}_{\mu\nu})\}_{k\in\mathbb{N}}\) is a sequence of globally hyperbolic spacetimes converging in the sense of Definition~\ref{def:quantumfoam} to a Gaussian Quantum Foam element. Then
if the shift vector \(\beta^i_{(k)}\) is gauged to zero by the foliation-preserving diffeomorphism \eqref{eq:dtransformation} and \eqref{eq:dflow}, the dependence of scalar quantities such as the trace of the extrinsic curvature on \(\beta^i_{(k)}\) is preserved in the \(3{+}1\) decomposition. In particular, the stretching and contraction of spatial grids, expressed through the relative velocity, remain effects of the shift vector. Consequently, the shift vector functions as a proxy for the Quantum Foam and, in the classical limit, represents a remnant of the quantum information contained in Wheeler’s Geon as a self gravitating quantity—or here, a Gaussian Quantum Foam element as defined in Definition~\ref{def:quantumfoam}.

In passing, we remark that the form of the extrinsic curvature, and the fact that it is proportional to the local Hubble parameter, implies that there exists a class of solutions to general relativity in which the local Hubble parameter, or equivalently the extrinsic curvature, is enormous at the Planck scale but becomes negligible at macroscopic scales aligned with the statement of Carlip \cite{carlip2019}. This follows immediately from its dependence on the divergence of the shift vector and from the fact that the shift vector, or strictly its components, converge in the sense of distributions to a Dirac measure.

From this classical role, we now turn to the quantum perspective. Returning to Definition~\ref{def:quantumfoam}, we note that it naturally admits a Gelfand triple on each hypersurface, comprising a Hilbert space \(\mathcal{H}(U_{\kappa^{(k)}_t})\), densely embedded between a test function space \(\mathcal{S}(U_{\kappa^{(k)}_t})\) and its dual \(\mathcal{S}^{\prime}(U_{\kappa^{(k)}_t})\) of linear distributions. This structure is compatible with the fact that the sequence of shift vectors \(\beta^i_{(k)}\), regarded as test functions in each component, converges in the sense of distributions. The shift vector can therefore serve as a proxy not only for the residual quantum information in a classical, geometric setting, but also in a fully quantum one. Within the framework of the Gelfand triple, a Hilbert space can be defined that provides the structure needed to quantise the shift vector as a bosonic field on the hypersurfaces.

We remark that, given the construction of a Gaussian Quantum Foam element in Definition~\ref{def:quantumfoam} as a sequence of smooth, homotopic and globally hyperbolic spacetimes with a shift vector or more precisely, whose components of the shift vector converge to a Dirac measure, then, once a Gelfand triple is admitted, any quantum field theory constructed on this basis is not merely a theory on a single spacetime, but a simultaneous specification of a single quantum field theory on an embedding of globally hyperbolic spacetimes \cite{kay2023}.

To make this statement precise, it is sufficient to consider that the sequence \((M_{(k)}, g^{(k)}_{\mu\nu})_{k\in\mathbb{N}}\) in Definition~\ref{def:quantumfoam} provides a causality-preserving embedding. Secondly, since the sequence admits a Gelfand triple, it realises a Hilbert space for each \(k\) and consequently a well-defined Fock space on which the quantised, operator-valued shift-vector distribution acts.  This means that the quantisation prescription is aligned with the algebraic viewpoint on quantum field theory, see e.g. \cite{kay2023}, and hence provides a single, simultaneous theory over the whole embedded sequence.

Let us now proceed with the actual construction of the quantum field theory. In the Schrödinger picture, we can expand the shift vector on each Cauchy surface as an ``infinite sum of decoupled harmonic oscillators'' in the limit represented by a Fourier integral. Each oscillator mode is quantised in the standard way as a harmonic oscillator, with creation and annihilation operators satisfying the canonical commutation relations. This expansion necessarily promotes the shift vector to an operator-valued distribution; see~\cite{haag1996} for the broader context of local quantum fields.
The fact that each shift vector component is only depending on one spatial coordinate means that the Hilbert space is a direct tensor product of three independent vacuum Hilbert spaces,
\[
\mathcal{H}^{(k)}=\mathcal{H}^{(k)}_1\otimes\mathcal{H}^{(k)}_2\otimes \mathcal{H}^{(k)}_3,
\]
with \(\mathcal{H}^{(k)}_{i}=\mathcal{H}(U^i_{\kappa_{t}^{(k)}})\) where \(U^i_{\kappa_{t}^{(k)}}\) is the subspace in the direction of the basis element \(\partial_i\). That is, each Hilbert space represents an independent ray for each spatial direction spanning the Cauchy surface. This construction ensures a smooth transition of the proxy from the microscopic Quantum Foam regime to a broad class of classical globally hyperbolic spacetimes. In doing so, we also establish a Correspondence Principle for Quantum Foam.

To construct a quantum field theory for the Gaussian Quantum Foam, we thus begin by establishing the appropriate Hilbert space structure within the framework of a Gelfand triple. Therefore, consider the sequence \(\{(M_{(k)},g^{(k)}_{\mu\nu})\}_{k\in\mathbb{N}}\) of globally hyperbolic and homotopic spacetimes in Definition \ref{def:quantumfoam}. Bernal and Sánchez's theorem \cite{bernal2005} guarantees the existence of a smooth and regular global time function \(t_{(k)}: M_{(k)} \to \mathbb{R}\), which \(\forall t\in \mathbb{R}\), induces a foliation of the manifold with spacelike Cauchy surfaces \(\Sigma_t^{(k)}=\{p\in M_{(k)}:t_{(k)}(p)=t\}\). Let \(\kappa_{(k)}: O_{\kappa_{(k)}}\subset M_{(k)}\to U_{\kappa_{(k)}} \subset \mathbb{R}^4\) be a local coordinate chart on \((M_{(k)}, g^{(k)}_{\mu\nu})\), and denote by \(\kappa^{(k)}_t = \kappa|_{\Sigma_t^{(k)}}\) its restriction to a fixed hypersurface.

A Hilbert space \(\mathcal{H}^{(k)}_t\) can be defined on each hypersurface as the space of square-integrable functions over \(U^{(k)}_{\kappa_t}\), equipped with the \(k^{-3}\)–weighted inner product
\begin{equation}
\langle u_{(k)}, v_{(k)}\rangle_{\mathcal{H}^{(k)}_t} 
:= \int_{U^{(k)}_{\kappa_t}} \frac{1}{k^3}\,u_{(k)}^{*}(x)\,v_{(k)}(x)\,d^3x,
\end{equation}
where \(u_{(k)}, v_{(k)}\in \mathcal{S}(U_{\kappa^{(k)}_t})\) are localised to the subspace of modulated Gaussians and converge in the sense of distributions to \(u, v \in \mathcal{S}^{\prime}(U_{\kappa_t})\). This restriction ensures that the integrand remains regular in the distributional limit and that the inner product is well defined in both the pre-Hilbert and limiting sense.

Since each spatial component \(\beta_{(k)}^i\) depends only on the corresponding coordinate \(x^i\), the associated Hilbert space \(\mathcal{H}^{(k)}_i\) is effectively defined over the one-dimensional subspace. The corresponding weighted inner product on \(\mathcal{H}^{(k)}_i\), in the Gelfand triple \(\mathcal{S}^{(k)}_i \subset \mathcal{H}^{(k)}_i \subset \mathcal{S}_i^{\prime \,(k)}\), is then
\begin{equation}
\langle u_{(k)}^i, v_{(k)}^i \rangle_{\mathcal{H}^{(k)}_i} 
:= \int \frac{1}{k}\, u_{(k)}^{i*}(x^i)\, v_{(k)}^i(x^i)\,dx^i,
\end{equation}
where the time index has been omitted for clarity. This inner product reflects the separable structure of the quantisation scheme while preserving mathematical consistency in the distributional setting. 

The appearance of the \(k^{-3}\) scaling in the three-dimensional case (and \(k^{-1}\) in the one-dimensional reduction) is introduced here without further justification, except that it is required for distributional consistency. A rigorous foundation for this choice will be provided in Section~\ref{sec:algebra}, where we construct a renormalised distribution algebra based on a scaled restriction of Schwartz space with localised Gaussian kernels converging, in the sense of distributions, to a Dirac measure.

To quantise the shift vector field \(\beta^i_{(k)}\) as the sequence index \(k \to \infty\), corresponding to the Gaussian Quantum Foam contracting toward the Planck scale, harmonic oscillator annihilation and creation operators are introduced. Since each component \(\beta^i_{(k)}\) is a function of its corresponding spatial coordinate \(x^i\), the quantisation proceeds independently for each spatial direction. Thus, each shift vector component is promoted to an operator-valued distribution, \(\hat{\beta}_{(k)}^i\), and in the Schr\"odinger picture, expanded in a local Gaussian wave packet basis:
\begin{equation} \label{eq:qgaussian}
\hat{\beta}_{(k)}^i:=\hat{\beta}_{(k)}(x^i) := \int_{\mathcal{F}(U^i_{\kappa^{(k)}_t})} d p^i \biggl( \psi_{(k)}(p^i, x^i)\hat{a}_{(k)}^i(p^i)  + \psi^*_{(k)} (p^i,x^i)\hat{a}_{(k)}^{i\dagger}(p^i)  \biggr).
\end{equation}
Here, the annihilation operator \(\hat{a}_{(k)}^i(p^i)\) and creation operator \(\hat{a}_{(k)}^{i\dagger}(p^i)\) satisfy the bosonic commutation relations:
\begin{equation}
[\hat{a}_{(k)}^i(p^i), \hat{a}_{(k)}^{j\dagger}(p^{\prime j})] = \delta^{ij} \delta(p^i - p^{\prime j}),
\end{equation}
\begin{equation}
[\hat{a}_{(k)}^i(p^i), \hat{a}_{(k)}^{j}(p^{\prime j})] = 0, \quad [\hat{a}_{(k)}^{i\dagger}(p^i), \hat{a}_{(k)}^{j\dagger}(p^{\prime j})] = 0.
\end{equation}
The mode functions are chosen as Gaussian wave packets, defined via their momentum-space representation:
\begin{equation}\label{psi}
\psi_{(k)}(p^i, x^i) = \left( \frac{\sigma^2}{\pi k^2} \right)^{1/4} e^{- \frac{\sigma^2}{2k^2} (p^i)^2} e^{i p^i x^i}.
\end{equation}
\begin{Remark}
Here Einstein's summation convention is not used unless otherwise stated.
\end{Remark}
The state space and thus the Fock space is then constructed as a tensor product of three independent bosonic Fock spaces:
\begin{equation}
\mathcal{F}(\mathcal{H}^{(k)}) = \mathcal{F}(\mathcal{H}^{(k)}_1) \otimes \mathcal{F}(\mathcal{H}^{(k)}_2) \otimes \mathcal{F}(\mathcal{H}^{(k)}_3),
\end{equation}
where each individual Fock space is built as
\begin{equation}
\mathcal{F}(\mathcal{H}^{(k)}_i) = \mathbb{C} \oplus \mathcal{H}^{(k)}_i \oplus (\mathcal{H}^{(k)}_i\otimes_s \mathcal{H}^{(k)}_i)\oplus \dots.
\end{equation}
The vacuum state for the complete Fock space is given by:
\begin{equation}
\ket{0} := \ket{0}_1 \otimes \ket{0}_2 \otimes \ket{0}_3,
\end{equation}
with vacuum states \(\ket{0}_i\) satisfying:
\begin{equation}
\hat{a}^i_{(k)}(p^i) \ket{0}_i = 0, \quad \forall p^i, i.
\end{equation}
One-particle states are given by:
\begin{equation}
|1^i(p^i)\rangle = \hat{a}^{i\dagger}_{(k)}(p^i) \ket{0}_i.
\end{equation}
Multi-particle states follow as:
\begin{equation}
|p_1^i, p_2^i, \dots, p_m^i\rangle = \hat{a}^{i\dagger}_{(k)}(p_1^i) \hat{a}^{i\dagger}_{(k)}p_2^i) \dots \hat{a}^{i\dagger}_{(k)}(p_m^i) \ket{0}_i.
\end{equation}
To describe the quantised spacetime geometry in a way that remains as close as possible to classical physics, we employ coherent states. The choice of coherent states is motivated by several fundamental considerations:

First, the Correspondence Principle encoded in the Gelfand triple structure dictates that the transition from the quantum to the classical regime should be smooth. Coherent states, being eigenstates of the annihilation operator with smooth eigenvalues, provide the closest quantum analogue to classical fields, ensuring that quantum fluctuations do not disrupt the macroscopic structure of spacetime.

Second, Schrödinger’s 1926 work \cite{schrodinger1926} introduced what are now termed “minimum uncertainty wave packets” to describe quantum systems that transition smoothly to classical behaviour. These wave packets, later recognised as coherent states, evolve in a manner closely resembling classical motion. As detailed in Steiner’s work on Schrödinger’s discovery of coherent states \cite{steiner1988} these states provide the optimal representation of quantum systems with well-defined classical limits.

Thirdly, the coherent state naturally describes a system where the ground-state wave packet is displaced from the origin. In the context of the shift vector, this aligns perfectly with its role in encoding spatial displacements between hypersurfaces. Thus, the coherent state provides an optimal quantum representation of the shift vector, ensuring that its expectation values follow the classical behaviour predicted in the Correspondence Principle. This fundamental role of coherent states in quantum mechanics and quantum optics was developed by Glauber \cite{glauber1963}, whose work established their significance in describing quantum states that exhibit classical-like properties.

Finally, and perhaps most importantly, Wheeler essentially interpreted fluctuations in spacetime geometry as displacements of the vacuum state, a process that occurs throughout space at all times \cite{mtw1973}. In the context of Quantum Foam, this suggests that the displacement is to be understood in terms of the shift vector on spatial hypersurfaces and with respect to a global time parameter. Accordingly, we must consider quantum states that naturally exhibit this displacement property, expressed through the fluctuating behaviour of the shift vector field.

A coherent state  \(|\alpha_{(k)}^i\rangle\) for each separate Fock space \(\mathcal{F}(\mathcal{H}^{(k)}_i)\) is defined as an eigenstate of the annihilation operator  \( \hat{a}_{(k)}^i(p^i)\):
\begin{equation} \label{coherentstate}
\hat{a}_{(k)}^i(p^i) |\alpha_{(k)}^i\rangle := \alpha_{(k)}^i(p^i) |\alpha_{(k)}^i\rangle.
\end{equation}
By expressing the annihilation operator \(\hat{a}_{(k)}^i(p^i)\) in terms of canonical field operators, it becomes clear that in the coherent state \(|\alpha_{(k)}^i(p)\rangle\), the quantity \(\sqrt{2}\alpha_{(k)}^i(p^i)\) encodes the displacement of the quantum field configuration from the vacuum. This displacement, when interpreted through the quantised shift vector \eqref{eq:qgaussian}, is not merely a formal construct; it corresponds to the geometric displacement between neighbouring hypersurfaces, given by \(\delta x^i = \beta_{(k)}^i dt\).

The classical profile \(\beta_{(k)}^i\), which is in the restricted Schwartz space, is realised in the quantum perspective as the expectation value \(\langle \alpha_{(k)}^i | \hat{\beta}_{(k)}^i | \alpha_{(k)}^i \rangle\). To maintain this correspondence, the eigenvalue \(\alpha_{(k)}^i(p^i)\) must be chosen accordingly.

Using the fact that the annihilation operator is defined as a linear combination of the canonical field operators that constitute phase space, it follows that each coherent mode‐amplitude wavefunction satisfies a first‐order ODE whose unique (up to overall phase) normalisable solution is a Gaussian wavepacket modulated by a plane‐wave phase. Hence, as expected, the coherent states are in the Schwartz space, represented by (possibly modulated) Gaussians.

Therefore, within the framework of the Gelfand triple \(\mathcal{S}^{(k)}_i \subset \mathcal{H}^{(k)}_i \subset \mathcal{S}_i^{\prime \ (k)}\),  where all admissible test functions and mode functions reside in the Schwartz space, it is natural to choose the displacement eigenvalue \(\alpha_{(k)}^i(p^i)\) as a Gaussian:
\begin{equation}\label{eq:eigenalpha}
\alpha_{(k)}^i(p^i) = \left( \frac{k^2}{256\pi^3\sigma^2} \right)^{1/4} e^{- \frac{\sigma^2 {p^i}^2}{2k^2}}.
\end{equation}
While this choice is not uniquely fixed by the operator algebra, it is canonical in the sense that it respects the minimal uncertainty property of coherent states, aligns with the structure of the Gelfand triple, and yields the correct classical limit.\\

Next, the complete coherent state, \(|\alpha_{(k)}^{hij}\rangle\), for the shift vector, is given by the tensor product:
\begin{equation}\label{eq:tensorcoherentstate}
|\alpha_{(k)}^{hij}\rangle= |\alpha^h_{(k)}\rangle \otimes |\alpha^i_{(k)}\rangle\otimes |\alpha^j_{(k)}\rangle
\end{equation}
Before proceeding with the analysis of the expectation value of the shift vector, the trace of the extrinsic curvature and building an understanding of the number of bosonic quanta in the limit of Gaussian Quantum Foam, it is worth noting that, given the state space construction, the shift field operators \(\hat{\beta}_{(k)}^i\) act only on their respective sectors \(\mathcal{F}^{(k)}_i\) of the total Fock space \(\mathcal{F}^{(k)} = \mathcal{F}^{(k)}_1 \otimes \mathcal{F}^{(k)}_2 \otimes \mathcal{F}^{(k)}_3\).  Consequently, their combined action factorises as
\begin{equation}
\left( \hat{\beta}_{(k)}^h \otimes \hat{\beta}_{(k)}^i \otimes \hat{\beta}_{(k)}^j \right) |\alpha_{(k)}^{hij}\rangle = \left( \hat{\beta}_{(k)}^h |\alpha_{(k)}^h\rangle \right) \otimes \left( \hat{\beta}_{(k)}^i |\alpha_{(k)}^i\rangle \right) \otimes \left( \hat{\beta}_{(k)}^j |\alpha_{(k)}^j\rangle \right).
\end{equation}
The expectation value of the shift vector field, in each of its components \(\hat\beta^i_{(k)}\), in the coherent state is:
\begin{equation}
\langle \alpha_{(k)}^{xyz}| \hat{\beta}_{(k)}^i| \alpha_{(k)}^{xyz} \rangle = \langle \alpha_{(k)}^{i}| \hat{\beta}_{(k)}^i | \alpha_{(k)}^{i} \rangle=\int dp^i \biggl( \alpha_{(k)}^i(p^i) \psi_{(k)}(p^i, x^i) + \alpha_{(k)}^{i*}(p^i) \psi_{(k)}^*(p^i, x^i)\biggr).
\end{equation}
Here we have used that the expectation values of the annihilation operator \(\hat{a}^i_{(k)}\), and the creation operator \(\hat{a}^{\dagger i}_{(k)}\), in the coherent state (\ref{coherentstate}), are related by complex conjugation and therefore that: 
\begin{equation}
   \langle\alpha^i_{(k)}|\hat{a}^{\dagger \ i}_{(k)}|\alpha^i_{(k)}\rangle= (\langle\alpha^i_{(k)}|\hat{a}^i_{(k)}|\alpha^i_{(k)}\rangle)^* = (\alpha^{i}_{(k)})^*,
\end{equation}
since 
\begin{equation}
   \langle\alpha^i_{(k)}|\hat{a}^i_{(k)}|\alpha^i_{(k)}\rangle=\alpha^i_{(k)}.
\end{equation}
Using the fact that both the mode function  \( \psi_{(k)}(p^i, x^i)\), and the coherent eigenvalue  \( \alpha_{(k)}^i(p^i)\), given in (\ref{psi}) and (\ref{eq:eigenalpha}) are Gaussians and that the Fourier transform of a Gaussian is also a Gaussian, we get
\begin{equation} \label{eq:expbeta}
\langle \alpha^i_{(k)} | \hat{\beta}_{(k)}^i| \alpha^i_{(k)} \rangle = \frac{1}{\sqrt{4 \sigma^2 / k^2 \pi}} \exp\left(-\frac{(x^i)^2}{4 \sigma^2 / k^2}\right)=\beta^i_{(k)}.
\end{equation}
Proceeding to the expectation value of the trace of the extrinsic curvature \eqref{eq:tracek}, the stretching and contraction of any local grid, that in its quantised form is taken as
\[
\hat{K}^{(k)} = \frac{1}{N_{(k)}} \left(\partial_x\hat{\beta}_{(k)}^x+\partial_y\hat{\beta}_{(k)}^y+\partial_z\hat{\beta}_{(k)}^z\right)
\]
and using the expectation value (\ref{eq:expbeta}) then we have
\begin{equation}\label{eq:exph}
\langle \alpha_{(k)}^{xyz}| \hat{K}^{(k)}| \alpha_{(k)}^{xyz} \rangle
= \frac{1}{N_{(k)}}\left(\partial_x\langle\hat{\beta}_{(k)}^x\rangle
+ \partial_y\langle\hat{\beta}_{(k)}^y\rangle
+ \partial_z\langle\hat{\beta}_{(k)}^z\rangle\right)
= \frac{1}{N_{(k)}}\left(\partial_x\beta_{(k)}^x+\partial_y\beta_{(k)}^y+\partial_z\beta_{(k)}^z\right)
=K^{(k)}.
\end{equation}
Thus, we have concluded that the expectation values of the quantised shift vector field~\eqref{eq:qgaussian} and the quantised trace of the extrinsic curvature~\eqref{eq:exph} precisely reproduce their classical counterparts~\eqref{eq:gaussian} and~\eqref{eq:tracek}, respectively. This agreement holds not only for finite values of the sequence index  \(k\), but also in the distributional limit as  \(k\to\infty\). Both pointwise and smeared expectation values preserve the functional and distributional structure of the classical fields, confirming the internal consistency of the quantum formulation and that using the shift vector as a proxy is justified.

Before proceeding further, let us also consider the energy content of the quantised field~\eqref{eq:qgaussian} in the coherent state~\eqref{coherentstate} or~\eqref{eq:tensorcoherentstate}, by calculating the expected number of quanta for each element in the sequence of globally hyperbolic spacetimes in Definition~\ref{def:quantumfoam} that converges in distribution to a Gaussian Quantum Foam element. That is, consider the occupation number operator
\[
\mathcal{N}_{(k)}^i = a_{(k)}^{i \ \dagger}a_{(k)}^i.
\]
Then the energy content, i.e. the Hamiltonian on each hypersurface, is given by
\begin{equation} \label{eq:qhamilton}
  \langle \mathcal{H}_{(k)} \rangle 
  := \sum_{i=1}^3 \int \langle \alpha_{(k)}^{xyz}| \mathcal{N}_{(k)}^i| \alpha_{(k)}^{xyz} \rangle \, dp^i
  = \sum_{i=1}^3 \int |\alpha_{(k)}(p^i)|^2 \, dp^i
  = \frac{3k^2}{16\pi\sigma^2}.
\end{equation}
This follows since the expectation value of the number operator in a coherent state is the modulus squared of the coherent eigenvalue  \(\alpha_{(k)}^i\)~\eqref{eq:eigenalpha}, which is a Gaussian, and thus we can use that
\[
\int_{-\infty}^{\infty} e^{-\frac{\sigma^2 (p^i)^2}{k^2}} dp^i
= \frac{\sqrt{\pi}\,k}{\sigma}.
\]
As expected, the energy content~\eqref{eq:qhamilton}, obtained by weighting over all field modes, diverges in the Quantum Foam limit. This is not an anomaly: as we will later see in Section~\ref{sec:waveequation}, the non-linear field equation for the shift vector field is driven by a Dirac measure and its second-order distributional derivative in a well-defined distribution geometry. The Dirac measure represents the bare mass monopole whose coefficient diverges quadratically in the quantum limit. After renormalisation, which subtracts this universal divergence, we are left with the observable finite mass in the emerging classical spacetimes. Thus, the quadratic divergence in the Hamiltonian~\eqref{eq:qhamilton} as  \(k\to\infty\) is treated by standard renormalisation.

These results demonstrate that the classical spacetime structure is recovered exactly from the underlying quantum theory, without any modification of general relativity. In particular, the coincidence of the expectation value of the quantised shift vector with its classical counterpart, both pointwise and in the distributional limit, implies that a broad class of globally hyperbolic spacetimes, including solutions to the Einstein field equations, can emerge directly from the quantum domain.

Together with the earlier classical discussion concerning adopting the shift vector as a proxy for the Quantum Foam, we have shown that this identification is supported both classically and from a quantum field perspective. Even if the shift vector is gauged to zero in a Gaussian Quantum Foam, its observable effects persist and are encoded in the geometry across all scales, including the Planck scale.

After this introduction to Gaussian Quantum Foam, we proceed to construct a renormalised distribution algebra tailored to distributional operations, including non-linear products. This algebra will allow us to formulate and analyse the underlying non-linear wave equation. It also furnishes a self-consistent framework for distribution geometry in general relativity within the Quantum Foam setting and, crucially, provides a means to assess whether the notion of a singularity has any operational meaning in Quantum Foam.

\section{Renormalised Algebra for Gaussian Quantum Foam} \label{sec:algebra}

In this section, we develop a renormalised distributional algebra tailored to Gaussian Quantum Foam. The construction is based on sequences of smooth, scaled  localised Gaussian functions (modulated by compactly supported cut-offs, see Section \ref{sec:review}) that converge, in the sense of distributions, to singular geometric structures. Central to the formulation is the  \(1/k^{\,n-1}\) scaling factor in the definition of  \(n\)-fold Gaussian products. Although this scaling may at first seem unconventional, it ensures that non-linear operations such as multiplication and differentiation are well defined at the level of smooth representatives and, at the same time, provides a natural mechanism for renormalisation in the distributional limit. Without this scaling there exist products, particularly those involving higher-order derivatives, that diverge or fall outside the distribution space. Taken together, these features yield an internally consistent calculus in which the non-linear wave operator governing the shift vector field in Gaussian Quantum Foam is both mathematically well defined and physically meaningful, and they likewise enable an informal distribution geometry for analysing invariant scalar quantities on the singular support of Quantum Foam.

We conclude the construction of the renormalised distribution algebra by situating the model within both Nigsch–Vickers’ distribution geometry \cite{nigsch20201, nigsch20202} and with respect to the Colombeau theory \cite{colombeau2000} and its relation to model delta net. This in conjunction with a basic analysis of how the algebra relates to the spectral analysis of singularities and hence to microlocal analysis, with emphasis on characteristics. This is necessary because, as we will see, in the renormalised algebra, all non-linear operations are performed on smooth elements in the restricted Schwartz space, introduced in Section \ref{sec:review}, prior to taking the limit. These operations preserve causality, are separated from the singular support, and lie inside the null cone. This means that all points are regular directed points. Consequently, the wave front set is empty at finite  \(k\), whereas the wave front set is non-empty in the limit. Accordingly, singularities propagate along the bi-characteristics of the principal symbol (in the sense of the linearised operator).

\subsection{A Scaled Gaussian Restriction of Schwartz Space and its Dual} \label{sec:restriction}

To begin, we briefly return to Gaussian Quantum Foam as defined in Definition~\ref{def:quantumfoam}, which is based on the sequence  \( \{(M_{(k)}, g_{\mu\nu}^{(k)})\}_{k \in \mathbb{N}}\) of globally hyperbolic and homotopic spacetimes converging in the sense of distributions. For each  \( k\), global hyperbolicity guarantees the existence of a smooth and regular global time function  \( t_{(k)}: M_{(k)} \to \mathbb{R}\), which, for all  \( t \in \mathbb{R}\), induces a foliation of the manifold by spacelike Cauchy surfaces  \( \Sigma_t^{(k)} = \{ p \in M_{(k)} : t_{(k)}(p) = t \}\); see~\cite{bernal2005} for a proof that global hyperbolicity is sufficient for the existence of such a foliation.

As in Definition~\ref{def:quantumfoam}, let  \( \kappa^{(k)}: O_{\kappa^{(k)}} \subset M_{(k)} \to U_\kappa^{(k)} \subset \mathbb{R}^4\) be a local coordinate chart on  \( M_{(k)}\), and denote by  \( \kappa^{(k)}_t = \kappa^{(k)}|_{\Sigma_t^{(k)}}\) its restriction to any fixed hypersurface.

On any such Cauchy surface, and for each spatial coordinate  \( x^i\) (with the index  \( i\) suppressed when convenient), we introduce a Gaussian restriction of the Schwartz space to allow for scaled Gaussian products. Here, as in Section \ref{sec:review} we modulate, that is localise, the Gaussian kernel with a compactly supported cut-off function so that the restriction makes sense on the open sets and in the limit are local distributions on the spacetime manifolds as in definition 6.3.3 in \cite{hormander1983}.

\begin{Definition}[Gaussian Cauchy Surface Restriction of  \( \mathcal{S}(U_{\kappa^{(k)}_t})\)] \label{def:restriction}
\[
\begin{split} \label{eq:restrictionspace}
\mathcal{S_G}(U_{\kappa^{(k)}_t}) := \biggl\{ g^n_{(k)}(x,\alpha) = \frac{\prod\limits_{i=1}^{n} g_{(k)}(x,\alpha_i)}{k^{n-1}} \,\bigg|\, n\in\mathbb{N}, \,g_{(k)}(x,\alpha_i) \in \mathcal{S}(U_{\kappa^{(k)}_t}),\, \alpha_i > 0, \,  \int_{ U_{\kappa^{(k)}_t}}g_{(k)}(x,\alpha_i)dx= \frac{1}{\sqrt{\alpha_i}} \biggr\},
\end{split}
\]
where \(\alpha = \sum\limits_{i=1}^n \alpha_i\) is the total scale.
\end{Definition}
\begin{Remark}
Throughout this work,  \(g_{(k)}\)or  \(g^{(k)}\) denotes the Gaussian generator sequence of the restricted space. 
This should not be confused with the metric field  \(g^{(k)}_{\mu\nu}\), which always carries indices, 
except when it appears in the determinant  \(\sqrt{|g^{(k)}|}\) or in the wave operator  \(\Box_{g^{(k)}}\).
\end{Remark}

\noindent It should be noted that the restricted space of localised Gaussians defined here and generated by a Gaussian kernel \(g_{(k)}\) is not unique. It is introduced specifically as a bespoke construction, tailored to support non-linear operations and to provide the algebraic setting for a field equation in Gaussian Quantum Foam and the analysis of the operational meaning of singularities in a Gaussian Quantum Foam.

We now introduce a multiplication operation that will be well-defined in distributions and that will be needed for the analysis of the wave operator and later in a discussion of the notion of singularities in Quantum Foam.
\begin{Definition}[Multiplication in  \( \mathcal{S_G}(U_{\kappa^{(k)}_t})\)]\label{def:multiplication}
For any  \( g_{(k)}^{m}(x, \alpha), g_{(k)}^{n}(x, \gamma) \in \mathcal{S_G}(U_{\kappa^{(k)}_t})\), multiplication is defined by
\[
g_{(k)}^{m}(x, \alpha) \cdot g_{(k)}^{n}(x, \gamma) := g_{(k)}^{m+n}(x, \alpha + \gamma) 
:= \frac{\prod\limits_{i=1}^{m} g_{(k)}(x, \alpha_{i}) \cdot \prod\limits_{j=1}^{n} g_{(k)}(x, \gamma_{j})}{k^{m+n - 1}}
= \frac{\prod\limits_{i=1}^{m+n} g_{(k)}(x,\zeta_i)}{k^{m+n-1}}
= g_{(k)}^{m+n}(x, \zeta) \in \mathcal{S_G}(U_{\kappa^{(k)}_t}),
\]
where  \(\alpha = \sum_{i=1}^{m} \alpha_{i}\),  \(\gamma = \sum_{j=1}^{n} \gamma_{j}\), and  \(\zeta = \sum_{i=1}^{m+n}\zeta_i\), with
\[
\zeta_i =
\begin{cases}
    \alpha_i, & 1 \leq i \leq m, \\[4pt]
    \gamma_{i-m}, & m+1 \leq i \leq m+n.
\end{cases}
\]
\end{Definition}

\noindent This multiplication is bilinear, commutative, associative, and closes \(\mathcal{S_G}(U_{\kappa^{(k)}_t})\).

In the setting of the Gaussian Quantum Foam element in Definition \ref{def:quantumfoam} with the shift vector components given by \eqref{eq:gaussian} it is convenient to take  \( g_{(k)}(x,\alpha_i)\) as:
\begin{equation} \label{eq:scaledgaussian}
g_{(k)}(x,\alpha_i) := \beta_{(k)}(\sqrt{\alpha_i}x)= \frac{1}{\sqrt{4 \sigma^2 / k^2 \pi}} \exp\left(-\frac{\alpha_i x^2}{4 \sigma^2 / k^2}\right),
\end{equation}
where \(\alpha_i > 0\) and as before  \( \sigma\geq 1\) is a localisation parameter controlling the range where the effect of the shift vector cannot be neglected. As  \( k \to \infty\), the Gaussian~\eqref{eq:scaledgaussian} converges to a scaled Dirac measure:
\begin{equation} \label{eq:scaleddiracmeasure}
\lim_{k \to \infty} \langle g_{(k)}(x, \alpha_i), \phi(x) \rangle = \frac{1}{\sqrt{\alpha_i}} \phi(0), \quad \forall \phi \in \mathcal{S}(U_{\kappa^{(k)}_t}).
\end{equation}
We now introduce the corresponding restricted distributional space, defined as
\begin{Definition}[Gaussian Cauchy Surface Restriction of  \( \mathcal{S}^\prime(U_{\kappa^{(k)}_t})\)]\label{def:dalgebraspace}
\[
\mathcal{S_G}^{\prime}(U_{\kappa^{(k)}_t}) := \biggl\{ \lim_{k \to \infty} g^n_{(k)}(x,\alpha) \,\bigg|\, g_{(k)}^n(x, \alpha) \in \mathcal{S_G}(U_{\kappa^{(k)}_t}) \biggr\} \subset \mathcal{S}^{\prime}(U_{\kappa^{(k)}_t}).
\]
\end{Definition}
\noindent In the following section, we show that the space  \( \mathcal{S_G}^\prime\) is closed under multiplication, and first-order differentiation, with all non-linear operations performed at finite sequence index  \( k\), prior to taking the distributional limit. For each operation, a precise meaning is assigned through formal definitions. The restricted Schwartz space~\eqref{def:restriction} and its dual~\eqref{def:dalgebraspace}, together with the operations introduced below, hence define a distributional algebra  \( (\mathcal{S_G}^\prime(U_{\kappa^{(k)}_t}), \cdot, \partial)\).

However, while this algebra is closed under multiplication, and first-order differentiation, product operations involving second-order distributional derivatives or operations with a first order derivative multiplied with another first order derivative in distributions, must be handled separately. Such terms result in a linear combination of the distribution and its second-order derivative. The treatment of this case is deferred to Section~\ref{sec:algebraoperations2}.

\subsection{Multiplication, and Differentiation} \label{sec:algebraoperations} 

In this section, we introduce basic operations within the Gaussian restriction in the sense of distributions, presented as a series of formal definitions. The operations defined here, and in Section~\ref{sec:algebraoperations2} are sufficient for analysing the wave equation in the context of Gaussian Quantum Foam and provides for the distribution geometric tools to analyse curvature and stress-energy scalars in Quantum Foam. We begin by showing that the algebra is well defined and closed under linear combinations of the Gaussian kernel in the restricted space~\ref{def:restriction} and in its dual space~\ref{def:dalgebraspace}, and that the same holds for distributional multiplication in~\ref{def:multiplication}.

To demonstrate that the distribution space  \( \mathcal{S_G}^\prime\) defined in Definition~\ref{def:dalgebraspace} is closed under linear operations in the single generator  \(g_{(k)}\) of the space, we consider an arbitrary test function  \(\phi\in\mathcal{S}(U_{\kappa^{(k)}_t})\) in the admissible space of test functions in definition \ref{def:test-functions}, and scalars  \(a_1,\dots,a_p\in\mathbb{R}\). Using the definition, equation~(\ref{eq:scaledgaussian}) and its distributional limit~\eqref{eq:scaleddiracmeasure}, we observe that
\begin{align}
    \lim_{k\to\infty} \Big\langle \sum_{i=1}^p a_i g_{(k)}^{n}(x, \alpha), \phi(x)\Big\rangle
    &= \lim_{k\to\infty}\Big\langle\frac{\sum\limits_{i=1}^{p}a_i\prod\limits_{j=1}^{n}g_{(k)}(x, \alpha_{j})}{k^{n-1}}, \phi(x)\Big\rangle \notag \\ 
    &= \lim_{k\to\infty}\Big\langle\sum_{i=1}^p a_i \frac{1}{(2\sigma\sqrt{\pi})^{n-1}}\,g_{(k)}(x,\alpha), \phi(x)\Big\rangle \notag \\ 
    &= \sum_{i=1}^p\frac{a_i}{(2\sigma\sqrt{\pi})^{n-1}\sqrt{\alpha}}\,\phi(0),
\end{align}
where  \(\alpha= \alpha_1 + \alpha_2+\dots + \alpha_n\).

Thus we are led to the following definition for linear combinations in the restricted Gaussian space and its dual space of distributions.

\begin{Definition}[Linear combinations in Gaussian Distributions]\label{def:dlinearcomb}
Let  \(g_{(k)}^n(x,\alpha) \in \mathcal{S_G}(U_{\kappa^{(k)}_t})\) with fixed total scale  \(\alpha\), and let  \(\phi \in \mathcal{S}(U_{\kappa^{(k)}_t})\) be an arbitrary test function (in the admissible space of test functions in definition \ref{def:test-functions}). Then the distributional linear combination is defined by
\[
\lim_{k \to \infty} \left\langle \sum_{i=1}^p a_i g_{(k)}^n(x,\alpha),\, \phi(x) \right\rangle
:= \sum_{i=1}^p \frac{a_i}{(2\sigma\sqrt{\pi})^{\,n-1}\sqrt{\alpha}}\, \phi(0),
\]
for any scalars  \(a_i \in \mathbb{R}\).
\end{Definition}
\noindent In conjunction with this, observe that given (\ref{eq:scaledgaussian}), any product of  \( n\) scaled Gaussians in the restricted space of Definition~\ref{def:restriction} can be expressed as
\begin{equation} \label{eq:nscaledgaussian}
g_{(k)}^n(x, \alpha) = \frac{1}{(2\sigma\sqrt{\pi})^{\,n-1}}\, g_{(k)}\!\left(x, \alpha_1+\alpha_2+\dots+ \alpha_n\right),
\end{equation}
which converges in the distributional sense to a rescaled Dirac measure,
\begin{equation} \label{eq:nscaleddiracmeasure}
\lim_{k \to \infty} \langle g_{(k)}^n(x, \alpha), \phi(x) \rangle
= \frac{1}{(2\sigma\sqrt{\pi})^{\,n-1} \sqrt{\alpha_1+\dots+ \alpha_n}}\, \phi(0).
\end{equation}
For the multiplication operation introduced earlier in Definition~\ref{def:multiplication}, we have
\begin{equation}
g_{(k)}^{m}(x, \alpha)\, g_{(k)}^{n}(x, \gamma)
= g_{(k)}^{m+n}(x, \alpha+ \gamma)
=\frac{\prod\limits_{i=1}^{m} g_{(k)}(x, \alpha_{i}) \cdot \prod\limits_{j=1}^{n} g_{(k)}(x, \gamma_{j})}{k^{m +n- 1}} \in \mathcal{S_G}(U_{\kappa^{(k)}_t}),
\end{equation}
for any  \( g_{(k)}^{m}, g_{(k)}^{n} \in \mathcal{S_G}(U_{\kappa^{(k)}_t})\).  
It follows from \eqref{eq:nscaledgaussian} that the multiplication in Definition~\ref{def:multiplication} is
\begin{equation} \label{eq:simplifiedscaled}
g_{(k)}^{m}(x, \alpha)\, g_{(k)}^{n}(x, \gamma)
= \frac{1}{(2\sigma\sqrt{\pi})^{\,m+n - 1}}\, g_{(k)}(x, \alpha + \gamma),
\end{equation}
and hence in the sense of distributions, using \eqref{eq:nscaleddiracmeasure}, for any  \( \phi \in \mathcal{S}(U_{\kappa^{(k)}_t})\),
\begin{equation}
\lim_{k \to \infty} \langle g_{(k)}^{m}(x, \alpha)\, g_{(k)}^{n}(x, \gamma), \phi(x) \rangle
= \frac{1}{(2\sigma\sqrt{\pi})^{\,m+n - 1} \sqrt{\alpha + \gamma}}\, \phi(0),
\end{equation}
where \(\alpha = \sum\limits_{i=1}^{m} \alpha_{i}\) and \(\gamma = \sum\limits_{j=1}^{n} \gamma_{j}\).
Thus we are led to the following definition:
\begin{Definition}[Multiplication in Gaussian Distributions]\label{def:dmultiplication}
Let  \( g_{(k)}^{m}(x, \alpha), g_{(k)}^{n}(x, \gamma) \in \mathcal{S_G}(U_{\kappa^{(k)}_t})\) be scaled Gaussian products of orders  \(m\) and  \(n\), respectively, and let  \( \phi \in \mathcal{S}(U_{\kappa^{(k)}_t})\) be an arbitrary test functions. Then multiplication in the restricted Gaussian distribution space is defined by
\[
\lim_{k \to \infty} \langle g_{(k)}^{m}(x, \alpha)\, g_{(k)}^{n}(x, \gamma), \phi(x) \rangle
:= \frac{1}{(2\sigma\sqrt{\pi})^{\,m+n - 1} \sqrt{\alpha + \gamma}} \, \phi(0),
\]
where  \( \alpha = \sum\limits_{i=1}^{m} \alpha_i\) and  \( \gamma = \sum\limits_{j=1}^{n} \gamma_j\) are the total scales of the respective Gaussian products.
\end{Definition}
\noindent By repeated use of the reasoning above, we are led to the following definition.
\begin{Definition}[Gaussian Distributional Repeated Multiplication]\label{def:dmultop}
Let  \( g_{(k)}^{n_i}(x, \alpha_{n_i}) \in \mathcal{S_G}(U_{\kappa^{(k)}_t})\) for  \( i = 1, \dots, p\), and let  \( \phi \in \mathcal{S}(U_{\kappa^{(k)}_t})\). The repeated distributional multiplication is defined by
\[
\lim_{k \to \infty} \left\langle \prod_{i=1}^{p} g_{(k)}^{n_i}(x, \alpha_{n_i}), \phi(x) \right\rangle
:= \frac{\phi(0)}{(2\sigma\sqrt{\pi})^{\,n_1 + \dots + n_p - 1} \sqrt{\alpha_{n_1} + \dots + \alpha_{n_p}}},
\]
with  \( \alpha_{n_i} = \sum\limits_{j=1}^{n_i} \alpha_{(i)j}\) for  \( i = 1, \dots, p\).
\end{Definition}
\noindent If we use the definition\ref{def:multiplication} and the form \eqref{eq:simplifiedscaled} in conjunction with differentiation in distributions from the linear distribution theory, then it motivates the following definition:
\begin{Definition}[Gaussian Distributional Differentiation]\label{def:ddiff}
The distributional  \(n\)-th derivative of a product of localised Gaussians is defined by
\[
\lim_{k \to \infty} \left\langle \partial_x^n \prod_{i=1}^{p} g_{(k)}^{n_i}(x, \alpha_{n_i}), \phi(x) \right\rangle
:= (-1)^n \frac{\phi^{(n)}(0)}{(2\sigma \sqrt{\pi})^{\,n_1 + \dots + n_p - 1} (\alpha_{n_1} + \dots + \alpha_{n_p})^{\frac{n}{2}}},
\]
with  \( \alpha_{n_i} = \sum\limits_{j=1}^{n_i} \alpha_{(i)j}\) for  \( i = 1, \dots, p\).
\end{Definition}

\begin{Definition}[Gaussian Products with First-Order Derivative] \label{def:dprod1}
Let  \( \phi \in \mathcal{S}(U_{\kappa^{(k)}_t})\). Then the distributional product of a Gaussian and the derivative of another is defined by
\[
\lim_{k \to \infty} \left\langle g_{(k)}^m(x, \alpha)\, \partial_x g_{(k)}^n(x, \gamma), \phi(x) \right\rangle
:= -\frac{\gamma}{(2\sigma \sqrt{\pi})^{\,m+n-1} (\alpha + \gamma)^{3/2}}\, \phi'(0),
\]
where  \( \alpha = \sum\limits_{i=1}^{m} \alpha_i\) and  \( \gamma = \sum\limits_{j=1}^{n} \gamma_j\).
\end{Definition}
\noindent We are now in a position to define the basic structure of the renormalised Gaussian distribution algebra, with the exception of product operations that involve products of first order or second-order distributional derivatives (discussed separately in Section~\ref{sec:algebraoperations2}). Before stating the definition, we clarify that \(\partial\) appearing in the operations is understood as the distribution obtained by differentiating the smooth representatives at finite \(k\) and then taking the limit; the resulting distributions are included in \(\mathcal{S}^\prime_\mathcal{G}\) by definition. The microlocal perspective of this is that all operations are performed on a set where all points are regular directed points. Consequently, the wave front set is empty at finite  \(k\), whereas the wave front set is non-empty in the limit.

\begin{Definition}[Differential Algebra for Gaussian Quantum Foam,  \((\mathcal{S}_\mathcal{G}'(U_{\kappa^{(k)}_t}), \cdot, \partial)\)] \label{def:algebra}
The Gaussian restriction space defined in~\eqref{eq:restrictionspace}, together with its dual~\eqref{def:dalgebraspace}, and the operations defined in Definitions~\ref{def:dlinearcomb}–\ref{def:dprod1}, form a restricted renormalised distribution algebra for Gaussian Quantum Foam, denoted
\[
(\mathcal{S}_\mathcal{G}'(U_{\kappa^{(k)}_t}), \cdot, \partial).
\]
\end{Definition}
\subsection{Products of First- and Second-Order Gaussian Derivatives}\label{sec:algebraoperations2}
The restricted algebra  \((\mathcal{S}_\mathcal{G}'(U_{\kappa^{(k)}_t}), \cdot, \partial)\), defined in \ref{def:algebra}, supports pointwise linear combinations, multiplication, and arbitrary-order formal differentiation of smooth representatives prior to taking the distributional limit. However, when extended to include product operations involving a first- or second-order derivative, the resulting objects are not, in general, purely second-order distributions. Rather, they appear as linear combinations of distributions up to second-order.

\noindent To illustrate the obstruction, consider the following product of two Gaussians  \(g_{(k)}(x, \alpha_1)\) and  \(\partial_x^2 g_{(k)}(x, \alpha_2)\):
\[
g_{(k)}(x, \alpha_1)\,\partial_x^2 g_{(k)}(x, \alpha_2).
\]
The second-order derivative of the Gaussian introduces an even,  \(k\)-dependent, second-order polynomial factor:
\[
\partial_x^2 g_{(k)}(x, \alpha_2) = \left( -\frac{\alpha_2 k^2}{2\sigma^2} + \frac{\alpha_2^2 k^4 x^2}{4\sigma^4} \right) g_{(k)}(x, \alpha_2).
\]
Multiplication by the other Gaussian gives:
\[
g_{(k)}(x, \alpha_1)\,\partial_x^2 g_{(k)}(x, \alpha_2)
= \frac{k}{2\sigma\sqrt{\pi}}\left( -\frac{\alpha_2 k^2}{2\sigma^2} + \frac{\alpha_2^2 k^4 x^2}{4\sigma^4} \right) g_{(k)}(x,\alpha_1+\alpha_2).
\]
This is clearly neither an element of a sequence converging to a scaled Dirac measure nor a linear combination of the Dirac measure and its derivatives, due to the divergent  \(k\)-dependent prefactor. Now, if we scale by  \(k^{-1}\), the resulting product is of second-order type with support  \(\{x^i=0\}\) across the coordinates and—whether considered componentwise or jointly at the singular support  \(\{0\}\)—it converges, by Hörmander’s Theorem 2.3.4~\cite{hormander1983}, to a linear combination of a Dirac measure and its derivatives up to the order of the derivative appearing in the product:
\begin{equation}
\lim_{k \to \infty} \left\langle \tfrac{1}{k} g_{(k)}(x, \alpha_1)\,\partial_x^2 g_{(k)}(x, \alpha_2), \phi(x) \right\rangle
= a_0\phi(0)+a_1\phi^\prime(0)+ a_2\phi^{\prime\prime}(0),
\end{equation}
for all  \(\phi \in \mathcal{S}(U_{\kappa^{(k)}_t})\), with real constants  \(a_0, a_1, a_2\). However, since the product is even in  \(x\), all odd-order derivatives of test functions in the admissible test-function space \ref{def:test-functions} vanish in the limit, giving:
\begin{equation}
\lim_{k \to \infty} \left\langle \tfrac{1}{k} g_{(k)}(x, \alpha_1)\,\partial_x^2 g_{(k)}(x, \alpha_2), \phi(x) \right\rangle
= a_0 \phi(0)+a_2\phi^{\prime\prime}(0).
\end{equation}

This motivates us to include the generalisation below to  \(n\)-products involving second-order distributional derivatives in the restricted algebra  \((\mathcal{S}_\mathcal{G}'(U_{\kappa^{(k)}_t}), \cdot, \partial)\), defined in \ref{def:algebra}. This type of product will appear not only in the quantum foam wave equation but also in scalar expressions for the curvature and, via Einstein’s field equations, in scalar expressions for the stress–energy–momentum tensor and hence in expressions describing the distribution of matter.

\begin{Definition}[Gaussian Products with Second-Order Derivative]\label{def:prodsecondorder}
Let  \(g_{(k)}^{m}, g_{(k)}^{n}\in \mathcal{S}_\mathcal{G}(U_{\kappa^{(k)}_t})\) and let  \(\phi \in \mathcal{S}(U_{\kappa^{(k)}_t})\).  
The product of a  \(m\)-scaled Gaussian and the second-order derivative of a  \(n\)-scaled Gaussian are defined by
\[
\lim_{k \to \infty} \left\langle g_{(k)}^{m}(x, \alpha_m)\,\partial_x^2 g_{(k)}^{n}(x,\gamma_n),\phi(x) \right\rangle
:= a_0\,\phi(0) + a_2\,\phi^{\prime\prime}(0),
\]
where
\[
a_0 = \lim_{k \to \infty} \langle g_{(k)}^{m}\,\partial_x^2 g_{(k)}^{n},\, 1\cdot\phi^{(k)}_c\rangle,
\qquad
a_2 = \lim_{k \to \infty} \langle g_{(k)}^{m}\,\partial_x^2 g_{(k)}^{n},\,\frac{x^2}{2}\cdot\phi^{(k)}_c\rangle,
\]
and where, as in Section~\ref{sec:review}, measurements with  \(1\) and  \(x^2/2\) are understood through the cut-off convention:  \(\phi^{(k)}_c \in C^\infty_0(U_{\kappa^{(k)}_t})\) is a compactly supported, positive test function equal to  \(1\) in a neighbourhood of  \(K_{\kappa^{(k)}_t} \subset U_{\kappa^{(k)}_t}\).
\end{Definition}

This definition is motivated by the scaling introduced in Definition~\ref{def:restriction} in the previous section and by the use of Theorem 2.3.4 in \cite{hormander1983}. The restriction to normalised and scaled Gaussians converging as sequences to scaled Dirac measures ensures closure under multiplication, while the theorem guarantees that second-order products remain in the space and in its dual space (Definition~\ref{def:dalgebraspace}) as a linear combination of the Dirac measure and its second derivative.

Nevertheless, one remaining issue concerns the coefficients  \(a_0\) and  \(a_2\). By integration by parts it follows that, for any finite  \(k\), one has  \(a_0 < 0\) and  \(a_2 > 0\). In the limit  \(k \to \infty\), however, the coefficient  \(a_0\) diverges quadratically in  \(k\).

Since this product appears directly in the scalar curvature expressions and thus, via Einstein’s equations, in the matter distribution, the divergence of  \(a_0\) must be addressed. The situation is closely analogous to the quadratic divergence of the bare electron mass in QED: the observed finite value arises only after renormalisation, where quantum fluctuations effectively screen the bare mass to yield the physical electron mass.
In the present case, the coefficient of the measure term may be interpreted as the bare matter monopole in the frozen configuration when the level surfaces converge to characteristics and the slowness goes to zero in Quantum Foam. The physically observed matter distribution, obtained as the foam evolves through the non-linear wave equation, is recovered by renormalising, that is, by subtracting the divergent  \(k^2\)-contribution from  \(a_0\). Concretely, one has
\begin{align*}\
    a_0 &= \lim_{k \to \infty} \langle g_{(k)}^{m}(x,\alpha_m)\,\partial_x^2 g_{(k)}^{n}(x,\gamma_n),\, 1\cdot\phi^{(k)}_c\rangle \\
        &= \lim_{k \to \infty} \int g_{(k)}^{m}(x,\alpha_m)\,\partial_x^2 g_{(k)}^{n}(x, \gamma_n)\,dx
         = - \lim_{k \to \infty} \int \partial_x g_{(k)}^{m}(x,\alpha_m)\,\partial_x g_{(k)}^{n}(x,\gamma_n)\,dx \\
        &= -\lim_{k \to \infty}\frac{\alpha\gamma}{(2\sigma\sqrt{\pi})^{n+m-1}(\alpha+\gamma)^2}
           \int\!\Bigg[\Big(-\frac{(\alpha+\gamma)k^2}{2\sigma^2}
           +\frac{(\alpha+\gamma)^2k^4x^2}{4\sigma^4}\Big)\beta_{(k)}(\sqrt{\alpha+\gamma}\,x) \\
         &  \qquad\qquad\qquad\qquad + \frac{(\alpha+\gamma)k^2}{2\sigma^2}\,\beta_{(k)}(\sqrt{\alpha+\gamma}\,x)\Bigg]\,dx\\
         &= -\lim_{k \to \infty}\frac{\alpha\gamma}{(2\sigma\sqrt{\pi})^{n+m-1}(\alpha+\gamma)^2}
           \int\!\Bigg[\partial^2_x\beta_{(k)}(\sqrt{\alpha+\gamma}\,x) + \frac{(\alpha+\gamma)k^2}{2\sigma^2}\,\beta_{(k)}(\sqrt{\alpha+\gamma}\,x)\Bigg]dx\\
           &= -\lim_{k \to \infty}\frac{\alpha\gamma}{(2\sigma\sqrt{\pi})^{n+m-1}(\alpha+\gamma)}
           \int\!\Bigg[\frac{k^2}{2\sigma^2}\,\beta_{(k)}(\sqrt{\alpha+\gamma}\,x)\Bigg]dx,
\end{align*}
where we have used that the second-order derivative term vanishes at the boundary of the support and that \( \alpha = \sum_{i=1}^{m} \alpha_i\) and  \( \gamma = \sum_{j=1}^{n} \gamma_j\). Clearly the remaining integral is divergent. Nevertheless, using that the integral of a scaled Gaussian  \(\beta_{(k)}(\sqrt{\alpha+\gamma}\,x)\) equals  \(1/\sqrt{\alpha+\gamma}\) for all finite  \(k\), we obtain for finite \(k\)
\begin{equation}
\begin{aligned} \label{eq:geonmassa0}
a_0^{(k)}=m^{(k)}_B:=-\frac{\alpha\gamma}{2^{m+n}\sigma^{m+n+1}\sqrt{\pi}^{\,n+m-1}(\alpha+\gamma)}\int\beta_{(k)}(\sqrt{\alpha+\gamma}\,x)dx\\
=-\frac{\alpha\gamma\sqrt{\alpha+\gamma}}{2^{m+n}\sigma^{m+n+1}\pi^{\frac{m+n-1}{2}}(\alpha+\gamma)^2}k^2<0.
\end{aligned}
\end{equation}
This represents a contribution to the bare mass monopole which we renormalise to zero in distribution geometry, but for finite  \(k\) leaves a finite renormalised  \(a_0\), interpreted as the observable mass density, together with a finite  \(a_2>0\) representing the universal curvature impulse. To see that  \(a_2\) is finite and positive, we use integration by parts and the fact that the localised Gaussian that generates the algebra is compactly supported. We then find: 
\begin{equation}\label{eq:a2}
\begin{aligned}
a_2^{(k)}&=\lim_{k \to \infty} \left\langle g_{(k)}^{n}\,\partial_x^2 g_{(k)}^{m},\,\frac{x^2}{2}\cdot x\,\phi^{(k)}_c\right\rangle \\
&=
-\frac{1}{(2\sigma\sqrt{\pi})^{m+n-2}}\biggl [\int\frac{1}{k}\beta_{(k)}(\sqrt{\alpha}\,x)\,\partial_x\beta_{(k)}(\sqrt{\gamma}\,x)\cdot x\,\phi^{(k)}_c\,dx \\
&\qquad\qquad\qquad\qquad\quad+\int \frac{1}{k}\partial_x\beta_{(k)}(\sqrt{\alpha}\,x)\,\partial_x\beta_{(k)}(\sqrt{\gamma}\,x)\cdot \frac{x^2}{2}\,\phi^{(k)}_c\,dx\biggr]\\
&=-\frac{2\sigma^2\gamma}{(2\sigma\sqrt{\pi})^{m+n-1}(\alpha+\gamma)^2}\int \frac{1}{k^2}\biggl(\partial^2_x\beta_{(k)}(\sqrt{\alpha+\gamma}\,x)+\frac{k^2(\alpha+\gamma)}{2\sigma^2}\beta_{(k)}(\sqrt{\alpha+\gamma}\,x)\biggr)\,dx\\
&\qquad+\frac{\alpha\gamma}{(2\sigma\sqrt{\pi})^{m+n-1}(\alpha+\gamma)^2}\int\partial^2_x\beta_{(k)}(\sqrt{\alpha+\gamma}\,x)\cdot \frac{x^2}{2}\,dx
= \frac{\gamma^2\sqrt{\alpha+\gamma}}{(2\sigma\sqrt{\pi})^{m+n-1}(\alpha+\gamma)^3}.
\end{aligned}
\end{equation}

\begin{Remark}[Signs at finite  \(k\) and renormalisation of  \(a_0\)]\label{re:delta_coefficients}
To reinforce and summarise the reasoning above, with the cut-off convention of Section~\ref{sec:review}, for each fixed  \(k\) one has
\[
\begin{aligned}
&a_0^{(k)}=\big\langle g_{(k)}^{m}\,\partial_x^2 g_{(k)}^{n},\,1\cdot\phi^{(k)}_c\big\rangle
= -\!\int (\partial_x g_{(k)}^{m})(\partial_x g_{(k)}^{n})\,dx =- \frac{\alpha\gamma\sqrt{\alpha+\gamma}}{2^{m+n}\sigma^{m+n+1}\pi^{\frac{m+n-1}{2}}(\alpha+\gamma)^2}k^2<0, \\
& a_2^{(k)}=\big\langle g_{(k)}^{m}\,\partial_x^2 g_{(k)}^{n},\,\tfrac{x^2}{2}\cdot\phi^{(k)}_c\big\rangle= \frac{\gamma^2\sqrt{\alpha+\gamma}}{(2\sigma\sqrt{\pi})^{m+n-1}(\alpha+\gamma)^3} >0.
\end{aligned}
\]
As  \(k\to\infty\),  \(a_0^{(k)}\) exhibits a quadratic UV divergence, whereas  \(a_2^{(k)}\to a_2>0\) is finite. We define a renormalised coefficient by removing the universal divergence:
\[
[a_0]_r := \lim_{k\to\infty}\big(a_0^{(k)}+Ck^2\big), \quad C\in\mathbb{R},
\]
and therefore fix  \([a_0]_r\) by a normalisation condition (e.g. ``vacuum scheme''  \([a_0]_r=0\) or a ``physical mass'' scheme that matches the observed monopole density). The surviving  \(a_2>0\) is scheme independent and represents the curvature impulse. To avoid unnecessary notation clutter, going forward, we will remove the superscript  \([]_r\) from the coefficient  \(a_0\).
\end{Remark}

Before closing this section, let us also consider a product of derivatives involving two differentiated Gaussian factors that appears in the context of the distribution geometry of the Gaussian Quantum Foam, each a first-order derivative, and show that it can be handled using a similar strategy to Definition~\ref{def:prodsecondorder}.  
That is, for simplicity, consider the following product:
\[
\partial_x g_{(k)}(x, \alpha)\,\partial_x g_{(k)}(x, \gamma).
\]
In scaled form, we have 
\[
\begin{aligned}
\frac{1}{k}\,
\partial_x g_{(k)}(x, \alpha)\,
\partial_x g_{(k)}(x, \gamma)
  &= \frac{\alpha\gamma}{2\sigma\sqrt{\pi}(\alpha + \gamma)^2} \left(\frac{k^4}{4\sigma^4}(\alpha + \gamma)^2x^2\right)
g_{(k)}(x,\alpha+\gamma)\\
&=\frac{\alpha\gamma}{2\sigma\sqrt{\pi}(\alpha+\gamma)^2}\biggl[\partial^2_x g_{(k)}(x,\alpha+\gamma)+\frac{(\alpha+\gamma)k^2}{2\sigma^2}g_{(k)}(x, \alpha+\gamma)\biggr].
\end{aligned}
\]
and hence, in a vacuum renormalised form,
\[
\partial_x g_{(k)}(x, \alpha)\,
\partial_x g_{(k)}(x, \gamma)
\xrightarrow{\mathcal{S}_\mathcal{G}^{\prime}}
\frac{\alpha\gamma}{2\sigma\sqrt{\pi}(\alpha+\gamma)^{5/2}}\,\delta^{\prime\prime}_x.
\]
This motivates the following general definition:
\begin{Definition}[Renormalised Vacuum Gaussian Products with First-Order Derivative Factors]\label{def:prodfirstfirstorder}
Let  \(g_{(k)}^{n}, g_{(k)}^{m} \in \mathcal{S}_\mathcal{G}(U_{\kappa^{(k)}_t})\) and take  \(\phi \in \mathcal{S}(U_{\kappa^{(k)}_t})\) to be an admissible test function. The renormalised vacuum product of the first-order derivative of a  \(n\)-scaled Gaussian and the first-order derivative of a  \(m\)-scaled Gaussian is defined by
\[
\lim_{k \to \infty}
\partial_x g_{(k)}^{m}(x, \alpha_m)\,
\partial_x g_{(k)}^{n}(x, \gamma_n)
= \frac{\alpha\gamma}{(2\sigma\sqrt{\pi})^{m+n-1}(\alpha+\gamma)^{5/2}}\,
\delta^{\prime\prime}_x.
\]
Here  \( \alpha = \sum_{i=1}^{m} \alpha_i\) and  \( \gamma = \sum_{j=1}^{n} \gamma_j\).
\end{Definition}

\subsection{A Smooth Transition to Distribution Geometry}\label{sec:wavefrontset}

Here we concisely place the renormalised algebra of Gaussian Quantum Foam in relation to both the Nigsch–Vickers programme on distributions in a non-linear setting of general relativity \cite{nigsch20201, nigsch20202} and Colombeau theory \cite{colombeau2000}.

We also, initially, position the algebra within the context of the spectral analysis of singularities and thus provide a microlocal perspective. In this respect we will mainly lean on Hörmander’s text \cite{hormander1983} on distribution theory, differential operators, and microlocal analysis. For the benefit of readers who have not previously been exposed to microlocal analysis, we refer to the pedagogical treatise by Brouder et al.\ \cite{brouder2014} and Strichartz’s guide to both distribution theory and microlocal analysis \cite{strichartz2003}, both of which also have been used as background here.

\medskip

To start, there is a common view that renormalisation “solves” the problem of multiplying distributions with coincident singular supports (see, e.g., \cite{brouder2014}). A more accurate formulation, and the one adopted in this work, is that genuinely fundamental (non-derived) observables first appear as \emph{bare} quantities that are generically infinite, and physical (finite) values arise only through vacuum polarisation effects, which effectively shield the bare observable; the shielded quantity becomes measurable only after renormalisation. In other words, vacuum polarisation renders the observable finite, and mathematically we handle this by renormalisation.

This perspective leads directly to the following programme: if we wish to understand physics at the singular support, hence the nature of time and the fabric of spacetime, we need an algebra that allows well-defined products of distributions whose singular supports are not disjoint. There are several technical routes to such products. Many begin with localisation through compactly supported cut-off functions (whose existence is guaranteed by Theorem 1.4.1 in \cite{hormander1983}) pass to Fourier space (placing the problem in a Fréchet setting), and return (via the inverse transform) through convolution to define the product \cite{brouder2014}. A caveat is that these routes often rely on absolute convergence and therefore break down for distributions supported at a point, such as the Dirac measure ( \(\delta:=\delta_{x}\)) and its derivatives. Indeed, if  \(\phi\) is a cut-off with  \(\phi\equiv 1\) near the support point, then the Fourier transform of the modulated Dirac measure is  \(\widehat{\phi\,\delta}=1\) (since  \(\phi\,\delta=\delta\)), so the convolution criterion fails to be absolutely convergent. In fact, by Hörmander’s Theorem~2.3.4 \cite{hormander1983}, any finite-order distribution  \(u\) on an open set  \(U\subset\mathbb{R}^n\) with point support  \(\mathrm{supp}(u)=\{x\}\) and hence a distribution that can be expressed as a finite linear combination of  \(\delta\) and its distributional derivatives, is not covered by that approach. More fundamentally, from the perspective of spectral analysis, the obstruction to defining products on intersecting singular supports is microlocal. Recall that the wave front set  \(WF(u)\subset T^*U\setminus\{0\}\) collects those  \((x,\xi)\) that are \emph{not} regular directed points (i.e. no cut-off at  \(x\) yields a Fourier transform of the distribution that is smooth and rapidly decreasing in a conic neighbourhood of  \(\xi\)). Hörmander’s wave front set criterion, Theorem~8.2.10 \cite{hormander1983}, states that the product  \(uv\), where  \(u\) and  \(v\) are linear distributions on an open set  \(U\subset \mathbb{R}^n\), is well defined as the pullback of the tensor product  \(u\otimes v\) unless there exist  \((x,\xi)\in WF(u)\) and  \((x,-\xi)\in WF(v)\) for some  \((x,\xi)\in U\times(\mathbb{R}^n\setminus\{0\})\). Returning to the Dirac measure with its  \(\operatorname{singsupp}(\delta)=\{0\}\), we have that its product with itself is not defined. As noted above, if  \(\phi\) is a cut-off with  \(\phi\equiv 1\) near the origin, then  \(\phi\,\delta=\delta\) and the Fourier transform of the modulated measure is
\[
\widehat{\phi\,\delta}(\xi) = 1 \quad \text{for all } \xi\in\mathbb{R}^n.
\]
Thus there is no conic neighbourhood in which the Fourier transform is rapidly decreasing, so there are no regular directed covectors  \(\xi\neq 0\). The resulting wave front set is
\[
WF(\delta_x)=\{(0,\xi):\xi\in \mathbb{R}^n\setminus\{0\}\},
\]
and Hörmander’s criterion shows that  \(\delta_x\cdot\delta_x\) is not defined, since for any  \(\xi\neq 0\) we simultaneously have  \((0,\xi)\in WF(\delta_x)\) and  \((0,-\xi)\in WF(\delta_x)\).

Now consider the localised Gaussian sequences in Definition~\ref{def:quantumfoam}, converging to the Dirac measure in  \(\mathcal{S_G}'\). By themselves they do not remove the obstruction: for an  \(n\)-fold product the smooth representatives grow like  \(k^{\,n-1}\), so the naïve product diverges as  \(k\to\infty\). In the present construction, the explicit  \(1/k^{\,n-1}\) scaling of  \(n\)-fold Gaussian products, together with subtraction of universal  \(k^2\)-type terms representing the bare mass monopole in a vacuum renormalisation scheme is essential: it cancels the growth, yielding a distribution supported on the common singular set which, by Hörmander’s Theorem~2.3.4 \cite{hormander1983}, is a finite linear combination of  \(\delta\) and its derivatives (odd coefficients vanish for even representatives). Thus we do not contradict the wave front set criterion; rather—and this remark is crucial—we define a renormalised product at the level of scaled smooth representatives, where the Leibniz rule is preserved by construction and all operations preserve causality and lie inside the null cone, where all points are regular directed and the wave front set is empty. Only then do we pass to the distributional limit within the restricted algebra. Without this scaling, absolute convergence and Fourier–convolution constructions already fail for products involving  \(\delta\) and  \(\partial^\alpha\delta\); with it, they can be rectified in the present Gaussian framework.

This approach thereby extends differential geometry to a distribution geometry by starting from the physical picture: the classical shift vector, which displaces spatial coordinates across the level sets of a global, smooth time function, is the remnant of its quantum counterpart, an operator-valued distribution that displaces the vacuum in coherent states. This allows adjacent regions of positive and negative energy density to emerge and, via non-linear harmonic oscillations, drives the emergence of classical spacetime. We have described this as a physically motivated, tailored, and easy-to-use framework. In this sense it naturally has practical advantages over more general and abstract frameworks such as the Colombeau algebra \cite{colombeau2000} or the fully diffeomorphism-invariant approach of Nigsch and Vickers \cite{nigsch20201, nigsch20202}. However, it is clearly a special case and amounts to a specific gauge choice relative to the Nigsch–Vickers diffeomorphic distribution theory—analogous, perhaps, to the relation between special and general relativity.

From a mathematical standpoint, the renormalised distribution algebra developed here is based on a smooth, unit-integral localised Gaussian kernel familiar to physicists, with Quantum Foam obtained as the limit of spacetimes equipped with such kernels. The construction is grounded in the Colombeau framework—specifically, the special case of a \emph{model delta net} (see Section~2 of \cite{nigsch20201})—but with one difference: we localise or restrict the  delta net using compactly supported cut-off functions. The existence of the limit of the delta net is guaranteed by Hörmander’s Theorem~4.1.5 on sequences of test functions converging, in the sense of distributions, to a prescribed distribution; microlocal consistency is ensured by Theorem~8.2.3 \cite{hormander1983}, which allows the approximants to be chosen so that their wave front sets (and that of the limit) lie in a prescribed closed conic set determined by the characteristics (for hyperbolic operators, the characteristic cone—e.g.\ the null cone). Crucially, the scaling built into the algebra (the  \(k\)-dependent normalisations) and the subsequent renormalisation are what make the distributional limits of non-linear products exist without violating the wave front set criterion.

The algebra is well defined locally on each open set of each Lorentzian spacetime in the sequence of Definition~\ref{def:quantumfoam}, because the kernel is localised or restricted by compactly supported cut-off functions, as described in Section~\ref{sec:review}. Together with Hörmander’s Definition~6.3.3 of distributions on a differentiable manifold, this guarantees that the algebra is globally well defined on spacetime.

By contrast, Nigsch and Vickers develop a fully diffeomorphism-invariant distribution geometry on Lorentzian manifolds using smoothing kernels that extend to the tensor setting, providing a framework aimed at full general covariance. The algebra here is instead tailored specifically to globally hyperbolic spacetimes whose singular support, in the Quantum Foam setting, is a single point; accordingly, the relevant distributions are finite linear combinations of the Dirac measure and its distributional derivatives. Furthermore, the low regularity of the coefficients in that linear combination can be identified with the geon’s bare mass, which is set to zero in the vacuum renormalisation scheme.

\section{The Wave Equation in Gaussian Quantum Foam}\label{sec:waveequation}
The main motivation for compiling these notes has been the identification of a missing component in the formulation of a quantum theory of Gaussian Quantum Foam, namely a non-linear wave equation that provides dynamical content to the shift vector field  \(\beta_{(k)}\). In this section, we now have the tools to both formulate such an equation and establish its consistency within a distributional geometric framework.

The construction relies on the distributional renormalisation algebra  \((\mathcal{S}_\mathcal{G}'(U_{\kappa^{(k)}_t}), \cdot, \partial)\) developed in Sections~\ref{sec:restriction}–\ref{sec:algebraoperations2}, built from sequences of smooth Gaussian functions in a restriction  \(\mathcal{S_G}(U_{\kappa^{(k)}_t})\) of Schwartz space, converging in the sense of distributions. Within this algebra, non-linear operations such as multiplication and differentiation are defined at the level of smooth representatives of the sequences prior to taking the distributional limit. This approach ensures that the non-linear wave operator is well defined and closed on the space of admissible test sequences.

We show that, in the weak-field regime, the non-linear wave equation for the shift vector components reduces to the classical Klein--Gordon equation for a massless scalar field in vacuum. In the singular limit, the source becomes a linear combination of a Dirac delta measure  \(\delta_x\) and its second derivative  \(\delta_x^{\prime\prime}\). This identifies the shift vector as a singular geometric source, capturing the curvature response of a collapsing Quantum Foam element within a consistent distributional framework.

\medskip

For each spacetime element in the Gaussian Quantum Foam spacetime defined in Definition~\ref{def:quantumfoam},
with metric field in compact matrix form
\begin{equation}\label{eq:metricmatrix}
g_{\mu\nu}^{(k)} =
\begin{pmatrix}
\beta^{(k)}_i\beta_i^{(k)}-N_{(k)}^{2}& \beta_j^{(k)} \\
\beta_i^{(k)} & \eta_{ij}
\end{pmatrix},
\end{equation}
and inverse
\begin{equation} \label{eq:inversemetric}
g^{\mu\nu}_{(k)} = \frac{1}{N_{(k)}^{2}}
\begin{pmatrix}
-1 & \beta^j_{(k)} \\
\beta^i_{(k)} & N_{(k)}^{2}\eta^{ij}-\beta^i_{(k)}\beta^j_{(k)}
\end{pmatrix},
\end{equation}
the non-linear wave operator acting on  \(\beta_{(k)}^x\) is
\begin{equation}
\begin{aligned} \label{eq:waveequation1}
\Box_{g^{(k)}} \beta^x_{(k)}
&= \frac{1}{\sqrt{|g^{(k)}|}}\,\partial_\mu\!\left( \sqrt{|g^{(k)}|}\,g_{(k)}^{\mu\nu}\,\partial_\nu \beta_{(k)}^x \right) \\[6pt]
&= -\frac{1}{N_{(k)}^2} \biggl(
  2\,\beta_{(k)}^x \big(\partial_x\beta_{(k)}^x\big)^2
  + \beta_{(k)}^x \,\partial_x \beta_{(k)}^x \,\partial_y \beta_{(k)}^y \\
&\qquad\qquad\qquad\quad
  + \beta_{(k)}^x \,\partial_x \beta_{(k)}^x \,\partial_z \beta_{(k)}^z
  - \big(N_{(k)}^2 - (\beta_{(k)}^x)^2\big)\,\partial_x^2 \beta_{(k)}^x
\biggr).
\end{aligned}
\end{equation}

\begin{Remark}[Same form for \(\beta_{(k)}^y\) and \(\beta_{(k)}^z\)]
The wave equation for the other components of the shift vector has the same form. Hence the discussion can be restricted to a single component without loss of generality.
\end{Remark}

To analyse the singular structure of this equation, for the lapse scaling  \(N_{(k)} = k\) and test functions in the admissible space of test functions in Definition~\ref{def:test-functions}, each term in~\eqref{eq:waveequation1} is evaluated in the distributional algebra  \((\mathcal{S}_\mathcal{G}^\prime, \cdot, \partial)\) defined in Definition~\ref{def:algebra}, using the rules given in Sections~\ref{sec:algebraoperations} and~\ref{sec:algebraoperations2}.

\noindent
The cross-terms with  \(\partial_y \beta_{(k)}^y\) and  \(\partial_z \beta_{(k)}^z\) vanish when tested against an even, positive test function in the admissible space (Definition~\ref{def:test-functions}). The remaining non-vanishing terms in the wave equation \eqref{eq:waveequation1} are 
\begin{equation} \label{eq:wvequad}
\frac{2}{k^2}\,\beta_{(k)}^x \big(\partial_x\beta_{(k)}^x\big)^2,
\end{equation}
and
\begin{equation}  \label{eq:wavesecond}
\left(1 - \frac{(\beta_{(k)}^x)^2}{k^2} \right) \partial_x^2 \beta_{(k)}^x
= \partial_x^2 \beta_{(k)}^x - \frac{(\beta_{(k)}^x)^2}{k^2}\,\partial_x^2 \beta_{(k)}^x.
\end{equation}

The former of these two terms can be handled using Definition~\ref{def:prodfirstfirstorder} by absorbing the factor  \(\beta_{(k)}^x\) into the derivative factor. Nevertheless, we find it useful to work out the details here so that the renormalisation becomes transparent. Using the Gaussian defined in Definition~\ref{def:quantumfoam} and the standard second derivative of a Gaussian, we have, in a functional sense, that
\[
\begin{aligned}
\lim_{k\to\infty}\frac{2}{k^2}\,\beta_{(k)}^x \big(\partial_x\beta_{(k)}^x\big)^2
&= \frac{2}{36\pi\sigma^2}\lim_{k\to\infty}\biggl[\frac{9k^4x^2}{4\sigma^4}\beta_{(k)}^{\sqrt{3}x}
-\frac{3k^2}{2\sigma^2}\beta_{(k)}^{\sqrt{3}x}
+\frac{3k^2}{2\sigma^2}\beta_{(k)}^{\sqrt{3}x} \biggr] \\
&=\frac{2}{36\pi\sigma^2}\lim_{k\to\infty}\biggl[\partial^2_x\beta_{(k)}^{\sqrt{3}x}
+\frac{3k^2}{2\sigma^2}\beta_{(k)}^{\sqrt{3}x} \biggr]
=\biggl[\frac{\sqrt{3}}{54\pi\sigma^2}\delta^{\prime\prime}_{x^i}\biggr]_{r}.
\end{aligned}
\]
Here, the first term within the brackets in the equality before the renormalised result \([\cdot]_{r}\) converges to a scaled second-order distributional derivative, while the second term represents the infinite bare mass contribution that we remove by renormalisation in the limit as  \(k\to\infty\), so that the result makes sense in distribution geometry. On the other hand, for finite values of the sequence index  \(k\), this term represents a contribution, \(m_{B+}^{(k)}\), to the observable mass of a geon:
\begin{equation} \label{eq:mbplus}
m_{B+}^{(k)}=\frac{\sqrt{3}}{36\pi\sigma^4}k^2.
\end{equation}

Continuing with \eqref{eq:wavesecond}, we observe that it consists of a standard second-order derivative, converging to the second-order distributional derivative of the Dirac measure, and a term that can be worked out in a functional sense and in the algebra simply by computing the second derivative and rearranging the terms before the limit is taken. We then obtain (in a functional sense of smooth localised sequences converging in distributions)
\begin{equation} \label{eq:tempwaveterm}
\lim_{k\to\infty} \frac{1}{k^2}(\beta_{(k)}^x)^2\partial_x^2 \beta_{(k)}^x =\lim_{k\to\infty} \frac{1}{(2\sigma\pi)^2}\biggl[-\frac{k^2}{2\sigma^2}+\frac{k^4x^2}{4\sigma^4}\biggr]\beta^{\sqrt{3}x}_{(k)}
=\lim_{k\to\infty}\frac{1}{36\pi\sigma^2}\biggl[\partial^2_x\beta^{\sqrt{3}x}_{(k)}-\frac{3k^2}{\sigma^2}\beta^{\sqrt{3}x}_{(k)}\biggr].
\end{equation}
The first term in this result converges to a second-order distributional derivative expression of the Dirac measure,
\begin{equation} \label{expr:a2delta2}
\frac{\sqrt{3}}{108\pi\sigma^2}\delta^{\prime\prime}_{x},
\end{equation}
while the remaining term represents part of the observable bare mass contribution,  \(m_{B-}^{(k)}\), in the field equation. Using that the integral of a scaled Gaussian is the reciprocal of the scaling factor, we find
\begin{equation} \label{eq:mbminus}
m_{B-}^{(k)}=-\frac{\sqrt{3}}{36\pi\sigma^4}k^2.
\end{equation}
To have a sensible result in distribution geometry, this term must be renormalised in \eqref{eq:tempwaveterm}.

\medskip

\noindent Changing perspective, we can arrive at the same result using the rule in Definition~\ref{def:prodsecondorder}:
\[
\lim_{k\to\infty} \frac{1}{k^2}(\beta_{(k)}^x)^2\partial_x^2 \beta_{(k)}^x
= a_0\,\delta_x + a_2\,\delta_{x}^{\prime\prime}.
\]
As discussed earlier, the coefficient  \(a_0^{(k)}\) must be renormalised since it contributes to a bare mass term:
\begin{equation}
 a_0^{(k)}
=\big\langle\tfrac{1}{k^2}(\beta^x_{(k)})^2\partial^2\beta^x_{(k)}, 1\cdot\phi^{(k)}_c\big\rangle
=\frac{1}{36\pi\sigma^2}\int \biggl[\partial^2_x\beta^{\sqrt{3}x}_{(k)}-\frac{3k^2}{\sigma^2}\beta^{\sqrt{3}x}_{(k)}\biggr]dx
=-\frac{\sqrt{3}}{36\pi\sigma^4}k^2=m_{B-}^{(k)}
\end{equation}
Now, adding the bare mass contributions \eqref{eq:mbplus} and \eqref{eq:mbminus} to the field equation \eqref{eq:waveequation1}
for the observable, emerging spacetime and for each spacetime element
\((M^{(k)}, g_{\mu\nu}^{(k)})\), we obtain
\begin{equation} \label{eq:total_field_bare_mass}
    m_{B}^{(k)} = m_{B+}^{(k)} + m_{B-}^{(k)} = 0.
\end{equation}
This is an expected result: in the emerging spacetime, a region of positive energy
density is separated from a region of negative energy density by a vacuum shell,
and their contributions must cancel.

\medskip

\noindent To determine the finite coefficient  \(a_2\) above, we can use the result in \eqref{eq:a2}, but for the reader’s convenience we work out some of the details here. Thus, formally—as an integral in the functional sense—using integration by parts and the fact that the localised Gaussian lies in the restricted Schwartz space, we have
\[
\begin{aligned}
a_2^{(k)}
&=\big\langle \tfrac{1}{k^2}(\beta_{(k)}^x)^2\partial_x^2 \beta_{(k)}^x,\tfrac{x^2}{2}\cdot\phi^{(k)}_c\big\rangle \\
&=
-\int\tfrac{1}{k^2}(\beta^x_{(k)})^2\partial_x\beta^x_{(k)}\,x\,\phi^{(k)}_c\,dx
-\int \tfrac{1}{k^2}\beta^x_{(k)}(\partial_x\beta^x_{(k)})^2\,x^2\,\phi^{(k)}_c\,dx. 
\end{aligned}
\]
Notice that the first integral involves a quadratic polynomial factor arising from the product  \(x\,\partial_x\beta_{(k)}^x\). This allows us to rearrange and rewrite the expression as a second-order Gaussian derivative plus an additional term, where that additional term is bounded and, in fact, constant due to the scaling with  \(k^{-2}\). Hence,
\[
\big\langle \tfrac{1}{k^2}(\beta^x_{(k)})^2\partial_x\beta^x_{(k)},x\big\rangle
=\frac{1}{18\pi\sigma^2}\int \biggl( -\partial^2_x\beta_{(k)}^{\sqrt{3}x} - \frac{3}{2\sigma^2}\beta_{(k)}^{\sqrt{3}x}\biggr)dx
=-\frac{\sqrt{3}}{36\pi\sigma^2}.
\]
Here the cut-off function  \(\phi^{(k)}_c\) has been suppressed, and we have used that the Gaussian vanishes on the boundary, that the integral of the scaled Gaussian equals  \(1/\sqrt{3}\) for each finite  \(k\), and that we are working in the distributional setting. Similarly, we have
\[
\big\langle\tfrac{1}{k^2}\beta^x_{(k)}(\partial_x\beta^x_{(k)})^2,x^2\cdot\phi^{(k)}_c\big\rangle = \frac{\sqrt{3}}{54\pi\sigma^2}.
\]
Collecting the terms, we get
\[
a_2 = \frac{\sqrt{3}}{108\pi\sigma^2}\quad \text{for all } k,
\] 
which is the same coefficient as in earlier expression \eqref{expr:a2delta2}.
To summarise, in a functional sense we have in vacuum that
\[
2\,\beta_{(k)}^x \big(\partial_x\beta_{(k)}^x\big)^2-
\left(1 - \frac{(\beta_{(k)}^x)^2}{k^2} \right)\partial_x^2 \beta_{(k)}^x 
\xrightarrow{\mathcal{S_G}^{\prime}} \biggl(\frac{\sqrt{3}}{36\pi\sigma^2}-1\biggr)\delta_x^{\prime\prime}.
\]
Substituting this into the wave equation \eqref{eq:waveequation1}, we conclude that, in renormalised form, the vacuum field equation for the (unstable) Gaussian Quantum Foam element, or geon, is
\[
\Box_{g_{(k)}} \beta_{(k)}^x \xrightarrow{\mathcal{S_G}^{\prime}}\biggl(1-\frac{\sqrt{3}}{36\pi\sigma^2}\biggr)\delta_x^{\prime\prime}.
\]

In the weak-field regime  \(|\beta_{(k)}|\ll N_{(k)}\) (and for slowly varying spatial dependence), the non-linear terms are negligible and the wave equation reduces to
\[
\Box_{g^{(k)}} \beta_{(k)}^x = \partial_x^2 \beta_{(k)}^x + \mathcal{O}((\beta_{(k)}^x)^2)= 0,
\]
that is, the massless vacuum Klein–Gordon equation. This motivates the following general and renormalisation-scheme-independent definition of the non-linear wave equation for the shift vector components:
\begin{Definition}[Non-linear Wave Operator Equation in Distribution Geometry]\label{def:wave-equation}
Let  \(\{\beta_{(k)}^{x^i}\}_{k\in\mathbb{N}} \subset \mathcal{S}_\mathcal{G}(U_{\kappa^{(k)}_t})\) be the sequence of shift vector components, in Definition~\ref{def:quantumfoam} of a Gaussian Quantum Foam element, converging in the sense of distributions to a Dirac measure in  \(\mathcal{S}'(U_{\kappa^{(k)}_t})\). Then the renormalised non-linear wave equation for the Gaussian Quantum Foam is given by:
\begin{equation}
\lim_{k \to \infty} \Box_{g^{(k)}} \beta_{(k)}^{x^i} := a_i\,\delta_{x^i}+ b_i\,\delta^{\prime\prime}_{x^i},
\end{equation}
where  \(i = 1, 2, 3\),  \(a_i\in \mathbb{R}\) are renormalised, scheme-dependent constants, and  \(b_i>0\) are universal, scheme-independent constants.
\end{Definition}
The second-order distributional derivative  \(\delta^{\prime\prime}_{x^i}\) appearing in the wave equation acts as a singular curvature source: a sharp, localised response of the quantum geometry to the collapse of a Quantum Foam element. In analogy with signal theory,  \(\delta^{\prime\prime}_{x^i}\) may be interpreted as an idealised acceleration impulse—the limit of a sharply peaked force gradient that sets the vacuum displacement and oscillations in motion, generating regions of positive and negative energy separated by a vacuum shell. As classical spacetime emerges and the quantum aspects of the oscillations around vacuum (and hence vacuum fluctuations) fade in accordance with the correspondence principle in~\cite{cramer20251}, only remnants of the quantum dynamics remain, exhibited in the geometric shifting of coordinates across the hypersurfaces.

\medskip

\noindent The critical reader may note that the field equation for Gaussian Quantum Foam was introduced in what could be regarded as an \emph{ad hoc} manner. However, this is not the case. In what follows we show that the wave equation in Definition~\ref{def:wave-equation} arises naturally from a variational principle applied to the Einstein–Hilbert action for a sequence of spacetimes converging, in the sense of distributions, to Gaussian Quantum Foam according to Definition~\ref{def:quantumfoam}. Thus, the field equation in Definition~\ref{def:wave-equation} is the field equation \emph{in the Gaussian gauge for the shift sector}. Since, as shown in Section~\ref{sec:review}, a well-defined Hilbert space can be constructed on each hypersurface and a bosonic quantum field theory formulated for the shift vector field~\eqref{eq:gaussian}, it follows that general relativity admits a natural quantum field-theoretic formulation within this framework.

To make this statement explicit, consider the sequence of spacetimes \(\{(M_{(k)}, g^{(k)}_{\mu\nu})\}_{k\in\mathbb{N}}\) in Definition~\ref{def:quantumfoam}. These are globally hyperbolic spacetimes on a fixed manifold, with a metric field \(g_{\mu\nu}^{(k)}\) as in \eqref{eq:metricmatrix} and hence with the line element
\[
ds^2_{(k)} = -N_{(k)}^2\,dt^2 + \eta_{ij}\big(dx^i+\beta_{(k)}^i dt\big)\big(dx^j+\beta_{(k)}^j dt\big).
\]
Here, as before, the lapse scales with the sequence index \(k\) but is otherwise constant,  \(N_{(k)}=k\), and the components of the shift vector are static in the Schrödinger picture. 

The Einstein–Hilbert action—used here specifically to show that general relativity contains its own quantum theory—can then be written as
\begin{equation}
S^{(k)}[\beta] =\frac{1}{16\pi}\int d^4x\, N_{(k)}\,g^{\mu\nu}_{(k)} R_{\mu\nu}^{(k)}.
\label{eq:ehaction}
\end{equation}
We have used \(\sqrt{|g_{(k)}|}=N_{(k)}\) for each element in the Gaussian Quantum Foam, so the integral is well defined in the Lebesgue sense. It should be noted that the Gibbons–Hawking–York boundary term vanishes in the present setting, since the shift is supported inside a compact region (restricted Schwartz space modulated by a compactly supported test function), and we impose standard falloff/compact-support conditions so that boundary contributions are zero.

Using the principle of variations and calculus of variations, we then have
\begin{equation} \label{eq:variationaction}
\delta S^{(k)}[\beta] = \frac{1}{16\pi}\int d^4x\left(
N_{(k)}\,\delta g^{\mu\nu}_{(k)}[\beta] \, R_{\mu\nu}^{(k)} \;+\;
N_{(k)}\,g^{\mu\nu}_{(k)}\,\delta R_{\mu\nu}^{(k)}[\beta]\right).
\end{equation}

To calculate the variation of the metric field, use the inverse metric \eqref{eq:inversemetric}, from which it follows that the independent variations with respect to the shift are
\begin{equation} \label{eq:variationmetric}
\delta g^{00}_{(k)} = 0,\qquad
\delta g^{0i}_{(k)} = \frac{\delta\beta^i}{N_{(k)}^{2}},\qquad
\delta g^{ij}_{(k)} = -\,\frac{\beta^i_{(k)}\,\delta\beta^j + \beta^j_{(k)}\,\delta\beta^i}{N_{(k)}^{2}}.
\end{equation}

Thus, the varied Einstein–Hilbert action takes the form
\begin{equation} \label{eq:variationaction2}
\delta S^{(k)}[\beta] = \frac{1}{16\pi}\int d^4x\,N_{(k)}\left[
\sum_{i=1}^3\left(R_{0i}^{(k)} -2\beta^j_{(k)}R_{ij}^{(k)}\right)\delta\beta^{i} \;+\;
g^{\mu\nu}_{(k)}\,\delta R_{\mu\nu}^{(k)}[\beta]\right].
\end{equation}
By the Palatini identity,  \(g^{\mu\nu}\delta R_{\mu\nu}=\nabla_\alpha(\cdots)\). Because the lapse  \(N_{(k)}\) is constant in spacetime and the shift lies in the localisation of the Schwartz space with compact support, this contribution is purely a boundary term and vanishes under the stated falloff conditions. Hence, only the first term in \eqref{eq:variationaction2} contributes to the equations of motion.

To continue from here we use \texttt{SageMath}~\cite{sagemath2022} to compute the non-zero components of the Ricci tensor in the first term of the varied action \eqref{eq:variationaction2}:
\begin{align}
R_{01}^{(k)}
&= \frac{1}{N^2_{(k)}}\biggl[
   \beta^{x}_{(k)}\,(\partial_x\beta^{x}_{(k)})^{2}
 + (\beta^{x}_{(k)})^{2} \,\partial^{2}_{x}\beta^{x}_{(k)}
 + \beta^{x}_{(k)}\,\partial_x\beta^{x}_{(k)}\,\partial_y\beta^{y}_{(k)}
 + \beta^{x}_{(k)}\,\partial_x\beta^{x}_{(k)}\,\partial_z\beta^{z}_{(k)}
\biggr], \\[6pt]
R_{02}^{(k)}
&= \frac{1}{N^2_{(k)}}\biggl[
   \beta^{y}_{(k)}\,\partial_x\beta^{x}_{(k)}\,\partial_y\beta^{y}_{(k)}
 + \beta^{y}_{(k)}\,(\partial_y\beta^{y}_{(k)})^{2}
 + (\beta^{y}_{(k)})^{2} \,\partial^{2}_{y}\beta^{y}_{(k)}
 + \beta^{y}_{(k)}\,\partial_y\beta^{y}_{(k)}\,\partial_z\beta^{z}_{(k)}
\biggr], \\[6pt]
R_{03}^{(k)}
&= \frac{1}{N^2_{(k)}}\biggl[
   \bigl(\partial_x\beta^{x}_{(k)} + \partial_y\beta^{y}_{(k)}\bigr)\,\beta^{z}_{(k)}\,\partial_z\beta^{z}_{(k)}
 + \beta^{z}_{(k)}(\partial_z\beta^{z}_{(k)})^{2}
 + (\beta^{z}_{(k)})^{2}\,\partial^{2}_{z}\beta^{z}_{(k)}
\biggr], \\[6pt]
R_{11}^{(k)}
&= \frac{1}{N^2_{(k)}}\biggl[
   (\partial_x\beta^{x}_{(k)})^{2}
 + \beta^{x}_{(k)}\,\partial^{2}_{x}\beta^{x}_{(k)}
 + \partial_x\beta^{x}_{(k)}\,\partial_y\beta^{y}_{(k)}
 + \partial_x\beta^{x}_{(k)}\,\partial_z\beta^{z}_{(k)}
\biggr], \\[6pt]
R_{22}^{(k)}
&= \frac{1}{N^2_{(k)}}\biggl[
   \partial_x\beta^{x}_{(k)}\,\partial_y\beta^{y}_{(k)}
 + (\partial_y\beta^{y}_{(k)})^{2}
 + \beta^{y}_{(k)}\,\partial^{2}_{y}\beta^{y}_{(k)}
 + \partial_y\beta^{y}_{(k)}\,\partial_z\beta^{z}_{(k)}
\biggr], \\[6pt]
R_{33}^{(k)}
&= \frac{1}{N^2_{(k)}}\biggl[
   \bigl(\partial_x\beta^{x}_{(k)} + \partial_y\beta^{y}_{(k)}\bigr)\,\partial_z\beta^{z}_{(k)}
 + (\partial_z\beta^{z}_{(k)})^{2}
 + \beta^{z}_{(k)}\,\partial^{2}_{z}\beta^{z}_{(k)}
\biggr].
\end{align}
Together, this gives
\begin{equation}
\sum_{i=1}^3\!\left(R_{0i}^{(k)} - 2\,\beta^j_{(k)}\,R_{ij}^{(k)}\right)\delta\beta^{i}
= -\sum_{i=1}^3 R_{0i}^{(k)}\,\delta\beta^{i}.
\end{equation}
Since  \(N_{(k)}\) is constant and the shift components have compact support, we first take the variation as written and then perform a single integration by parts in each spatial direction (boundary terms vanish). That is, we use that
\[
\begin{aligned}
\frac{1}{N^2_{(k)}}\int\partial_x \bigl(N^2_{(k)}-(\beta^x_{(k)})^2\bigr)\,\partial_x\beta^x_{(k)}\,dx
&=-\frac{2}{N^2_{(k)}}\int\beta^x_{(k)}\bigl(\partial_x\beta^x_{(k)}\bigr)^2\,dx.
\end{aligned}
\]
This together yields the wave equation \eqref{def:wave-equation} for the  \(x\)-component, and the  \(y\)- and  \(z\)-components follow by cyclic permutation.

This completes the demonstration that the theoretical framework of Gaussian Quantum Foam, which arises as the limit—in the sense of distributions—of sequences of globally hyperbolic spacetimes, is self-consistent and intrinsically equipped with a well-defined field equation. From this equation, a bosonic quantum field theory can be constructed without the difficulties that usually plague quantum gravity, such as problems in defining the volume-element measure in Lebesgue’s sense and, consequently, in constructing the Hilbert space (see, e.g., \cite{kiefer2025}) that forms the basis of the Fock space for the field, quantised as an operator-valued distribution represented by an infinite, weighted sum of harmonic oscillators.

We now leave the discussion about the wave equation and continue the application of the distributional algebra to examine whether the notion of a singularity retains any meaning in the context of Gaussian Quantum Foam, given that a well-defined distribution geometry can be formulated in which continuity and differentiability are preserved, albeit in the distributional sense.

\section{Singularities and the Quantum Foam Paradigm}\label{sec:singularities}

This section addresses whether the notion of a ``singularity'' retains operational meaning in Gaussian Quantum Foam, and if so, in what sense. Although continuity and differentiability remain meaningful in this framework, they are interpreted in the distributional sense. As reviewed in Section~\ref{sec:wavefrontset}, the renormalised distribution algebra and the resulting distributional geometry are compatible with the microlocal picture even though the wave front set is not empty in the Gaussian Quantum Foam limit. In the limit  \(k\to\infty\), the normals to the level sets become null, the level sets themselves become characteristic hypersurfaces, and the configuration realises an unstable self-gravitating geon with a divergent bare mass monopole. There is no microlocal inconsistency because all non-linear operations are performed at finite  \(k\), where every phase space point is a regular directed point (so  \(WF=\varnothing\)), and only then is the distributional limit taken; renormalisation removes the universal divergences so that the limit is a well defined distribution. Hence continuity and differentiability persist, but only in the distributional sense.

This introduces two distinct perspectives that must be reconciled in the transition from a smooth differential geometry to a non-linear distribution geometry. It is therefore essential that, throughout the forthcoming discussion, not only with respect to singularities but across all aspects of the Gaussian Quantum Foam framework, we clearly distinguish between these two closely related but fundamentally different perspectives.

On the one hand, we consider the local physics associated with each element in the sequence of smooth, globally hyperbolic, and homotopic spacetimes in Definition~\ref{def:quantumfoam}. These are classical spacetimes for finite values of the sequence index  \(k\), where all geometric quantities are smooth and well defined. On the other hand, the Quantum Foam element itself is defined as the distributional limit of this sequence. It consists of a singular support together with a well-defined distributional structure that includes curvature, shift vector fields, and extrinsic geometry. 

This distinction is crucial because, while finite- \(k\) spacetimes exhibit smooth variations in curvature, energy density, and congruence expansion, the limiting element encodes sharply supported impulses and singular features that drive the emergence of classical spacetime from vacuum displacement.

\medskip 

With this remark in mind, we are now ready to start the discourse on the notion of singularity in Quantum Foam, and we will do so by briefly recalling the notion from a classical perspective. In classical general relativity, singularities are inferred from the existence of incomplete and inextensible causal geodesics, provided that certain conditions hold, most notably the positivity of the scalar projection of the Ricci curvature along causal vectors. This condition, either for a projection along timelike or null vectors, is a crucial aspect in the formulation of the Penrose–Hawking singularity theorems~\cite{penrose1965, hawking1967, hawking1970}. See Chapter 9 of Wald's \emph{General Relativity}~\cite{wald1984} for a detailed treatment, Steinbauer's recent pedagogical review~\cite{steinbauer2023}, written in honour of Penrose, or Kontou and Sanders' treatment~\cite{kontou2020} of the energy conditions in a classical and semi-classical context and their applications to the singularity theorems.

However, as Wald notes, there exists no universally accepted, precise, and local definition of a singularity; see Section 9.1 in Chapter 9 of \cite{wald1984}. This is likely a consequence of the fact that geodesic incompleteness does not provide much information about the very nature of a singularity. Thus, the standard heuristic picture of a "point" where a null or timelike geodesic abruptly ends~\cite{steinbauer2023} is not sufficient in the quantum context. In the present framework, we replace this notion with that of a singular support: a sharply localised set within the distribution geometry where all geometric quantities remain well defined in the sense of distributions, even as classical notions of smoothness break down.

In the Gaussian Quantum Foam model, this singular support is accompanied by a second-order distributional curvature term that acts as a catalytic impulse: it drives the shift vector field into oscillatory motion and initiates the emergence of spacetime itself—namely, the finite- \(k\) elements in the sequence of smooth spacetimes in Definition~\ref{def:quantumfoam}. The shift vector is modelled as a sequence of scaled Gaussians  \( \beta^i_{(k)}\), converging in the distributional sense to a Dirac measure. Thus, the geometry of the Gaussian Quantum Foam is encoded in a sharply supported curvature impulse located at the origin, where time emerges from the non-linear dynamics of the shift vector field.
The timelike scalar projection of the curvature  \( R^{(k)}_{\mu\nu} u_{(k)}^{\mu} u_{(k)}^{\nu}\), which appears in the Penrose–Hawking singularity theorems, is related to the energy content of spacetime through the Einstein field equations. In particular, it can be expressed as a linear combination of the timelike scalar projection of the energy density and the trace of the stress-energy tensor.
\begin{equation}
R^{(k)}_{\mu\nu} u_{(k)}^{\mu} u_{(k)}^{\nu} = 8\pi \left( T^{(k)}_{\mu\nu} u_{(k)}^{\mu} u_{(k)}^{\nu} + \frac{1}{2} T^{(k)} \right),
\end{equation}
where
\[
T^{(k)} = g^{\mu\nu}_{(k)} T^{(k)}_{\mu\nu}= -\frac{1}{8\pi}R^{(k)}
\]
is the trace of the stress-energy tensor. We will therefore begin with some fundamental observations about these two quantities — the scalar projection of the stress-energy and the Ricci scalar — both of which remain well-defined throughout the sequence of globally hyperbolic and homotopic spacetimes \(\{(M_{(k)},g^{(k)}_{\mu\nu})\}_{k\in\mathbb{N}}\) converging to a Gaussian Quantum Foam element in Definition ~\ref{def:quantumfoam}.

In this context, the weak energy condition is taken to mean that the projected energy density along the worldlines of any Eulerian observer is non-negative. In the case of Gaussian Quantum Foam, this condition does not hold locally~\cite{cramer20251, cramer20252}. To see this, we follow ~\cite{cramer20252}, and use \texttt{SageMath}~\cite{sagemath2022} symbolic calculation engine to calculate the stress-energy tensor in the  \(3+1\) decomposition for a globally hyperbolic and homotopic spacetime:
\begin{equation} \label{eq:tuu}
T^{(k)}_{\mu\nu}u_{(k)}^{\mu}u_{(k)}^{\nu}= \frac{1}{8\pi N_{(k)}^2} \left(\partial_x\beta_{x}^{(k)} \partial_y\beta_{y}^{(k)} + {\left(\partial_x\beta_{x}^{(k)} + \partial_y\beta_{y}^{(k)}\right)} \partial_z\beta_{z}^{(k)}\right).
\end{equation}

In the setting of Gaussian Quantum Foam, where the shift vector is given by \eqref{eq:gaussian}, the projected stress-energy \eqref{eq:tuu} simplifies to:
\begin{equation}\label{eq:tuu2}
   T^{(k)}_{\mu\nu}u_{(k)}^{\mu}u_{(k)}^{\nu} = \frac{k^4}{32\sigma^4\pi N_{(k)}^2}\biggl(xy\beta_{(k)}^x\beta_{(k)}^y + xz\beta_{(k)}^x\beta_{(k)}^z + yz\beta_{(k)}^y\beta_{(k)}^z\biggr).
\end{equation}

For any finite  \( k\), the local projected energy density \eqref{eq:tuu2} takes both positive and negative values, separated by surfaces where the expression on the right-hand side vanishes. For example, it vanishes when  \( x = 0 \land z = 0\) or  \( y = 0 \land z = 0\), is positive in the first and third quadrants in the  \( x\text{--}y\) plane with \(z=0\) and negative in the second and fourth. Thus, locally is the weak energy condition not satisfied.

In contrast, in the distribution limit with  \( N_{(k)} = k\), and using that we only consider even, positive and locally concave test functions since any odd test function is in the orthogonal complement to the Gaussian sector and the Gaussian are locally peaked, that is, test functions in the admissible space of test functions \ref{def:test-functions}, then the projected stress-energy vanishes,
\begin{equation} \label{eq:inttuu}
   \lim_{k\to\infty} \int_{U_{\kappa^{(k)}}} T^{(k)}_{\mu\nu} u_{(k)}^{\mu} u_{(k)}^{\nu} \, \varphi \, \sqrt{|g_{(k)}|}\, d^4x = \lim_{k\to\infty}\int_{U_{\kappa^{(k)}_t}} T^{(k)}_{\mu\nu} u_{(k)}^{\mu} u_{(k)}^{\nu}\, \phi\, N_{(k)}\, d^3x=0.
\end{equation}
This statement is true, since each shift component \(\beta_{(k)}^i\) depends only on its own coordinate \(x^i\), so all terms of the form \(\partial_i\beta_{(k)}^j\) with \(i\neq j\) vanish. The remaining terms are products of first derivatives in different coordinates, i.e. \(\partial_i\beta_{(k)}^i\,\partial_j\beta_{(k)}^j\) with \(i\neq j\), which converge (in the sense of distributions) to the well-defined product \(\delta'_{x^i}\,\delta'_{x^j}\). For positive test functions \(\phi\) that are even in each coordinate, these terms evaluate to zero. In fact, due to the specific choice of the shift vector, no rule from the renormalised distribution algebra (Definition~\ref{def:algebra}) is required here. Moreover, \eqref{eq:inttuu} holds not only in the sense of distributions but for all finite \(k\), because each integrand is odd in the relevant coordinate(s) and hence integrates to zero against a positive and even test function. 

The claim that \eqref{eq:inttuu} holds for all \(k\) deserves clarification. In a purely classical setting (finite \(k\)), there exist local regions of positive energy density separated from regions of negative energy density by a vacuum shell, yet the spatial average vanishes. Thus, the averaged weak energy condition holds both classically and—more importantly—at the singular support of the spacetime sequence in Definition~\ref{def:quantumfoam}. For a more extensive discussion of the averaged weak energy condition and its relation to semi-classical gravity, see~\cite{kontou2020}. We note, however, that within distribution theory the term “averaged” can be misleading at the singular support: strictly speaking, one has a weak energy condition in the sense of distributions. On the other hand, any physical measurement entails averaging, so the wording remains sufficiently precise in physics.

The following remark clarifies why we are allowed to reduce the spacetime integral to a spatial integral:
\begin{Remark} \label{spatialint}
The reduction to a spatial integral in ~\ref{eq:inttuu} follows from the fact that, for any test function in the admissible space of test functions in definition \ref{def:test-functions}, we can write
\[
\varphi(t,\mathbf{x}) = \chi(t)\,\phi(\mathbf{x}), \quad \phi \in \mathcal{S}(U_{\kappa^{(k)}_t}),
\]
with  \(\chi \in C^\infty_0((0,1))\) satisfying  \(\int_0^1 \chi(t)\,dt = 1\).
This decomposition is possible because the lapse, shift and induced metric on the hypersurfaces are static and consequently time independent. Consequently,
\[
\int_{U_{\kappa^{(k)}}}T^{(k)}_{\mu\nu} u_{(k)}^{\mu} u_{(k)}^{\nu} \, \varphi\, \sqrt{|g_{(k)}|}\, d^4x
= \int_{U_{\kappa^{(k)}_t}} T^{(k)}_{\mu\nu} u_{(k)}^{\mu} u_{(k)}^{\nu}\, \phi\, N_{(k)}\, d^3x.
\]
\end{Remark}

We now turn to the Ricci scalar  \( R^{(k)}\), which is proportional to the trace of the stress-energy tensor via the Einstein equations and hence also enters into considerations of energy conditions. Recall from the discussion on the wave equation in the previous section that by using \texttt{SageMath} symbolic calculation engine, we can calculate the Ricci scalar in the Gaussian Quantum Foam setting:
\begin{equation}
\begin{split}
    R^{(k)}= \frac{2}{N_{(k)}^2}\biggl(
        (\partial_x\beta_{(k)}^{x})^{2} + \beta_{(k)}^{x}\partial^{2}_x\beta_{(k)}^{x}
        + \partial_x\beta_{(k)}^{x}\partial_y\beta_{(k)}^{y}
        + (\partial_y\beta_{(k)}^{y})^{2} + \beta_{(k)}^{y}\partial^{2}_y\beta_{(k)}^{y}\\
        + \left(\partial_x\beta_{(k)}^{x}
        + \partial_y\beta_{(k)}^{y}\right)\partial_z\beta_{(k)}^{z}
        + (\partial_z\beta_{(k)}^{z})^{2} + \beta_{(k)}^{z}\partial^{2}_z\beta_{(k)}^{z}
    \biggr).
    \end{split}
\end{equation}
It is relatively straightforward to see that for finite values of the sequence index, and given that the terms involving second derivatives of Gaussians necessarily introduce a second-order polynomial factor with two distinct real roots, then there exist local regions near the roots where the Ricci scalar changes sign. Continuing, and using the same arguments as in the analysis of the wave equation—thus working in the distributional algebra  \((\mathcal{S}_\mathcal{G}^\prime, \cdot, \partial)\) defined in (\ref{def:algebra}) and applying the rules from Sections~\ref{sec:algebraoperations} and~\ref{sec:algebraoperations2} with  \( N_{(k)} = k\), we find that, for any test function that is positive, even and locally concave (that is, a  test function \(\phi\) in the space of admissible test functions given in definition \ref{def:test-functions}), 
\begin{equation} \label{ricciscalar}
    \lim_{k\to\infty} \int_{U_{\kappa^{(k)}_t}} R^{(k)}\phi N_{(k)} d^3x= \sum_{i=1}^3 \left (a_i\langle\delta_{x^{i}},\phi\rangle + b_i\langle\delta^{\prime\prime}_{x^{i}},\phi\rangle\right),
\end{equation}
where the \(a_i<0\) and \(b_i>0\), \(i=1,2,3\), see Remark \ref{re:delta_coefficients}. As before, we have used that for any test function  \( \varphi \in \mathcal{S}(U_\kappa)\), the lapse, shift, and induced metric are time-independent. Moreover, the shift is positive, even and locally concave in a neigbourhood of the singular supprt. Thus, we only consider test functions in the admissible space of test functions in definition \ref{def:test-functions}
\[
\varphi(t,\mathbf{x}) = \chi(t)\,\phi(\mathbf{x})\  \text{with}\,\chi \in C^\infty_0((0,1))\, \text{satisfying} \int_0^1 \chi(t)\,dt = 1,
\]
so that
\[
\int_{U_\kappa}R^{(k)}\, \varphi\, \sqrt{|g_{(k)}|}\, d^4x = \int_{U_{\kappa^{(k)}_t}} R^{(k)} \phi\, N_{(k)}\, d^3x.
\]
\begin{Remark}[Vacuum state of Ricci]
Proceeding in an analogous way to the analysis of the wave equation, and
suppressing functional notation, we find that in the vacuum state the scalar
curvature is given by
\begin{equation}
R=\frac{\sqrt{2}}{8\sigma\sqrt{\pi}} \bigl(\delta^{\prime\prime}_x + \delta^{\prime\prime}_y + \delta^{\prime\prime}_z\bigr).
\end{equation}
For finite \(k\), the corresponding bare mass contributions are
\begin{equation}
  m_{B,\pm}^{(k)} = \pm\,\frac{\sqrt{2}}{16\sigma^3\pi}\,k^2,
\end{equation}
so that \(m_{B,+}^{(k)} + m_{B,-}^{(k)} = 0\). Each such signed contribution corresponds to a region of either positive or negative energy density. In particular, each term is divergent and requires renormalisation when considered individually, but taken together their contributions cancel in the renormalised theory. This is precisely the mechanism by which, in this model, the emergent spacetime carries net zero energy.
\end{Remark}
This result is structurally similar to the distributional curvature computed by Nigsch and Vickers~\cite{nigsch20202} in the context of conical spacetimes, where they showed that curvature is supported on a lower-dimensional set and can be represented as a Radon measure. Their framework, based on  \( C^{1,1}\) regularity, is tailored to ensure well-posedness of geodesic motion and the curvature tensor as a measure-valued object. In contrast, the Gaussian Quantum Foam model adopts a more flexible distributional approach, allowing for curvature distributions beyond measures, including second-order derivative terms such as  \( \delta^{\prime\prime}\). This is possible due to the construction of the algebra from smooth but scaled sequences in Schwartz space, and specifically by applying Hörmander’s Theorem 2.3.4~\cite{hormander1983}, which characterises all distributions with point support as finite sums of the Dirac measure and its derivatives. As in the wave equation, this reflects a sharply localised curvature impulse associated with vacuum displacement and inflationary dynamics. A similar regularity assumption appears in Steinbauer’s recent low-regularity treatment of the Penrose--Hawking singularity theorems~\cite{steinbauer2023}, which is also restricted to  \( C^1\) metrics and does not admit higher-order distributional curvature terms.

Having established that the weak energy condition fails locally for smooth spacetime elements in the sequence of globally hyperbolic and homotopic spacetimes converging to a Gaussian Quantum Foam but remains well-defined distributionally —and that similar structure appears in the scalar curvature—we now proceed to the timelike scalar projection of the Ricci curvature and state a precise theorem governing its structure in the Gaussian Quantum Foam 
\begin{Theorem}[Projected Curvature Structure and the Strong Energy Condition]\label{theorem:projected-curvature}
Let  \( \beta^i_{(k)}\) be the Gaussian components of the shift vector field on a sequence of globally hyperbolic and homotopic spacetimes  \( \{(M_{(k)}, g^{(k)}_{\mu\nu})\}_{k \in \mathbb{N}}\) forming a Gaussian Quantum Foam element according to Definition~\ref{def:quantumfoam}, with lapse  \( N_{(k)} = k\).  

Consider the admissible scalar projection of the Ricci curvature along the four-velocity  \( u_{(k)}^\mu\) of an Eulerian observer, that is, the curvature scalar
 \( R^{(k)}_{\mu\nu}u_{(k)}^{\mu}u_{(k)}^{\nu}\)
along timelike worldlines orthogonal to any Cauchy surface  \( \Sigma_t^{(k)}\), retaining only those terms whose contribution does not vanish when paired with admissible test functions in Definition \ref{def:test-functions}:
\begin{equation}\label{eq:projectedcurvature}
    R^{(k)}_{\mu\nu}u_{(k)}^{\mu}u_{(k)}^{\nu} 
    = -\frac{1}{N_{(k)}^2}\biggl(
(\partial_{x}{\beta^{x}_{(k)}})^2 + \beta^{x}_{(k)} \,\partial_x^{2}{\beta^{x}_{(k)}} 
+ (\partial_{y}{\beta^{y}_{(k)}})^2 + \beta^{y}_{(k)}\,\partial_y^{2}\beta^{y}_{(k)} 
+ (\partial_{z}{\beta^{z}_{(k)}})^2 + \beta^{z}_{(k)}\,\partial_z^{2}{\beta^{z}_{(k)}}
\biggr).
\end{equation}
Then the following hold:
\begin{enumerate}
    \item[(i)] (\textbf{Non-negativity at the singular support}) For all finite  \( k \in \mathbb{N}\), the scalar projection of the Ricci curvature is greater than zero at the origin:
    \[
    R^{(k)}_{\mu\nu}u_{(k)}^{\mu}u_{(k)}^{\nu}\Big|_{x=0} > 0.
    \]
    In the distributional limit  \( k \to \infty\), this scalar projection satisfies
    \[
    \begin{aligned}
    \lim_{k\to\infty} \int_{U_{\kappa^{(k)}}} R^{(k)}_{\mu\nu} u_{(k)}^\mu u_{(k)}^\nu\, \varphi\,\sqrt{|g_{(k)}|}\,d^4x \\
    = \lim_{k\to\infty} \int_{U_{\kappa^{(k)}_t}} R^{(k)}_{\mu\nu} u_{(k)}^\mu u_{(k)}^\nu\,\phi\, N_{(k)}\,d^3x\\
    = -\sum_{i=1}^3 \left(a_i\, \langle \delta_{x^i}, \phi\rangle + b_i\, \langle \delta^{\prime\prime}_{x^i}, \phi\rangle\right),
    \end{aligned}
    \]
    where is  \(\varphi(t,\mathbf{x})\) is a test function in the admissible space of test functions \ref{def:test-functions}, that is  \(\varphi(t,\mathbf{x}) = \chi(t)\,\phi(\mathbf{x})\) is a positive, even and in a neighbourhood of the singular support locally concave test function with  \(\phi \in \mathcal{S}(U_{\kappa^{(k)}_t})\) and  \(\chi \in C^\infty_0((0,1))\) satisfying  \(\int_0^1 \chi(t)\,dt = 1\), and where  \(a_i<0\) and \(\ b_i>0 \in \mathbb{R}\) are constants. This defines the distributional curvature content of the quantum foam element at its singular support  \(\mathrm{sing\,supp} = \{0\}\).
    \item[(ii)] (\textbf{Local sign structure for finite \(k\)}) For any finite  \( k \in \mathbb{N}\), there exist a set \(\{x_0^{(k)}|x_0^{(k)}\in U^{(k)}_{\kappa^{(k)}_t} \setminus \{0\}\}\) such that
    \[
     R^{(k)}_{\mu\nu}u_{(k)}^{\mu}u_{(k)}^{\nu}(x_0) = 0.
    \]
    Moreover, there exists an open ball  \( B_{\kappa^{(k)}_t} \subset U^{(k)}_{\kappa^{(k)}_t} \setminus \{0\}\) centered at  \( x_0^{(k)}\) such that within  \( B_{\kappa^{(k)}_t}\) there exist disjoint subregions  \( B_{\kappa^{(k)}_t}^+ \subset B_{\kappa^{(k)}_t}\) and  \( B_{\kappa^{(k)}_t}^- \subset B_{\kappa^{(k)}_t}\) satisfying:
    \[
    R^{(k)}_{\mu\nu}u_{(k)}^{\mu}u_{(k)}^{\nu} > 0 \ \text{in} \ B_{\kappa^{(k)}_t}^+ ,\quad R^{(k)}_{\mu\nu}u_{(k)}^{\mu}u_{(k)}^{\nu} < 0 \ \text{in} \ B_{\kappa^{(k)}_t}^-.
    \]
    That is, the local projected scalar Ricci curvature changes sign on open sets away from the singular support, and hence the strong energy condition is violated locally in those regions.
\end{enumerate}
\end{Theorem}
\begin{proof}
Locally, and for each coordinate \(x^i\), we have from \eqref{eq:projectedcurvature} 
\[
-\frac{1}{k^2} \left ((\partial_{x^i}\beta^{x^i}_{(k)})^2+\beta_{(k)}^{x^i} \, \partial^2_{x^i} \beta_{(k)}^{x^i} \right )= -\frac{1}{2k\sigma \sqrt{\pi}} \left(\frac{k^4}{2\sigma^4}(x^i)^2 - \frac{k^2}{2\sigma^2} \right) \beta_{(k)}^{\sqrt{2}x^i}.
\]
Thus we can write
\[
R^{(k)}_{\mu\nu}u_{(k)}^{\mu}u_{(k)}^{\nu}= \sum_{i=0}^3p_\sigma^{(k)}(x^i)\ \frac{\beta_{(k)}^{\sqrt{2}x^i}}{2k\sigma \sqrt{\pi}}
\]
where 
\[
p_\sigma^{(k)}(x^i)= -\left(\frac{k^4}{2\sigma^4}(x^i)^2 - \frac{k^2}{2\sigma^2} \right).
\]
Clearly, for any finite  \( k\), \(R^{(k)}_{\mu\nu}u_{(k)}^{\mu}u_{(k)}^{\nu}>0\) at the origin since \(p_\sigma^{(k)}(0)>0\) and \(\beta_{(k)}(0)>0\). 
\newline 
\noindent The polynomial \(p_\sigma^{(k)}\) in each coordinate has two real roots located at
\[
x^i_{(k)}=\pm \frac{\sigma}{k}.
\]
In the neighbourhood of the negative root, there exists \(k\) such that the polynomial \(p_\sigma^{(k)}\) is strictly increasing, while in the neighbourhood of the positive root it is strictly decreasing. This follows since \(k\) can be chosen, for any fixed \(\sigma\), such that the derivative of the polynomial factor is positive for  \( x^i_{(k)}< 0\) and negative for  \( x^i_{(k)}> 0\). This establishes the sign-changing structure at finite  \( k\), occurring at both roots. Hence, the expression changes sign across these roots, and open regions of both positive and negative curvature scalar projection exist in  \( U^{(k)}_{\kappa^{(k)}_t} \setminus \{0\}\).
\newline
\newline
Finally, in the limit as \(k\to\infty\), as in the analysis of the wave equation in Section~\ref{sec:waveequation}, we only consider positive, even and locally concave test functions. This is justified by the fact that odd test functions lie in the orthogonal complement of the Gaussian sector and therefore yield zero when probing the Quantum Foam. The positivity and concavity requirement is enforced due that the Gaussians are positive and locally concave in a neighbourhood of the singular support. The term in~\eqref{eq:projectedcurvature} is then evaluated in the distributional algebra  \( (\mathcal{S}_\mathcal{G}^\prime, \cdot, \partial)\) defined in~\ref{def:algebra}, using the rules given in Sections~\ref{sec:algebraoperations} and~\ref{sec:algebraoperations2}. Specifically, the terms involving products of first order derivatives of the shift vector with the shift vector itself vanish when evaluated against positive and even test functions. From the distributional analysis in Section~\ref{sec:algebraoperations2}, and by Definition~\ref{def:prodsecondorder}, it follows that the remaining second order derivative terms converge, as  \( k \to \infty\), in the sense of distributions to the linear combination
\[
a_i\,\delta_{x^i} + b_i\,\delta''_{x^i},
\]
where  \(a_i < 0\) and  \(b_i > 0\) are real constants; see Remark~\ref{re:delta_coefficients}. Consequently, the projected curvature at the singular point  \(\{0\}\) has positive support, since it is given by the negative of the sum of these linear combinations. This is consistent with the earlier result for the limit~\eqref{ricciscalar} and with the fact that the projected stress-energy vanishes in the distributional limit; see~\eqref{eq:inttuu}.
\end{proof}

The conclusion that the strong energy condition holds at the singular support, but that there exist finite values of the sequence index \(k\), and hence classical spacetime elements, where it fails on open sets in its complement, shows that Quantum Foam inherently includes a mechanism for the emergence of spacetime. As the vacuum is displaced by the oscillations of the shift vector, distant regions are brought into causal contact through local stretching and contraction, which warp the hypersurfaces. 
This behaviour arises from the existence of spatial regions of positive and negative energy density, while the total energy at all times remains zero. 
Such dynamics in the finite- \(k\) setting suggest that, although spacetime curvature is generated, light rays will not be globally trapped in the hypersurfaces.

To analyse whether this is indeed the case, let us briefly return to the sequence of globally hyperbolic and homotopic spacetimes  \(\{(M_{(k)},g^{(k)}_{\mu\nu})\}_{k\in\mathbb{N}}\) in Definition~\ref{def:quantumfoam} and to the volume expansion
\begin{equation} \label{eq:volexpansion}
\Theta_{(k)} = \nabla^{(k)}_\mu u_{(k)}^\mu = -\frac{1}{N_{(k)}} \partial_i \beta_{(k)}^i,
\end{equation}
as observed by an Eulerian observer with four-velocity \(u^\mu_{(k)}\), and discussed in relation to the quantisation of Gaussian Quantum Foam in~\cite{cramer20252} and in relation to the Correspondence Principle for Quantum Foam in \cite{cramer20251}. 
Defining the extrinsic curvature of the hypersurfaces by 
\begin{equation}\label{eq:ext-curvature}
K^{(k)}_{\mu\nu} := -h^\gamma_{(k)\,\mu} \nabla^{(k)}_\gamma u^{(k)}_\nu,
\end{equation}
we immediately observe that its trace is given by
\begin{equation} \label{eq:ext-trace}
K_{(k)} = -\Theta_{(k)}=\frac{1}{N_{(k)}} \partial_i \beta_{(k)}^i,
\end{equation}
which is in agreement with the result \eqref{eq:tracek} from Section \ref{sec:review} where the notion of Gaussian Quantum Foam was reviewed. Thus, the Eulerian observer is not only measuring the local expansion of matter—interpreted as the local Hubble parameter—but also the bending and stretching of the hypersurfaces. It is therefore worthwhile to pause and clarify the physical interpretation of the volume expansion~\eqref{eq:volexpansion}.

When the expansion is strictly positive,  \( \Theta_{(k)} > 0\), nearby matter elements are moving apart. According to~\eqref{eq:ext-trace}, this corresponds to a negative trace of the extrinsic curvature,  \( K_{(k)} < 0\), indicating that the hypersurfaces  \( \Sigma_t^{(k)}\) are locally expanding. If the expansion vanishes,  \( \Theta_{(k)} = 0\), the congruence is momentarily stationary, and the geometry of the hypersurfaces remains unchanged as time progresses. Conversely, if the expansion is negative,  \( \Theta_{(k)} < 0\), matter elements are converging, and the hypersurfaces are locally contracting.

To examine whether light rays are trapped within the hypersurfaces, we extend the notion of expansion from timelike congruences—associated with matter—to null congruences, corresponding to the behaviour of photons.
For any closed spacelike two-surface \(S_t^{(k)} \subset \Sigma_t^{(k)}\), let \(s^i_{(k)}\) denote the outward unit normal to \(S_t^{(k)}\) in \(\Sigma_t^{(k)}\). 
Define the outward and inward null normals by
\begin{equation}
\ell^\mu_{(\pm),k} = u_{(k)}^\mu \pm s^\mu_{(k)}.
\end{equation}
Then, the outward and inward null expansions in \(\{(M_{(k)},g^{(k)}_{\mu\nu})\}_{k\in\mathbb{N}}\) are defined by
\begin{equation}\label{eq:cov-theta}
\theta^{(k)}_{(\pm)} := q^{\mu\nu}_{(k)}\,\nabla^{(k)}_\mu \ell^{\ }_{(\pm),(k)\,\nu},
\end{equation}
where \(q_{\mu\nu}^{(k)} = h_{\mu\nu}^{(k)} - s^{(k)}_\mu s^{(k)}_\nu\) is the induced metric on \(S_t^{(k)}\), and 
 \(h_{\mu\nu}^{(k)} = g_{\mu\nu}^{(k)} + u^{(k)}_{\mu} u^{(k)}_\nu\) is the spatial metric on each hypersurface \(\Sigma_t^{(k)}\) in the \(3{+}1\) decomposition.

We now express the expansions~\eqref{eq:cov-theta} more explicitly as
\begin{equation}
\theta_{(\pm)}^{(k)} = \pm H^{(k)} - K_{(k)} + K_{ij}^{(k)} s^i_{(k)} s^j_{(k)},
\label{eq:theta-3p1}
\end{equation}
where
\begin{equation}
H^{(k)} = q^{ij}_{(k)}\,\partial_i s_{(k)\,j}
\end{equation}
is the mean curvature of the embedded two-surface in the hypersurface.

The physical interpretation of the sign of the null expansions  \( \theta^{(k)}_{(\pm)}\) mirrors that of the timelike expansion, but in the context of light rays orthogonal to the closed two-surface  \( S_t^{(k)} \subset \Sigma_t^{(k)}\). When  \( \theta^{(k)}_{(\pm)} > 0\), the corresponding null congruence is expanding—outgoing or ingoing light rays diverge away from the surface. If  \( \theta^{(k)}_{(\pm)} = 0\), the congruence is momentarily stationary, corresponding to a marginally trapped or anti-trapped configuration. Finally, if  \( \theta^{(k)}_{(\pm)} < 0\), the light rays converge, which is a necessary condition for the formation of trapped surfaces. In the Gaussian Quantum Foam framework, and as will be shown in Theorem~\ref{thm:null-not-trapped}, such configurations do not occur at the singular support but may arise locally for finite  \( k\), reflecting the transient nature of curvature compression during the inflationary phase.

On a historical note, the concept of trapped surfaces was introduced by Penrose in his work on singularity theorems and gravitational collapse~\cite{penrose1965}.
We now state the central result:

\begin{Theorem}[Null Expansion in Gaussian Quantum Foam]\label{thm:null-not-trapped}
Let \(\beta^i_{(k)}\) be the Gaussian components of the shift vector field on a sequence of globally hyperbolic and homotopic spacetimes \(\{(M_{(k)},g^{(k)}_{\mu\nu})\}_{k\in\mathbb{N}}\) forming a Gaussian Quantum Foam element according to Definition~\ref{def:quantumfoam}, with the lapse \(N_{(k)} = k\).  
Let \(u_{(k)}^\mu\) denote the four-velocity of an Eulerian observer. For any closed spacelike two-surface \(S_t^{(k)} \subset \Sigma_t^{(k)}\), let \(s^i_{(k)}\) be the outward unit normal to \(S_t^{(k)}\) in \(\Sigma_t^{(k)}\), and define the local null expansion by
\begin{equation}\label{eq:cov-theta_theorem}
\theta_{(\pm)}^{(k)} = \pm H^{(k)}-K_{(k)} + K_{ij}^{(k)}\,s^i_{(k)} s^j_{(k)},
\end{equation}
with \(H^{(k)} = q^{ij}_{(k)}\,\partial_i s_{(k)\,j}\) representing the mean curvature.  
Then, in the singular support \(\{0\}\) in the limit  \(k \to \infty\),
\[
\lim_{k\to\infty}\int_{U_{\kappa^{(k)}}} \theta^{(k)}_{\pm}\,\varphi\,d^4x = 0,  
\quad \text{for all } \varphi \in \mathcal{S}(U_\kappa),
\]
while locally, there exists an open ball \(B_r\) centered at the origin with radius \(r > 0\) such that
\[
\theta^{(k)}_{\pm} < 0, \quad \forall x \in B_r(0)\setminus \{0\}, \quad \text{for all sufficiently large } k.
\]
\end{Theorem}
\begin{proof}
Let \(\varphi(t,\mathbf{x}) = \chi(t)\,\phi(\mathbf{x})\) with \(\chi \in C^\infty_0((0,1))\), \(\int_0^1 \chi = 1\), and \(\phi \in \mathcal{S}(U_{\kappa^{(k)}_t})\) be an admissible test function in definition \ref{def:test-functions}. Since the lapse, the shift, and the induced spatial metric on \(\Sigma^{(k)}_t\) are static and \(\sqrt{|g_{(k)}|} = N_{(k)}\), we obtain
\[
\int_{U_{\kappa^{(k)}}} \theta^{(k)}_{(\pm)}\,\varphi\,\sqrt{|g_{(k)}|}\,d^4x
= \int_{U_{\kappa_t^{(k)}}} \theta^{(k)}_{(\pm)}\,\phi\,N_{(k)}\,d^3x.
\]
Using the line element~\eqref{eq:gaussianmetric} with flat induced spatial metric \(\eta_{ij}\), the extrinsic curvature and its trace are
\[
K^{(k)}_{ij} = \frac{1}{2N_{(k)}}\left(\partial_i\beta^{(k)}_{j} + \partial_j\beta^{(k)}_{i}\right),
\qquad
K^{(k)} = \frac{1}{N_{(k)}}\,\partial_i \beta_{(k)}^i.
\]
Substituting into \eqref{eq:cov-theta_theorem} yields
\[
\theta^{(k)}_{(\pm)} = \pm H^{(k)} -\frac{1}{N_{(k)}}\left( \partial_i \beta_{(k)}^i - s^i_{(k)} s^j_{(k)} \,\partial_i \beta^{(k)}_{j} \right)
= \pm H^{(k)} - \frac{1}{N_{(k)}}\,q^{ij}_{(k)}\,\partial_i \beta^{(k)}_{j}.
\]
By Definition~\ref{def:quantumfoam}, each \(\beta^{(k)}_{i}\) is even in \(x^i\) and depends only on \(x^i\), so \(\partial_i \beta_{(k)}^i\) is odd, while \(q^{ij}_{(k)}\) and \((s^i_{(k)})^2\) are smooth and even. Moreover, \(\partial_i \beta^{(k)}_{j} = 0\) for \(i \neq j\).  

\noindent Therefore, for any even and positive test function \(\phi\),
\[
\lim_{k\to\infty} \int_{U_{\kappa_t^{(k)}}}K ^{(k)}\,\phi\,N_{(k)}\,d^3x = 0,
\qquad
\lim_{k\to\infty} \int _{U_{\kappa_t^{(k)}}}K_{ij}^{(k)}\,s^i_{(k)} s^j_{(k)}\,\phi\,N_{(k)}\,d^3x = 0.
\]
Hence,
\[
\lim_{k\to\infty}\int_{U_{\kappa_t^{(k)}}} \theta^{(k)}_{(\pm)}\,\phi\,N_{(k)}\,d^3x
= \pm \lim_{k\to\infty}\int_{U_{\kappa_t^{(k)}}} H^{(k)}\,\phi\,N_{(k)}\,d^3x.
\]
From here we continue using integration by parts. On \(\Sigma_t^{(k)}\) with flat induced metric \(\eta_{ij}\), we can write \(H^{(k)}=\partial_i s^{i}_{(k)}\), since 
\[
H^{(k)}= q^{ij}_{(k)}\,\partial_i s_{(k)\,j}=(\eta^{ij}- s_{(k)}^is_{(k)}^j)\partial_is^{(k)}_j=\partial_is^i_{(k)}-\frac{1}{2}s^i_{(k)}\partial_is_j^{(k)}s^j_{(k)}=\partial_is^i_{(k)},
\]
where we have used that \(s_j^{(k)}s^j_{(k)}=1\). Integrating by parts and using the compact support of \(\phi\) we have 
\[
\int_{U_{\kappa_t^{(k)}}} H^{(k)}\,\phi\,N_{(k)}\,d^3x
= -\,N_{(k)}\int_{U_{\kappa_t^{(k)}}} s^{i}_{(k)}\,\partial_i \phi\,d^3x.
\]
Thus, the integrand is odd and integrates to zero (for every finite \(k\)) over the symmetric domain of integration. Consequently,
\[
\lim_{k\to\infty}\int_{U_{\kappa_t^{(k)}}} H^{(k)}\,\phi\,N_{(k)}\,d^3x=0,
\]
and therefore together with the earlier result it follows that
\[
\lim_{k\to\infty}\int_{U_{\kappa^{(k)}}} \theta^{(k)}_{(\pm)}\,\varphi\,d^4x = 0.
\]

\noindent Finally, to establish local negativity away from the singular support and hence the local negativity of \(\theta^{(k)}_{\pm}\), note that for any ball \(B_\epsilon(0)\) we have \(H^{(k)}\) bounded on \(B_\epsilon(0)\), while \(\beta_{(k)}\) becomes sharply peaked.  
Since \(|\partial_i\beta^{\ }_{(k)j}|\) grows like \(k^3\) in a neighbourhood of \(\{0\}\) as \(k\to\infty\) (from the Gaussian scaling), we can find \(k_0\) such that for all \(k>k_0\),
\[
\left| \frac{1}{N_{(k)}}\,q^{ij}_{(k)}\,\partial_i \beta^{(k)}_{j} \right| > H^{(k)}
\quad \text{on } B_\epsilon(0)\setminus\{0\}.
\]
Hence \(\theta^{(k)}_{(\pm)}<0\) in \(B_\epsilon(0)\setminus\{0\}\) for all sufficiently large \(k\), which proves the claim.
\end{proof}

\noindent These two theorems reveal a central feature of Quantum Foam geometry.  
In line with the strong \emph{distributional} energy condition proposed by Steinbauer~\cite{steinbauer2023},  
the requirements of the classical singularity theorems are upheld at the singular support and in the distributional limit.  
Theorem~\ref{theorem:projected-curvature} establishes that the projected Ricci curvature is non-negative at the singular support and converges to a well-defined, positive distribution involving both the Dirac measure and its second derivative.  However, locally and for finite values of the sequence index, for the sequence of globally hyperbolic spacetimes converging to Gaussian Quantum Foam, the curvature scalar projection necessarily changes sign across open subsets. This local violation of the strong energy condition is not a pathology—it is a defining characteristic. It reflects the dipolar structure of vacuum fluctuations which, under the non-linear dynamics of the shift vector, drive inflationary displacement and create the necessary conditions for the emergence of classical spacetime at scales much larger than the Planck scale.

Theorem~\ref{thm:null-not-trapped} addresses the null expansions and shows that, in the distributional limit, both the extrinsic-curvature terms and the mean curvature integrate to zero.  
Consequently, the total expansion of null congruences vanishes at the singular support, rendering spacetime effectively frozen at that point.  
However, for finite values of the sequence index—when spacetime evolves away from its frozen null configuration with a non zero slowness and spacelike hypersurfaces, and the shift vector is set into oscillatory motion by the curvature impulse described by the wave equation in Definition~\ref{def:wave-equation}—  
open neighbourhoods emerge in which both the ingoing and outgoing null expansions are strictly negative.  
This indicates that local, transient trapped regions can occur during the inflationary phase.  
One may speculate that such trapping acts as a thermodynamic compression mechanism during reheating,  
signalling the transition from a coherent quantum foam phase to the emergence of classical spacetime geometry.

Together, these results suggest that the singular support is not a breakdown of predictability.  
Rather, it marks the precise location of vacuum displacement: a sharply supported curvature impulse that initiates the emergence of time.  
This reinterpretation honours the legacy of Penrose and Hawking, whose singularity theorems~\cite{penrose1965,hawking1967,hawking1970}  
demonstrated the inevitability of curvature accumulation under classical assumptions.  

In Penrose’s 1965 formulation~\cite{penrose1965}, if a spacetime singularity is to be avoided, then at least one of the following conditions must hold: the local energy density is negative, Einstein’s field equations are violated, spacetime is geodesically incomplete, or the notion of spacetime loses its meaning, at high curvature and possibly due to quantum phenomena.

In the present model, none of these conditions holds at the singular support. Both the weak and strong energy conditions are satisfied in distribution (with the weak vanishing and the strong remaining positive in distributions), and the congruence expansions for both timelike and null observers vanish at the singular support. By contrast, for finite values of \(k\)—and hence in an emerging and classical context—the smooth sequence exhibits local violations of the energy conditions and the formation of transient trapped surfaces.

The quantum foam limit is well-defined in the distributional sense, and in the distributional convergence from the sequence of smooth globally hyperbolic and homotopic spacetimes, the notions of continuity and differentiability remain valid in the distributional framework. Thus, and altogether, Einstein’s field equations hold in a distributional geometric sense at the singular support and the notion of spacetime remains intact in a distributional geometric context that encompasses the notion of classical spacetime.

In conjunction, this supports a rigorous claim that Gaussian Quantum Foam provides a singularity-free completion of early-universe geometry in the distributional framework.

\section{Gaussian Quantum Foam Beyond the Reliability Horizon}\label{sec:time_machines}

In this section, we revisit the open question of whether a civilisation with arbitrarily advanced technology and engineering skills could manufacture a time machine. That is, in layman’s terms, whether the engineers of such a civilisation could locally warp spacetime so that a region of closed timelike curves arises, separated from the strongly causal region they occupy. In a more general sense, a time machine is a spacetime model with an initial globally hyperbolic region separated from a region of closed timelike curves by a compactly generated Cauchy horizon, the chronology horizon. For these models, Hawking formulated his Chronology Protection Conjecture, arguing that any attempt to build a time machine would necessarily fail because the back-reaction of the renormalised expectation value of the stress–energy tensor on the spacetime would become unbounded at the chronology horizon \cite{hawking1992}.

Addressing the conjecture by investigating whether such spacetimes could arise within a semi-classical context in a neighbourhood of, and at, the chronology horizon immediately introduces technical challenges, due to the non-linearity of general relativity and the distributional nature of quantum field theory on curved spacetime (i.e., the field in its quantised form is represented by an operator-valued distribution and hence expectation values in any state necessarily involve products of distributions).

Nevertheless, as reviewed in Section~\ref{sec:wavefrontset}, the renormalised distribution algebra and the resulting distributional geometry are compatible with the microlocal picture even though the wave front set is not empty in the Gaussian Quantum Foam limit. In the limit  \(k\to\infty\), the normals to the level sets become null, the level sets themselves become characteristic hypersurfaces, and the configuration realises an unstable self-gravitating geon with a divergent bare mass monopole. There is no microlocal inconsistency because all non-linear operations are performed at finite  \(k\), where every phase space point is a regular directed point (so  \(WF=\varnothing\)), and only then is the distributional limit taken; renormalisation removes the universal divergences so that the limit is a well defined distribution

Given the instability of the geon, we do expect that a similar conclusion as in the KRW theorem of Kay, Radzikowski, and Wald (KRW) to hold. The trio used H\"ormander’s classical work on the spectral analysis of singularities \cite{hormander1983}—that is, microlocal analysis—to prove that the bi-distribution (two-point distribution) in any Hadamard state is singular on the chronology horizon. Moreover, Cramer and Kay \cite{cramer1996,cramer1998} proved, without using microlocal analysis, that for two specific models, one by Krasnikov \cite{krasnikov1996} and another by Sushkov \cite{sushkov1997}, with fields and states explicitly not covered by the KRW theorem on Misner space, the conclusions still hold, even if the renormalised stress–energy is zero all the way up to the chronology horizon.

While the KRW result is a no-go theorem in a semi-classical domain of quantum gravity, it is not clear whether it extends into the realm of a full, as-yet unknown theory of quantum gravity. In fact, Visser introduced the notion of a reliability horizon to argue that one cannot push semi-classical gravity—and the KRW result—beyond that point and into the region where metric fluctuations are expected to become significant \cite{visser1997}. To make this more precise, Visser introduced the notion of a reliability region as the region of spacelike geodesics shorter than a Planck length, where vacuum fluctuations—and hence metric fluctuations—become significant, and pointed out that ultimately the question of chronology protection must be decided by the physics inside the boundary of the reliability region, since quantum field theory in curved spacetime is an approximation to a yet-unknown theory of quantum gravity in which the metric field itself is quantised (in contrast to quantum field theory in curved spacetime, where only the matter fields are quantised). That is, the question of chronology protection must ultimately be decided by understanding the physics of vacuum fluctuations.

\medskip

\noindent However, in the context of a Gaussian Quantum Foam, we will show that no region of closed timelike curves can be engineered within the domain of the model. In this, we will find it convenient to work in the relative velocity gauge of the Gaussian Quantum Foam discussed in Section~\ref{sec:review} (see \eqref{eq:diffeomorphism}--\eqref{eq:dflow}). Therefore, let us recall that the line element in Definition~\ref{def:quantumfoam} in a coordinate chart  \(\kappa_{(k)}\) with coordinates  \((t,x,y,z)\) is given by
\[
ds_{(k)}^2 = -N_{(k)}^2\, dt^2 + \eta_{ij} \big(dx^i + \beta_{(k)}^i\, dt\big)\big(dx^j + \beta_{(k)}^j\, dt\big).
\]
The relative velocity gauge mentioned above is represented by the diffeomorphism  \(\Phi_{(k)}\) 
\[
 \kappa_{(k)} \circ \Phi_{(k)}^{-1} :
 \Phi_{(k)}(O_{\Phi_{(k)}}\cap O_{\kappa_{(k)}}) \;\longrightarrow\;
 \kappa_{(k)}(O_{\Phi_{(k)}}\cap O_{\kappa_{(k)}}),
\]
\[
\Phi_{(k)} : (t,x^1,x^2,x^3)\;\mapsto\;(t',x'^1,x'^2,x'^3),
\]
explicitly given by \eqref{eq:diffeomorphism}--\eqref{eq:dflow}, that is,
\[
t'=t, \qquad x'^i =  \Phi^i_{(k)}:=\Phi_{(k)}(x^i, t),\quad i=1,2,3,
\]
with the flow equation
\[
\partial_t \Phi^{-1 \, i}_{(k)} =
 -\beta^i_{(k)}\big(\Phi_{(k)}^{-1 \, i}(x^{\prime},t)\big).
\]
Under this gauge, the line element takes the form
\[
ds_{(k)}^2 = -N_{(k)}^2\, dt^2 + \eta_{mn}\,\partial_{x^{\prime i}}\Phi_{(k)}^{-1\ m}\,\partial_{x^{\prime j}}\Phi_{(k)}^{-1\ n}\,dx^{\prime i}dx^{\prime j}.
\]
As stated in Section~\ref{sec:review}, this is of the form used in the theorem of Bernal and S\'anchez~\cite{bernal2005}, and the shift vector has been gauged to zero; the induced Riemannian metric  \(h^{\prime (k)}_{ij}\) is
\[
    h^{\prime (k)}_{ij}=\eta_{mn}\,\partial_{x^{\prime i}}\Phi_{(k)}^{-1\ m}\,\partial_{x^{\prime j}}\Phi_{(k)}^{-1\ n}.
\]
From now on, we drop the coordinate-system prime. In this gauge, the global, smooth, and regular time function  \(t_{(k)}\) satisfies the eikonal equation (see \eqref{eq:sqreikon})
\[
\|\nabla t_{(k)}\|_{g_{(k)}} = v_{(k)}^{-1},
\]
where  \(v^{-1}_{(k)} := 1/N_{(k)}\) is the slowness, inversely proportional to the lapse function  \(N_{(k)}=k\).

\noindent Next, let us introduce a notion of reliability region that aligns with Visser’s \cite{visser1997}:
\begin{Definition}[Planck–scale reliability cylinder]\label{def:reliability-region}
Consider a Gaussian Quantum Foam as in Definition~\ref{def:quantumfoam}, in the relative velocity gauge \eqref{eq:diffeomorphism}--\eqref{eq:dflow}. 
Fix  \(k\in\mathbb{N}\), a time  \(t\in\mathbb{R}\), and a point  \(p\in\Sigma^{(k)}_{t}\) with induced metric  \(h^{(k)}\). 
In this  \(3{+}1\) decomposition, define the (open) cylinder
\[
\mathcal C^{(k)}_{\mathrm P}(p)
\;:=\; B_{h^{(k)}}\!\big(p,\ell_{\mathrm P}\big)\times \big(t-t_{\mathrm P},\,t+t_{\mathrm P}\big)\ \subset\ \Sigma^{(k)}_{t}\times\mathbb{R},
\]
where  \(B_{h^{(k)}}(p,\ell_{\mathrm P})\) is the \emph{open}  \(h^{(k)}\)–geodesic ball of Planck radius  \(\ell_{\mathrm P}\) in  \(\Sigma^{(k)}_{t}\), and the temporal Planck “height” \(t_{\mathrm P}\) is measured as proper time along the unit normal flow.
For fixed \(k\), the \emph{reliability region} is the union of such cylinders over all \(p\in\Sigma^{(k)}_{t}\) and all \(t\in\mathbb{R}\):
\[
\mathcal R^{(k)}\;:=\;\bigcup_{t\in\mathbb{R}}\ \bigcup_{p\in\Sigma^{(k)}_{t}}\ \mathcal C^{(k)}_{\mathrm P}(p).
\]
\end{Definition}

\begin{Definition}[Planck–scale reliability tube (boundary)] \label{def:reliability-boundary}
Let \(S_{h^{(k)}}(p,\ell_{\mathrm P}) := \partial B_{h^{(k)}}(p,\ell_{\mathrm P})\) be the \(h^{(k)}\)–geodesic sphere and \(\overline{B}_{h^{(k)}}(p,\ell_{\mathrm P})\) the closed ball.
The (reliability) \emph{tube} is the boundary of the cylinder:
\[
\partial \mathcal C^{(k)}_{\mathrm P}(p)
=\big(S_{h^{(k)}}(p,\ell_{\mathrm P})\times [t-t_{\mathrm P},\,t+t_{\mathrm P}]\big)
\;\cup\; \big(\overline{B}_{h^{(k)}}(p,\ell_{\mathrm P})\times\{t-t_{\mathrm P},\,t+t_{\mathrm P}\}\big).
\]
\end{Definition}
In what follows we will show that the Gaussian Quantum Foam element (the geon) is not strongly causal, and it is therefore convenient to introduce a definition concerning convexity of the chronological future of the spacetime elements in the definition of the Gaussian Quantum Foam element. The definition is related to the notion of convex subsets in metric spaces: a subset is convex if the segment joining any pair of points in the subset remains in the subset. However, in our case, convexity need not be preserved for a subset when embedded in a larger chronological space:

\begin{Definition}[Non-convex nesting of chronological futures]
Let \((M^{(\ell)},g^{(\ell)})\) and \((M^{(k)},g^{(k)})\) be two elements of the Gaussian Quantum Foam sequence with \(k>\ell\).
For fixed \(t\), set
\[
U:=I^{+}_{(\ell)}\!\big(\Sigma^{(\ell)}_{t}\big),\qquad
V:=I^{+}_{(k)}\!\big(\Sigma^{(k)}_{t}\big),
\]
where sets are identified near the common limit \(\Sigma_{0}\) via the relative–velocity gauge.
We say that \(V\) is a \emph{non-convex} nesting of \(U\) if there exist points \(p,r\in U\), a point \(q\in V\setminus U\), and a future–directed \(g^{(k)}\)–timelike curve \(\gamma\subset V\) from \(p\) to \(r\) such that \(q\in \mathrm{im}\,\gamma\).
Equivalently, \(U\) is not causally convex as a subset of \((M^{(k)},g^{(k)})\).
We say the nesting is \emph{strong near \(\Sigma_{0}\)} if this property holds with \(p,r,q\) chosen in every neighbourhood of \(\Sigma_{0}\).
\end{Definition}

Now, consider the Gaussian Quantum Foam framework, which is fully quantum and well defined in distribution geometry, and is therefore well defined in the reliability region without any restrictions, i.e., in the region consisting of spacelike geodesics shorter than the Planck length—equivalently, geodesics in any \emph{open cylinder} (Definition~\ref{def:reliability-region}) bounded by the tube (Definition~\ref{def:reliability-boundary}).

Specifically, in the sequence of spacetimes converging to a Gaussian Quantum Foam, the eikonal equation is well defined not only outside the reliability region but equally within it:
\[
\|\nabla t_{(k)}\|_{g_{(k)}} = v_{(k)}^{-1}.
\]
We note that the slowness decreases as the index  \(k\) increases since  \(v_{(k)}^{-1}=k^{-1}\), and that the normal one–form
 \(\nabla t_{(k)} = (\partial_\mu t_{(k)})\,dx^\mu\) to each level surface—and hence to each spacelike Cauchy surface
 \(\Sigma_t^{(k)} = \{ p \in M_{(k)} : t_{(k)}(p) = t \in \mathbb{R} \}\)—is timelike for each finite  \(k\) and converges to null in the limit  \(k \to \infty\). The same is true for the level surfaces, which converge to characteristic (null) hypersurfaces. An alternative view of this, with the chronological  \(I^+\) and causal future  \(J^+\) understood relative to the level surface  \(\Sigma^{(k)}_{t}\), is that for any fixed  \(t\) and any  \(k>\ell\), the homotopy of spacetime elements and  \(\|\nabla t_{(k)}\|_{g_{(k)}}=k^{-1}\) give rise to a non-convex nesting
\begin{equation} \label{eq:characteristic-chronology}
I^{+}\big(\Sigma^{(\ell)}_{t}\big)\;\subset\;
I^{+}\big(\Sigma^{(k)}_{t}\big)\;\subset\;
J^{+}\big(\Sigma^{(k)}_{t}\big).
\end{equation}
Thus, as the index  \(k\) increases and the slowness  \(v^{-1}_{(k)}\) tends to zero, the level sets of the global time function become asymptotically characteristic (null), and there is a \emph{pile-up of almost characteristic surfaces} at the initial limiting null hypersurface  \(\Sigma_{0}\), where the null expansion vanishes, while  \(\Sigma_{0}\) is the boundary of the open set on which the null expansion is strictly negative, see theorem \ref{thm:null-not-trapped}. This transition from spacelike Cauchy surfaces, together with the homotopy of the spacetime elements in Definition~\ref{def:quantumfoam}, yields a continuous—but causally non-convex—warping of the hypersurfaces in the reliability region (Definition~\ref{def:reliability-region}), bounded by the reliability tube (Definition~\ref{def:reliability-boundary}), such that, in a limiting sense, the Cauchy development and the causal future accumulate at a set of past terminal accumulation points. Thus, we conclude that  \(\Sigma_{0}\) is a compactly generated Cauchy horizon—i.e., as close as one can get to a time machine. The lack of convexity in the chronology nesting \eqref{eq:characteristic-chronology} can be made more precise through the following:
\begin{Example}[Convexity breach from nested futures across two elements]
Fix two elements of the sequence, \((M^{(\ell)},g^{(\ell)}_{\mu\nu})\) and \((M^{(k)},g^{(k)}_{\mu\nu})\) with \(k>\ell\), as in the definition of Gaussian Quantum Foam. For a fixed \(t\), let \(\Sigma^{(\ell)}_{t}\subset M^{(\ell)}\) and \(\Sigma^{(k)}_{t}\subset M^{(k)}\) be the corresponding level surfaces, identified near the common limit \(\Sigma_{0}\) via the relative-velocity gauge.
We have that 
 \(
\|\nabla t_{(k)}\|_{g_{(k)}}=k^{-1}
\)
in conjunction with the homotopy yields a continuous (in \(k\)) deformation from spacelike level sets of \(t_{(k)}\) to a characteristic one, and therefore a non-convex nesting of chronological futures \eqref{eq:characteristic-chronology},
\[
I^{+}\big(\Sigma^{(\ell)}_{t}\big)\;\subset\;
I^{+}\big(\Sigma^{(k)}_{t}\big)\;\subset\;
J^{+}\big(\Sigma^{(k)}_{t}\big).
\]
Set \(U:=I^{+}\!\big(\Sigma^{(\ell)}_{t}\big)\).
Let \(b\in\Sigma_{0}\) be a \emph{base point} (a past terminal accumulation point on the Cauchy horizon). Pick \(q\in I^{+}\!\big(\Sigma^{(k)}_{t}\big)\setminus U\) arbitrarily close to \(b\) such that \(b\) and \(q\) are connected by a past-directed null curve. Choose  \(p,r\in U\) near \(b\) with \(p\ll_{(k)} q \ll_{(k)} r\) (openness of \(I^{+}_{(k)}\)).
Then, by the non-convexity of the chronological-future nesting, there exists a future-directed \(g^{(k)}\)-timelike curve \(\gamma\) from \(p\) to \(r\) passing through \(q\notin U\).
Hence \(\gamma\) leaves and re-enters \(U\), so \(U\) is not causally convex (with respect to \(g^{(k)}\));
equivalently, there is \emph{timelike re-entry} in arbitrarily small neighbourhoods of \(\Sigma_{0}\).
\end{Example}

In conclusion, strong causality \emph{fails at}  \(\Sigma_0\), in alignment with the conclusions of Kay–Radzikowski–Wald \cite{krw1997}. Consequently, within this framework, no region of closed timelike curves can develop within the domain of the model, since there exists  \(k_0\) such that, for all  \(k>k_0\), the level surfaces and the geodesics of the normal observers are contained in
\[
\bigcup_{k}\Big(\Sigma^{(k)}_t \times (t-t_{\mathrm P},\,t+t_{\mathrm P})\Big)\;\subset\;\bigcup_{k}\mathcal C^{(k)}_{\mathrm P},
\]
and pile up towards  \(\Sigma_0\), which is unstable and will collapse in the presence of any back-reaction of any matter field, evolving in the way prescribed by the field equation \ref{def:wave-equation}. 

As established above, the theory is effectively frozen at  \(\Sigma_0\). Hence even engineers in an arbitrarily advanced civilisation cannot manufacture a time machine within the model. For each fixed  \(k\), the spacetime  \((M_{(k)},g^{(k)}_{\mu\nu})\) is globally hyperbolic and the algebraic quantum field theory on it is well defined (see, e.g., Kay \cite{kay2023}). From a microlocal perspective, we have shown that each such spacetime element is microlocally regular, since all points in phase space are regular directed; consequently, for each finite  \(k\) in the sequence, the wave front set is empty. In fact, it is precisely this property that allowed us to introduce the renormalised distribution algebra and, consequently, a distribution geometry admitting non-linear operations such as products of distributions. Nevertheless, the non-empty wave front set in the limit is real and reflects the regime where the level sets become characteristic and the unit normal approaches null. Thus, the limiting distribution acquires a wave front set at the Cauchy horizon, in accordance with Section~\ref{sec:wavefrontset}, and the renormalised vacuum expectation value  \(\langle T_{\mu\nu}\rangle\), representing the spontaneous vacuum polarisation, supplies the back reaction that collapses the geon. To understand this, notice that:

In fact, the instability of the geon can be seen directly from the Einstein equations, without appealing explicitly to the Kay--Radzikowski--Wald theorem. Taking the trace of
\[
R_{\mu\nu} - \tfrac12 g_{\mu\nu} R = 8\pi T_{\mu\nu}
\]
in our signature convention yields
\[
T^\mu{}_\mu =-\frac{1}{8\pi}R.
\]
On the singular support of the Gaussian Quantum Foam we have shown, in the section on singularities (Section~\ref{sec:singularities}), that the scalar curvature is a distribution with support on the Cauchy horizon, given by \eqref{ricciscalar} as
\[
R = \sum_{i=0}^3\big(a_0^i\, \delta_{x^i}+a^i_2\,\delta_{x^i}^{\prime\prime}\big),
\]
with finite renormalised coefficients  \(a_0^i,a_2^i\) determined by the Gaussian algebra. Then given that in the coherent state \(|\alpha\rangle\), see Section \ref{sec:review}, the expected value is the same as the classical value, we consequently have in this state that the expected stress-energy trace in a coherent state is likewise distributional,
\[
\langle \alpha| T^\mu{}_\mu |\alpha\rangle
= -\frac{1}{8\pi }\, R
=-\frac{1}{8\pi }\sum_{i=0}^3\big(a_0^i\,\delta_{x^i}+a^i_2\,\delta_{x^i}^{\prime\prime}\big),\,a_0^i<0, \ a_2^i>0.
\]
Clearly, the stress–energy configuration, which is non-trivial (and positive in the admissible sense) on the singular support, cannot induce a static equilibrium with a smooth, globally hyperbolic metric in any neighbourhood of that support. The geon is in an unstable state, very much like a pen balanced on its tip, so any back-reaction will inevitably drive it away from this configuration. Thus, the stress–energy distribution necessarily leads to back-reaction in the form of radiation, which in turn causes the collapse of the self-gravitating configuration.

In the Gaussian Quantum Foam this back-reaction manifests itself in the oscillation of the shift vector, the formation of adjacent regions of positive and negative energy density, and the transient trapped surfaces discussed above, which together signal the collapse of the geon and the onset of the inflationary phase. We can describe this in the following way: the geon is unstable, and one is naturally led to a conclusion reminiscent of the Kay--Radzikowski--Wald theorem, namely that any measurement (in the sense provided in Section~\ref{sec:review}) will cause it to collapse. This will consequently set the shift vector into an oscillating state, displacing the vacuum so that separate regions of positive and negative energy density emerge, separated by a vacuum shell. In turn, this leads to exponential fluctuations in the extrinsic curvature, its trace (the local Hubble parameter), and hence huge fluctuations in the volume expansion rate in the neighbourhood of the Eulerian observers. That is, inflationary expansion and contraction effects emerge in conjunction with transient trapped surfaces acting as a reheating element representing primordial black holes that will eventually evaporate.

After this brief interlude concerning the collapse of the geon, we now prove that it will actually take place, but first we need to show that that the characteristic is a compactly generated Cauchy horizon, and then we use it to show that the KRW theorem \cite{krw1997} holds in Gaussian Quantum Foam. In proving the lemma, we find it useful to recall some basic set-theoretic notions that are commonly used in general relativity and quantum field theory in curved spacetimes, especially in relation to the Cauchy problem and the time-machine problem; we follow \cite{budzynski2001}. To start, let  \(S^{(k)}\subset M_{(k)}\), where  \(M_{(k)}\) is the manifold in the spacetime element  \((M_{(k)},g^{(k)}_{\mu\nu})_{k\in\mathbb{N}}\) in Definition~\ref{def:quantumfoam} of a Gaussian Quantum Foam element. The set  \(S^{(k)}\) is said to be \emph{achronal} if no two points in the set can be connected by a timelike curve, and \emph{acausal} if no two points can be connected by a causal curve. We will also need the notion of a \emph{partial Cauchy surface}, which is a connected, acausal, edgeless set  \(S^{(k)}\). The meaning of \emph{edgeless} is that, for the closure of the set and any point  \(p\) in the closure, and for every neighbourhood  \(O_{(k)}\) of  \(p\), there exists no timelike curve from the chronological past to the chronological future with respect to  \(O_{(k)}\). Finally, we will make use of the \emph{future Cauchy development}  \(D^+(S^{(k)})\), the set of points in  \(M_{(k)}\) such that any past-directed, inextendible causal curve intersects  \(S^{(k)}\).
\begin{Lemma}[Characteristic limit is a future Cauchy horizon]\label{lemma:cauchyhorizon}
For each \(k\), let \(t_{(k)}\) be the regular time function in Gaussian Quantum Foam with level sets
 \(\Sigma_t^{(k)} = \{ p \in M_{(k)} : t_{(k)}(p) = t \in \mathbb{R} \}\).
Introduce the set
\[
S^{(k)}_{0^{-}} := \Sigma^{(k)}_{0^{-}},
\]
just to the past of \(\Sigma^{(k)}_{0}\), and let
\[
\begin{aligned}
S_{0^{-}} := \lim_{k\to\infty} S^{(k)}_{0^{-}},\qquad
\Sigma_{0} := \lim_{k\to\infty} \Sigma^{(k)}_{0}.
\end{aligned}
\]
Then \(\Sigma_{0}\) is the future Cauchy horizon of \(S_{0^{-}}\).
\end{Lemma}
\begin{proof}
Clearly \(S^{(k)}_{0^{-}}\) is a connected, acausal, edgeless \emph{spacelike} hypersurface and hence a partial Cauchy surface.
From the eikonal relation \(\|\nabla t_{(k)}\|_{g^{(k)}}=v^{-1}_{(k)}\to 0\) as \(k\to\infty\) near the singular support,
the level surfaces \(\Sigma^{(k)}_{0}\) become asymptotically characteristic; hence the limit \(\Sigma_{0}\) is null and achronal.
For each fixed \(k\), the future Cauchy horizon of \(S^{(k)}_{0^{-}}\) is
\[
H^{+}(S^{(k)}_{0^{-}}) = \partial D^{+}(S^{(k)}_{0^{-}})\cap J^{+}(S^{(k)}_{0^{-}})
= \Sigma^{(k)}_{0}.
\]
Since each spacetime element is homotopic and globally hyperbolic, the surfaces converge; hence
\[
H^{+}(S_{0^{-}})
= \lim_{k\to\infty} H^{+}(S^{(k)}_{0^{-}})
= \lim_{k\to\infty}\Big(\partial D^{+}(S^{(k)}_{0^{-}})\cap J^{+}(S^{(k)}_{0^{-}})\Big)
= \lim_{k\to\infty}\Sigma^{(k)}_0
= \Sigma_{0}.
\]
\end{proof}
We now state the theorem that shows the KRW theorem \cite{krw1997} holds in a Gaussian Quantum Foam. In what follows we refer to Kay \cite{kay2023} for further details on quantum fields in curved spacetimes.
\begin{Theorem}[Hadamard behaviour on Gaussian Quantum Foam elements]\label{thm:GQF-Hadamard}
Let \(G_k\) be the anticommutator bi-distribution in a quasi-free Hadamard state \(\omega_k\) of the scalar Klein--Gordon field on any spacetime element \((M_{(k)},g^{(k)}_{\mu\nu})\) in Definition~\ref{def:quantumfoam}. Then \(G_k\) is symmetric, a bi-solution in the weak (distributional) sense, positive, and of Hadamard form on \((M_{(k)},g^{(k)}_{\mu\nu})\) for every finite \(k\in\mathbb{N}\).

In the Gaussian Quantum Foam limit of Definition~\ref{def:quantumfoam}, the limiting spacetime \((\tilde M,\tilde g_{\mu\nu})_{\mathcal{G}}\) is distributional with scalar curvature
\begin{equation}\label{eq:GQF-R-limit}
R = \sum_{i=0}^3\big(a_0^i\, \delta_{x^i}+a^i_2\,\delta_{x^i}^{\prime\prime}\big),
\end{equation}
supported on the Cauchy horizon \(\Sigma_0\). Consequently, the renormalised trace of the stress-energy tensor in a coherent state \(|\alpha\rangle\)
\[
\langle\alpha |T^\mu{}_\mu|\alpha\rangle = -\frac{1}{8\pi}\,R
\]
is supported on \(\Sigma_0\). There can not exist a stable equilibrium with any smooth, globally hyperbolic extension across \(\Sigma_0\). In this sense the semi-classical Hadamard description necessarily breaks down at the Gaussian Quantum Foam Cauchy horizon.
\end{Theorem}
\begin{proof}
For any finite \(k\), Definition~\ref{def:quantumfoam} gives that \((M_{(k)},g^{(k)}_{\mu\nu})\) is a smooth, homotopic, globally hyperbolic spacetime. It is standard that on any such spacetime there exists a minimal field algebra of the covariant Klein--Gordon equation for a scalar field, and that quasi-free states have symmetrised two-point functions \(G_k\) which are symmetric, bi-solutions in the weak sense, positive, and of local Hadamard form (see, e.g.,~\cite{kay2023} and references therein). This establishes the first part of the statement.

For the second part, we invoke the singularity analysis of Section~\ref{sec:singularities} and the discussion about the spontaneous collapse of the geon. There we showed that, in the Gaussian Quantum Foam limit, the scalar curvature becomes a distribution supported on \(\Sigma_0\), with representation \eqref{eq:GQF-R-limit} and finite renormalised coefficients \(a_0^i,a_2^i\) determined by the Gaussian algebra. Taking the trace of the Einstein equation,
\[
R_{\mu\nu} - \tfrac12 g_{\mu\nu} R = 8\pi T_{\mu\nu},
\]
yields
\[
T^\mu{}_\mu=-\frac{1}{8\pi}R,
\]
so that
\[
\langle \alpha |T^\mu{}_\mu| \alpha \rangle = -\frac{1}{8\pi}R
=-\frac{1}{8\pi}\sum_{i=0}^3\big(a_0^i\,\delta_{x^i}+a^i_2\,\delta_{x^i}^{\prime\prime}\big), 
\]
where \(a_0^i<0\) and  \(a_2^i>0\). Thus, the trace of the stress–energy at the geon, and hence on the singular set \(\Sigma_0\), is a non-vanishing, positive distribution supported on \(\Sigma_0\). Clearly, the positive back-reaction on the singular support in the admissible sense, will destabilise the wave equation in Definition~\ref{def:wave-equation}, whose geon configuration represents an unstable equilibrium when probed by the admissible test functions in Definition~\ref{def:test-functions}. In particular, the geon cannot remain in this frozen configuration: the induced back reaction necessarily drives the system away from the distributional equilibrium and into a collapsing, dynamically evolving regime.

In particular, any attempt to maintain a globally Hadamard quasi-free state across \(\Sigma_0\) would be incompatible with the distributional curvature and stress-energy obtained in the Gaussian Quantum Foam limit. Hence the Hadamard description necessarily breaks down at the Cauchy horizon of the Foam.
\end{proof}
Theorem~\ref{thm:GQF-Hadamard} is a rather technical statement, that may not resonate much outside mathematical physics. I therefore describe, in layman’s terms, what would happen if one tries to build a time machine. The reason for doing this, even in a technical text like this, is that we are dealing with questions about time machines, and hence it is reasonable to be somewhat lenient with formality, since the question of whether a time machine can be manufactured certainly has its place in folklore and science fiction.

Imagine, therefore, a future spacetime engineer fostering an idea to tamper locally with a globally hyperbolic spacetime, warping it to such an extent that self-gravitating objects—geons—begin to form. The advice to the engineer must be: do not attempt this; it is an unstable configuration—this follows directly from the field equation, since the distributional second-order term is negative near the singular support—and it will blow up. You are trying to compactly generate a Cauchy horizon of geons by warping hypersurfaces in confinement, and any matter (including quantum vacuum fluctuations) approaching the unstable confinement of the geon will cause vacuum polarisation effects (driven by back-reaction sufficient to produce an impulse of the order of the second derivative of the Dirac measure in the field equation \ref{def:wave-equation}) that are intense, since some spatial regions will expand while others contract exponentially to very large absolute values, driven by positive and negative energy densities but with a net-zero effect. Nearly naked black holes will emerge in the process as the warping attempts to unwind, until matter forms from the transient trapping (Theorem~\ref{thm:null-not-trapped}), self-gravitating massless objects constituting the geon.

This is the physics behind the theorem, that in the geon context, providing strong arguments for Hawking’s Chronology Protection Conjecture.

\section{The Preservation of the Equivalence Principle}

In these notes, a fully classical and quantum framework has been constructed for a Gaussian Quantum Foam. Using the localised Gaussian structure of the foam together with a scaling prescription, we have been able to introduce a renormalised distribution algebra that provides the means not only to handle products of distributions whose wave front sets are not disjoint in their singular supports, but also to derive a non-linear field equation in the associated distribution geometry, thereby completing the quantisation of the foam.

The construction of a quantum theory for the foam started from the observation that the definition of the Gaussian Quantum Foam naturally admits a Gelfand triple and consequently a well-defined Hilbert space, allowing us to consistently quantise the field on any level surface in the sequence of spacetimes converging to the Gaussian Quantum Foam element, or the geon, to use Wheeler's terminology \cite{wheeler1955} for the unstable self-gravitating entity in the distribution geometry. 

Within this programme, in a coherent state the expectation values coincide with their classical counterparts for each finite value of the sequence index, and hence for every smooth and globally hyperbolic spacetime element. In distribution geometry, the trace of the Ricci scalar, and hence the trace of the stress-energy tensor, is given by a linear combination of the Dirac measure and its second-order distributional derivative. Thus, the back-reaction from the quantised matter field will cause the geon, unstable in its self-gravitating frozen state at the the characteristic, where strong causality is broken, to collapse and set in motion the process leading to the emergence of globally hyperbolic classical spacetime, via inflationary curvature fluctuations driven by separate regions of positive and negative energy density. In this, likely radiation-dominated era, transient trapped surfaces will emerge and act as reheating sources. This process arises from the frozen state of the geon, where the strong energy condition is preserved in the distribution geometry but not in the emerging spacetime.

Furthermore, we have argued that Hawking's chronology protection holds in a quantum gravitational context, since the back-reaction effects from any matter will cause the geon to collapse.

We will now use this phenomenology to argue that the Gaussian Quantum Foam model naturally incorporates a mechanism by which the Equivalence Principle is preserved, and that the vacuum polarisation effects in the unstable geon will cause it to collapse. In conjunction with this, we will provide an estimate for when this happens.

This statement, based on the framework developed in this work, suggests a concrete realisation of the general idea that gravity can trigger an objective transition from a quantum to an effectively classical regime, in a way reminiscent of Penrose’s proposal of gravity-induced collapse \cite{penrose1996} (further discussed in e.g, \cite{penrose1998}), albeit via a different mechanism: here the driving ingredient is the distributional back-reaction of the self-gravitating geon rather than the instability of superpositions of distinct classical geometries. However, since we have shown that in the coherent-state model of the foam the expectation values of the quantised foam, as a bosonic field, in a coherent state \(|\alpha^i_{(k)}\rangle\) necessarily coincide with their classical counterparts, no tension arises between classical and quantum descriptions. This follows from \eqref{eq:expbeta}:
\[
\langle \alpha^i_{(k)} | \hat{\beta}_{(k)}^i| \alpha^i_{(k)} \rangle 
= \frac{1}{\sqrt{4 \sigma^2 / k^2 \pi}} 
\exp\left(-\frac{(x^i)^2}{4 \sigma^2 / k^2}\right)
= \beta^i_{(k)},
\]
and from the fact that the differential geometry and distribution geometry in the Gaussian gauge are completely determined by the triple \((N_{(k)}, \beta_{(k)}, \eta)\). As we have seen in Section~\ref{sec:review}, this statement continues to hold even if the shift vector is gauged to zero. Hence, in Gaussian Quantum Foam the geodesic of a test particle is identical whether the particle is quantum or classical. Thus, no question of causality arises. This also renders superfluous the notion of a local quantum inertial frame, as discussed by Giacomini and Brukner \cite{giacomini2022}: in each spacetime element, no geometrical superposition of states can be distinguished from a classical configuration at the level of expectation values.

Thus, in contrast to the viewpoint of \cite{giacomini2022}, where a suitable quantum generalisation of the Equivalence Principle allows spatial superpositions of massive bodies without invoking gravitational collapse, the Gaussian Quantum Foam provides an explicit and concrete dynamical realisation of a quantum-gravitational regime. In the singular regime of the foam, the distributional curvature and the associated renormalised stress-energy imply that self-gravitating geon configurations are intrinsically unstable, and the restoration of classical, globally hyperbolic spacetimes is tied to a gravity-induced spontaneous state reduction. Hence, in this setting the collapse is the very foundation of the Equivalence Principle and what makes it prevail as time goes on, via the global and regular time function that foliates each spacetime element according to the Bernal--Sánchez result \cite{bernal2005}. Thus, the restoration of the Equivalence Principle is an effect of the fact that the self-gravitating configuration of the geon is initially unstable, and that upon its collapse it evolves into emerging globally hyperbolic spacetimes; this transition is due to the collapse of the self-gravitating geon from the back reaction of spontaneous polarisation events. That this is possible at all is due to the Heisenberg uncertainty relation for time and energy, which allows the geon to emit a particle pair for some time  \(\Delta t\) that, in turn, can emit another photon or be reabsorbed by the geon.

We can easily see how this works, starting from the collapse induced by the spontaneous polarisation of the geon. As we have shown, in the Gaussian Quantum Foam limit the scalar curvature becomes a distribution supported on \(\Sigma_0\), with representation \eqref{eq:GQF-R-limit} and finite renormalised coefficients \(a_0^i,a_2^i\) determined by the Gaussian algebra. Taking the trace of the Einstein equation,
\[
R_{\mu\nu} - \tfrac12 g_{\mu\nu} R = 8\pi T_{\mu\nu},
\]
yields
\[
T^\mu{}_\mu = -\frac{1}{8\pi}R.
\]
Thus, we obtain the following expectation value, since it is identical with its classical counterpart:
\[
\langle \alpha |T^\mu{}_\mu| \alpha \rangle = -\frac{1}{8\pi}R
=-\frac{1}{8\pi}\sum_{i=0}^3\big(a_0^i\,\delta_{x^i}+a^i_2\,\delta_{x^i}^{\prime\prime}\big), 
\]
where \(a_0^i<0\) and \(a_2^i>0\). Thus, as stated in the section on the chronology protection \ref{sec:time_machines} the trace of the stress–energy at the geon, and hence on the singular set \(\Sigma_0\), is a non-vanishing, positive distribution supported on \(\Sigma_0\). Clearly, the positive back-reaction on the singular support in the admissible sense, will destabilise the wave equation in Definition~\ref{def:wave-equation}, whose geon configuration represents an unstable equilibrium when probed by the admissible test functions in Definition~\ref{def:test-functions}. In particular, the geon cannot remain in this frozen configuration: the induced back-reaction necessarily drives the system away from the distributional equilibrium and into a collapsing, dynamically evolving regime.

Having argued for a collapse of the geon through its inherent polarisation properties, we can estimate the timescale on which this collapse takes place using the uncertainty relation
\[
\Delta E\,\Delta t \gtrsim \frac{\hbar}{2},
\]
and the fact that we have previously shown, in the construction of the algebra and in Section~\ref{sec:algebraoperations2}, and in relation to the wave equation in Section \ref{sec:waveequation} as well as in the discussion about the Ricci scalar in Section \ref {sec:singularities}, that a characteristic mass or energy scale \(\Delta E \simeq |\Delta m_{B\pm}^{(k)}|c^2\) is associated with each spacetime element \((M_{(k)},g^{(k)}_{\mu\nu})\). In the limit \(k \to \infty\), the quantities \(m_{B\pm}^{(k)}\) diverge and are interpreted as bare mass (parameters). For any finite value of \(k\), by contrast, they describe an observable mass configuration for the geon field equation, in an absolute sense, as given by \eqref{eq:mbplus} and \eqref{eq:mbminus}, namely
\begin{equation}
  m^{(k)}_{B\pm} \;=\; \pm\,\frac{\sqrt{3}}{36\sigma^4\pi}\,k^2.
\end{equation}
Using this, we find that a spontaneous back reaction on the self-gravitating configuration takes place on a timescale of order
\begin{equation}
\Delta t \sim \frac{\hbar}{2\,|\Delta m_{B\pm}^{(k)}|}.
\end{equation}
This causes the oscillation of the shift vector and thus, as described earlier, displaces the vacuum so that regions of positive and negative energy density form, driving fluctuations in the extrinsic curvature, its trace, and the local Hubble parameter, thereby triggering inflationary expansion together with transient trapped surfaces and the emergence of classical spacetime.

In conclusion, this shows that there is no inherent conflict for quantum gravity in the representation of a Gaussian Quantum Foam: while, as we have seen in the previous section, the geon is not strongly causal, its instability will necessarily cause it to collapse under its own polarisation effects, and the Equivalence Principle is preserved in the emerging globally hyperbolic spacetimes. Moreover, the geometrical structure in the coherent-state representation of the foam, where the expectation values of any field necessarily coincide with their classical counterparts, shows that there is no need to introduce any notion of quantum local inertial frames as argued by Giacomini and Brukner \cite{giacomini2022} in order to reconcile quantum superpositions with local free fall in this framework. In this sense, the picture developed here is conceptually close to Penrose’s view that gravity may play an active role in state reduction \cite{penrose1996}, although the mechanism is different, since the collapse is driven by the distributional back-reaction of an unstable self-gravitating geon rather than by the ill-definedness of time evolution for superposed classical spacetimes. 

Nevertheless, within the present framework we have implicitly argued that the \emph{gravity-related facet} of the measurement problem is resolved: a wide class of classical spacetimes emerges from an unstable, self-gravitating quantum configuration without any external observers, provided the Heisenberg energy–time uncertainty relation is taken as fundamental in the Quantum Foam regime. In this sense, \emph{gravity is gravitating into a wide class of classical and globally hyperbolic spacetimes}, always respecting Galileo's and Einstein's heritage.

\section{Conclusion}

In the exploratory phase of developing the notion of Gaussian Quantum Foam, necessarily marked by tentative steps and an incomplete mapping of the terrain, the question of the underlying field equation for the shift vector field, both in its classical and quantised formulations, was left open \cite{cramer20251,cramer20252}. Here we have shown that, by developing an appropriate distribution geometry, a non-linear wave equation for the shift vector field can be formulated on the singular support of that geometry.

In the classical framework, the shift vector serves as the geometric imprint of an operator-valued distribution, or more precisely, as the classical remnant of such an operator. It governs the displacement of spatial coordinates across Cauchy hypersurfaces in a sequence of homotopic, globally hyperbolic spacetimes that converge, in the sense of distributions, to Quantum Foam. This purely geometric mechanism gives rise to inflation without invoking an external inflaton field. Instead, inflation emerges from vacuum fluctuations involving both positive and negative energy densities, while preserving overall energy neutrality.  

One might object that, since the shift vector can be set to zero by an appropriate foliation-preserving diffeomorphism in the \(3{+}1\) decomposition of spacetime, it cannot be assigned any elevated physical role. Nevertheless, if the shift vector is gauged away in Quantum Foam, the physical effect remains. Heuristically this must be the case: if the shift vector vanishes everywhere while the lapse is kept unchanged, and if the corresponding sequence of spacetimes converges in the distributional sense to a Gaussian Quantum Foam element, then the induced metric necessarily evolves so as to deform any spatial grid between successive hypersurfaces. In this gauge, the remnant of Quantum Foam in emerging classical spacetimes is expressed through the contraction and expansion of spatial volume elements---and hence of any spatial grid---across the hypersurfaces. This is equivalent to the original Gaussian Quantum Foam description, where the same effect arises from the shifting of coordinates across hypersurfaces. In both formulations, all observable scalars remain identical.  

To make this statement precise, we have shown that, given a sequence of globally hyperbolic spacetimes converging to a Gaussian Quantum Foam element, one can gauge the shift vector to zero for such an element by a foliation-preserving relative-velocity diffeomorphism on the whole spacetime minus the singular support. Under this diffeomorphism, the evolution of the spatial coordinates remains fixed and aligned with the vacuum, and hence with the singular support, as time evolves. However, the dependence on the shift vector in scalar quantities such as the trace of the extrinsic curvature is preserved in the \(3{+}1\) decomposition. In particular, the stretching and contraction of spatial grids remain effects of the shift vector. This conclusion holds both for each classical spacetime element in the sequence and, via coherent states in the bosonic quantisation of the shift vector, for the quantum field description of Quantum Foam.

This observation highlights that the shift vector, far from being a dispensable artefact, captures genuine effects both in classical differential geometry and in quantum distribution geometry. The latter follows from the fact that the classical construction naturally admits a Gelfand triple, enabling the quantisation of the shift vector as a bosonic field. In this quantum setting, the shift vector takes on the role of a vacuum-displacement operator-valued distribution, and classical spacetime emerges in the limit as a coherent superposition of quantum excitations.

The identification of a non-linear wave equation that provides dynamical content to the shift vector field \(\beta_{(k)}\) completes the Quantum Foam model. Owing to its non-linear structure, this equation requires a distributional framework in which non-linear operations are well defined. To construct the wave equation, a localised algebra of distributions was developed, based on sequences of scaled Gaussian functions within a restriction to the Schwartz space. These sequences converge to Dirac measures and, critically, their products remain within the same algebraic structure. All non-linear operations, including multiplication and differentiation, are performed at the level of smooth representatives before taking the distributional limit.

Within this framework, the wave equation for \(\beta_{(k)}^{i}\) was shown to be driven by a linear combination of the Dirac measure, \(\delta_{x^i}\), and its second-order derivative, \(\delta^{\prime\prime}_{x^i}\). The second-order distributional derivative, \(\delta^{\prime\prime}_{x^i}\), is of pivotal importance for the emergence of spacetime. This term serves as more than a singular curvature source: it functions as a catalytic impulse that sets the shift vector into oscillatory motion. In this singular limit, the impulse displaces the vacuum itself, triggering inflation through a sharply localised redistribution of energy--momentum, consistent with Wheeler’s notion of vacuum fluctuations~\cite{wheeler1955, mtw1973, wheeler1981}.

As the wave equation evolves, the resulting oscillations of  \(\beta_{(k)}^{i}\) stretch and deform the hypersurfaces, giving rise to curvature through distributional dynamics. The residual imprint of the initial impulse is a persistent geometrical shift across hypersurfaces---a displacement field that seeds classical spacetime from the quantum vacuum. In this way,  \(\delta^{\prime\prime}_{x^i}\) acts as the singular origin of the Quantum Foam: a mathematical and physical spark from which spacetime itself emerges.

Notably, this framework offers a concrete realisation of Wheeler’s assertion that \emph{“Time is not a primordial and precise concept; it must be secondary, derivative, and approximate”}~\cite{wheeler1981}. Since a smooth and regular time function exists if and only if the spacetime is globally hyperbolic~\cite{bernal2005}, and since global hyperbolicity in the present model arises only once the shift vector is set in motion by the distributional curvature impulse, it follows that time itself emerges as a consequence of quantum geometric excitation---not as a background parameter, but as a response to vacuum displacement.

In relation to the discussion of the origin of time, it is necessary to address the status of singularities. If the notion of a singularity could be made precise within a Quantum Foam model, it could not be regarded as fundamental, as the model would break down at that point. Nor could such a singularity be the source of time, as argued here. However, the Gaussian Quantum Foam model is well defined, and the notions of continuity and differentiability remain valid for all geometric quantities, even at Planck scales, meaning that the classical notion of a singularity does not retain operational meaning. Formally, this is true because the concept of a physical body is here taken to be the distributional limit of a sequence of globally hyperbolic and homotopic spacetimes. Each element of the sequence is endowed with smooth geometry and well-defined dynamics. The resulting Gaussian Quantum Foam element inherits a sharply localised but well-defined distributional structure. In this setting, curvature remains finite in the distributional sense, and the singular support serves not as an indicator of geometric breakdown but as a precise localisation of field content. 

Nevertheless, we have shown that at the singular support, the scalar projection of the Ricci curvature is non-negative for all finite \(k\), and in the distributional limit it converges to the negative of a linear combination of a Dirac measure and its second-order derivative where the coefficient for the measure is negative while positive for the second-order derivative. Away from the singular support, however, there necessarily exist open regions where the strong energy condition is locally violated. These regions are not problematic---they are required. It is precisely this structure of sign-changing curvature, governed by the non-linear dynamics of the shift vector, that enables the emergence of spacetime itself from the quantum foam. To strengthen this argument, it has also been shown that the null expansions vanish at the singular support. This suggests that the spacetime is in a frozen configuration with null level surfaces and zero slowness in the distribution geometry of Gaussian Quantum Foam. However, similarly to the scalar projection of the Ricci curvature, for finite values of \(k\) there exist open regions where both the inward and outward null expansions are strictly negative. This indicates that local trapped surfaces can form---effectively acting as transient heat sinks---which might be a source of the reheating phase in the early evolution of spacetime.

The limiting structure at the characteristic set, and consequently in the distribution geometry, which we interchangeably have referred to as a Gaussian Quantum Foam element or a geon, represents a self-gravitating but unstable configuration in a non-strongly causal context. In this context we have shown that a compactly generated Cauchy horizon will arise and that, due to back reaction from any matter field, the geon will collapse under any attempt to create a region of closed timelike curves. This provides support for the hitherto open question of whether Hawking's chronology protection conjecture holds in quantum gravity.

The phenomenology that causes the geon to collapse has been used to argue that the Gaussian Quantum Foam model naturally incorporates a mechanism by which the Equivalence Principle is preserved, and that vacuum polarisation effects in the unstable geon will cause it to collapse. This suggests a concrete realisation of the general idea that gravity can trigger an objective transition from a quantum to an effectively classical regime, in a way reminiscent of Penrose’s proposal of gravity-induced collapse~\cite{penrose1996}, albeit via a different mechanism: here the driving ingredient is the distributional back-reaction of the self-gravitating geon rather than the instability of superpositions of distinct classical geometries.

In conclusion, \emph{gravity is gravitating into a wide class of classical and globally hyperbolic spacetimes}, always respecting Galileo's and Einstein's heritage.

\vspace{0.5cm}
\noindent
In summary, a coherent framework for Quantum Foam has been proposed in which the classical notion of a singularity has no operational meaning and is replaced by the notion of a \emph{singular support}, where continuity and differentiability remain well defined in the distributional sense. The central idea is that coordinate displacement---diffeomorphically invariant in the sense that its effects persist even if the shift vector is gauged to zero---encodes the residual imprint of a Quantum Foam element. These effects are manifested as the stretching and contraction of spatial grids across hypersurfaces in globally hyperbolic spacetimes. The underlying driver is the sharply localised curvature impulse (a distributional second derivative of the Dirac measure), which displaces the vacuum and initiates oscillatory dynamics, thereby endowing time with operational meaning and restoring the Equivalence Principle. This mechanism, rooted in the inherent polarisation of the unstable geon, triggers inflation and the subsequent emergence of classical spacetime. While grounded in a rigorous mathematical construction and physical reasoning, these claims represent a significant departure from conventional approaches and will rightly be subject to scrutiny by the broader research community.
\enlargethispage{20pt}

\section*{Acknowledgement}

I am deeply grateful for the unwavering love and support of my family—Gustav, Noel, Alice, and Caroline—whose encouragement has been invaluable throughout this quest. It has been a path marked by moments of insight, but also by challenges of circular reasoning that often led into seemingly endless rabbit holes.
\newline
\newline
\noindent
I am also thankful to the anonymous referees who, in an earlier submission when much work still remained, maintained a highly professional manner, adopted a positive tone, and elaborated on both major and minor issues while suggesting constructive ways to address them. To the best of my knowledge, these have now been addressed within the scope of this project, providing a path forward in the long-standing quest to understand the foundations of gravity, at least in a minimal sense.
\newline
\newline
\noindent
Many thanks to Viggo Kraft, who has patiently spent a considerable amount of his valuable time reading my many mail and listening to me while I presented the technicalities in front of a whiteboard, always providing thoughtful remarks that forced me to be more rigorous than I might otherwise have been.
\newline
\newline 
\noindent
Finally, my unlimited gratitude goes to Bernard Kay for his endless patience in receiving my many e-mails with perceived Eureka moments, which invariably required further work and, in many cases, rightly belonged to the archive of wild explorations into the unknown. Most importantly, Bernard, thank you for your encouragement and for generously giving your time, even when I was acutely self-absorbed with this work.

\section*{Funding}

This research did not receive any specific grant from funding agencies in the public, commercial, or not-for-profit sectors.

\section*{Declaration of Generative AI and AI-Assisted Technologies in the Writing Process}

During the preparation of this work—based on notes and ideas dating back to 1994—the author utilised OpenAI’s language model to assist with spelling, grammar, notation, flow and propositional analysis. All content generated with the aid of this tool was reviewed and edited by the author, who takes full responsibility for the final publication.

\RaggedRight

\end{document}